\documentclass[11pt,dvipsnames]{article}

\usepackage{setspace}
\usepackage{jheppub} 
\usepackage[utf8]{inputenc}
\usepackage{epsfig}
\usepackage{bbm}
\usepackage{bbding}
\usepackage{amsfonts}
\usepackage{amssymb}
\usepackage{mathtools}
\usepackage{dsfont}
\usepackage{bm}
\usepackage{graphicx}
\usepackage{longtable}
\usepackage{pdflscape}
\usepackage{subcaption}
\usepackage{psfrag}
\usepackage[capitalise]{cleveref}
\usepackage{tikz}
\usepackage{pgfplots}
\usepackage{pgfplotstable}
\usepackage{subfiles}
\usepackage{longtable}
\usepackage{tabularx}
\usepackage{ltablex}
\usepackage{booktabs}
\usepackage[export]{adjustbox}
\usepackage{listings}
\usepackage{multirow} 
\usepackage{cancel,slashed}
\usepackage{bbm}
\usepackage{graphicx}
\usepackage{latexsym}
\usepackage{transparent}
\usepackage{mathtools}
\usepackage{array}
\usepackage{makecell}
\usepackage{graphics,psfrag}
\usepackage{placeins}
\usepackage{nowidow}
\usepackage{listings}
\usepackage[normalem]{ulem}
\usepackage{environ}
\usepackage{tikz}
\usepackage{enumitem}   
\usetikzlibrary{positioning,arrows.meta}
\usetikzlibrary{positioning,trees,decorations.pathmorphing,decorations.markings,decorations.pathreplacing,calc,shapes,shapes.geometric,shapes.symbols,patterns,arrows}

\usetikzlibrary{fadings}
\usetikzlibrary{decorations.shapes}
\usepackage{amsmath}
\usepackage{amssymb}
\usepackage{amsthm}
\usepackage{algorithm}
\usepackage{algpseudocode}
\usepackage{placeins}

\newcommand{\newshortstack}[1]
{\begingroup\renewcommand{\arraystretch}{1.1}
\ifmmode
\begin{array}{c}#1\end{array}%
\else
\begin{tabular}{c}#1\end{tabular}%
\fi
\endgroup}

\pgfplotsset{
        compat=1.9,
        compat/bar nodes=1.8,
    }

\theoremstyle{definition}

\newcommand{\be}{\begin{equation}}
	\newcommand{\ee}{\end{equation}}
\newcommand{\bea}{\begin{eqnarray}}
	\newcommand{\eea}{\end{eqnarray}}

\renewcommand{\epsilon}{\varepsilon}

\newcommand{\ben}{\begin{enumerate}}
	\newcommand{\een}{\end{enumerate}}
\newcommand{\bei}{\begin{itemize}}
	\newcommand{\eei}{\end{itemize}}
	
	\makeatletter
\tikzset{
    dot diameter/.store in=\dot@diameter,
    dot diameter=3pt,
    dot spacing/.store in=\dot@spacing,
    dot spacing=10pt,
    dots/.style={
        line width=\dot@diameter,
        line cap=round,
        dash pattern=on 0pt off \dot@spacing
    }
}

\usepackage{setspace}
\usepackage{graphicx} 
\usepackage{lmodern}
\usepackage{xspace}
\usepackage{array} 
\usepackage{etoolbox}
\usetikzlibrary{positioning,trees,decorations.pathmorphing,decorations.markings,decorations.pathreplacing,calc,shapes,shapes.geometric,shapes.symbols,patterns,arrows}
	
\tikzset{decorate sep/.style 2 args=
{decorate,decoration={shape backgrounds,shape=circle,shape size=#1,shape sep=#2}}}
\usepackage{tikz}
\usepackage{latexsym}
\usepackage{graphicx}
\usepackage{svg}
\usepackage{tikz-3dplot}
\usetikzlibrary{arrows.meta}
\usetikzlibrary{patterns}

\definecolor{bluetto}{HTML}{0088ff}
\definecolor{snowfl}{HTML}{00ace6}
\definecolor{dgreen}{HTML}{298A08}
\newtheorem*{theorem*}{Theorem}

\def\mpl{M_{\mathrm{pl}}}

\makeatletter
\renewcommand*\env@matrix[1][\arraystretch]{%
  \edef\arraystretch{#1}%
  \hskip -\arraycolsep
  \let\@ifnextchar\new@ifnextchar
  \array{*\c@MaxMatrixCols c}}
\makeatother

\usepackage{hyperref}

\begin{document}

	\pagestyle{plain}

	\makeatletter
	\@addtoreset{equation}{section}
	\makeatother
	\renewcommand{\theequation}{\thesection.\arabic{equation}}
	\pagestyle{empty}

\vspace{2cm}

\begin{flushright}
KCL-PH-TH/2024-75
\end{flushright}

\begin{center}
\phantom{a}\\
\vspace{0.8cm}
\scalebox{0.90}[0.90]{{\fontsize{24}{30} \bf{Fuzzy Axions and Associated Relics}}}\\
\end{center}

\vspace{0.4cm}
\begin{center}
\scalebox{0.95}[0.95]{{\fontsize{12}{30}\selectfont  
Elijah Sheridan,$^{a}$
Federico Carta,$^{b,c}$
Naomi Gendler,$^{d}$
Mudit Jain,$^{c}$
David J.~E.~Marsh,$^{c}$
}}\\
\scalebox{0.95}[0.95]{{\fontsize{12}{30}\selectfont  
Liam McAllister,$^{a}$
Nicole Righi,$^{c}$
Keir K. Rogers,$^{e}$ and
Andreas Schachner$^{a,f}$
}} 
\end{center}

\begin{center}
\vspace{0.25 cm}

\textsl{$^{a}$Department of Physics, Cornell University, Ithaca, NY 14853, USA}\\
\textsl{$^{b}$London Institute for Mathematical Sciences, Royal Institution, W1S 4BS, UK}\\ 
\textsl{$^{c}$Physics Department, King's College London, Strand, London WC2R 2LS, UK}\\
\textsl{$^{d}$Jefferson Physical Laboratory, Harvard University, Cambridge, MA 02138, USA}\\   
\textsl{$^{e}$Dunlap Institute, University of Toronto, Toronto, ON M5S 3H4, Canada}\\ 
\textsl{$^{f}$ASC for Theoretical Physics, LMU Munich, 80333 Munich, Germany}\\

  \vspace{1.1cm}
	\normalsize{\bf Abstract} \\[8mm]
\end{center}

\begin{center}
	\begin{minipage}[h]{15.0cm}

We study fuzzy axion dark matter in type IIB string theory, for axions descending from the Ramond-Ramond four-form in compactifications on orientifolds of Calabi-Yau hypersurfaces.
Such models can be tested by cosmological  
measurements if a significant relic abundance  of fuzzy dark matter 
arises, which  
we argue is 
most common
in models with small numbers of axions. 
We construct a topologically exhaustive ensemble of more than 
350{,}000 Calabi-Yau compactifications yielding up to seven axions,
and in this setting we perform a 
systematic analysis of misalignment production of fuzzy dark matter. In typical regions of moduli space, the fuzzy axion, the QCD axion, and other axions have comparable decay constants of $f_a\approx 10^{16}\text{ GeV}$.
We find that overproduction of heavier axions
is problematic, except at special loci in moduli space
where decay constant hierarchies can occur: 
without a contrived reheating epoch, it is necessary to fine-tune initial displacements.
The resulting dark matter is typically a mix of fuzzy axions and heavier axions, including the QCD axion.
Dark photons are typically present as a consequence of the orientifold projection. 
We examine the signatures of these models by simulating halos with multiple fuzzy axions, and by computing new cosmological constraints on ultralight axions and dark radiation.
We also give evidence that cosmic birefringence is possible in this setting.
Our findings determine the phenomenological  
correlates
of fuzzy axion dark matter in a corner of the landscape.

\end{minipage}
\end{center}

	\newpage

	\setcounter{page}{1}
	\pagestyle{plain}
	\setcounter{footnote}{0}
\tableofcontents
        \newpage
\section{Introduction}

Dark matter (DM) is one of the most profound mysteries of the cosmos. The astrophysical and cosmological evidence for DM shows that there is new fundamental physics beyond the Standard Model (SM) and general relativity~\cite{Marsh:2024ury}, and thus motivates the study of string theory, the leading candidate for a theory of quantum gravity and the fundamental forces.

Cosmological observations measure the DM relic density, $\rho_{\rm DM}$, to percent-level accuracy in the standard $\Lambda\mathrm{CDM}$  
cosmological model~\cite{Planck:2018vyg},
\begin{equation}
\Omega_{\rm DM} h^2 = \rho_{\rm DM}/(8.182 \cdot 10^{-11})\text{ eV}^4=0.1200\pm0.0012\,,  
\end{equation}  
and precision cosmology places limits on the DM   mass and   abundance for a wide range of models of DM~\cite{Amendola:2016saw,LSSTDarkMatterGroup:2019mwo,Dvorkin:2022bsc}.

Axions are one of the leading candidates to make up some or all of the DM.
A non-thermal population of axions can be produced in the early Universe by vacuum realignment~\cite{Preskill:1982cy,Abbott:1982af,Dine:1982ah,Turner:1983he}, in which the axion field is initially displaced from its potential minimum by some angle $\theta_a$,  
and undergoes damped harmonic oscillations when the Hubble rate drops below the axion mass $m_a$. 

An axion with $10^{-33}\text{ eV} \lesssim m_a \lesssim 10^{-18}\,\mathrm{eV}$ has effects
on DM structure formation that  
can be distinguished 
from those of standard cold DM (CDM) using current and near-future telescopes~\cite{LSSTDarkMatterGroup:2019mwo,Hotinli:2021vxg}: in particular, ultralight axion DM suppresses the formation of small-scale structures. 
Cosmological observations constrain axion DM lighter than $10^{-21}\text{ eV}$ to make up at most 
$\mathcal{O}(1-10)\%$ of the total DM abundance,  
depending on the mass. The higher mass range $10^{-21}\text{ eV}\lesssim m_a\lesssim 10^{-18}\text{ eV}$ is accessible to some astrophysical observables and to future cosmological measurements. 
Following \cite{Hu:2000ke}, we will call an axion that has $10^{-33}\text{ eV} \lesssim m_a \lesssim 10^{-18}\,\mathrm{eV}$,
and that makes up a significant part of the  
total DM abundance, 
a \emph{fuzzy axion}.\footnote{Including the relic abundance in the definition of fuzzy axions distinguishes them from the multitude of ultralight axions in string theory whose relic abundances are undetectably small.}

The fuzzy axion
relic abundance  
cannot be diluted by entropy production:
fuzzy axions begin to oscillate at temperatures below  
$1\text{ MeV}$,
and
the thermal history of the Universe is strongly constrained for such low temperatures.
The relic density of 
an axion with decay constant $f_a$ 
and mass $10^{-28}\text{ eV}\lesssim m_a \lesssim 10^{-15}\text{ eV}$ is
\begin{align}
   \label{eqn:fuzzy_relic_abundance}
   \Omega_{a}h^2 \approx 0.12\,\theta_a^2\left(\frac{m_a}{4.4 \cdot 10^{-19}\,{\rm eV}}\right)^{1/2}\left(\frac{f_a}{10^{16}\,{\rm GeV}}\right)^2\,.
\end{align}
Thus, if a theory possesses light axions with large $f_a$, 
they may --- for sufficiently large values of the parameters $m_a$, $f_a$, $\theta_a$ in \eqref{eqn:fuzzy_relic_abundance} --- constitute a significant fraction of the DM, and thus meet our definition of a fuzzy axion. 
A detailed discussion of the relic abundance and cosmology is given in Section~\ref{sec:cosmosetup}.

The low energy effective field theories (EFTs) describing four-dimensional solutions of string theory generically contain a number of light axion fields~\cite{Witten:1984dg,Svrcek:2006yi,Conlon:2006tq,Arvanitaki:2009fg,Cicoli:2012sz}, which
provide avenues for constraining such theories observationally. 
In this work we investigate fuzzy axion DM in type IIB superstring theory.
We search for compactifications in which the fuzzy DM abundance is near the observational limits, characterize 
the phenomenology of the resulting models, and    
discuss how current and future cosmological observations can be used to constrain these theories.

The setting for our work is the \emph{Kreuzer-Skarke (KS) Axiverse}~\cite{Kreuzer:2000xy,Demirtas:2018akl}: axion theories arising in compactifications of type IIB string theory on Calabi-Yau (CY) orientifold hypersurfaces in toric varieties obtained from triangulations of four-dimensional reflexive polytopes~\cite{batyrev1993dual}. 
In carrying out this work, we incorporate some new capabilities in the computational geometry of string compactifications. We make use of the public package \textsc{CYTools}~\cite{Demirtas:2022hqf}, and we employ the methods developed by Moritz in \cite{Moritz:2023jdb} for constructing explicit orientifolds.

The compactifications we study can be partially characterized by the Hodge number $h^{1,1}$, 
which counts the number of axions from Ramond-Ramond forms;\footnote{See Section~\ref{sec:CYandOrientifold} for details of our choice of orientifold action. We will restrict to orientifolds with $h^{1,1}_{-}=0$, which is the generic case in our setting, and in such cases $h^{1,1}=h^{1,1}_{+}$ counts the number of axions from the Ramond-Ramond four-form.} by the expectation values of the moduli determining the size and shape of the CY; and by certain discrete choices related to how the Standard Model is realized, as we discuss in detail below. Importantly, we will restrict attention to the \emph{geometric regime}, i.e.,~the region of moduli space where the $\alpha'$ expansion is well-controlled.

\begin{figure}
    \centering
    \includegraphics[width=\textwidth]{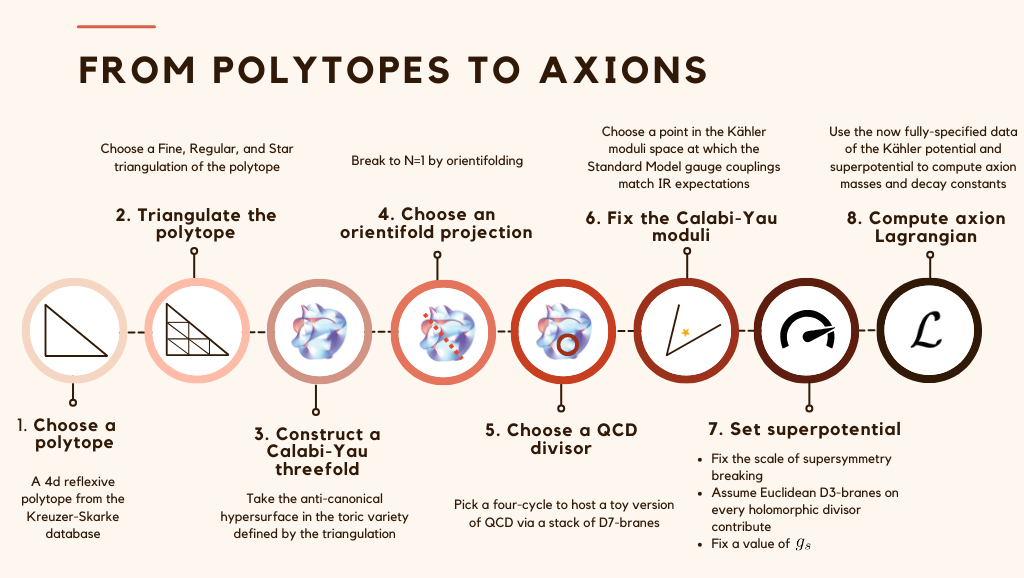}
    \caption{A representation of the pipeline that takes as input the points of a 4-dimensional reflexive polytope, and outputs the effective theory of axions in a given compactification of type IIB string theory. }
    \label{fig:cartoon}
\end{figure}

Prior work in this regime \cite{Demirtas:2018akl,Mehta:2021pwf} has found that axions with   
$f_a \gtrsim 10^{15}\text{ GeV}$,
and thus with 
potentially substantial $\Omega_a h^2$ by \eqref{eqn:fuzzy_relic_abundance}, are very rare unless  $h^{1,1}\lesssim 10$. Thus, in type IIB CY hypersurface compactifications in the geometric regime, models probed by cosmological constraints on fuzzy DM are rare, except in the corner of the landscape where $h^{1,1}\lesssim 10$. 
This corner is the 
subject of the present investigation, in which  
we characterize  
the fuzzy axiverse 
for all orientifolds that are induced on the Calabi-Yau by a holomorphic and isometric involution of the ambient toric variety, for all triangulations of polytopes in the Kreuzer-Skarke list with $2\le h^{1,1} \le 7$.\footnote{We require $h^{1,1} \ge 2$ so that there can be both a fuzzy axion and a QCD axion.}
We also construct and study a non-exhaustive sample at higher $h^{1,1}$: see Fig.~\ref{fig:h11_spread}.
The overall process is visualized in Fig.~\ref{fig:cartoon}, and a detailed explanation is given in \cref{sec:stsetup}. 

One of our main findings is that an observably-large fuzzy dark matter relic density is possible in this part of the landscape, but --- in the regions of moduli space that are generic in the Weil-Petersson  measure --- this abundance is typically accompanied by an \emph{overabundance} of heavier axion dark matter. This finding can be understood from the typical structure of axion spectra in CY compactifications, which spread over many decades in mass, but have more tightly clustered decay constants. Thus, having a fuzzy axion with a decay constant $f_{\text{fuzzy}}$ large enough to result in a large abundance is typically accompanied by other axions with similar decay constants $f_{\text{other}} \sim f_{\text{fuzzy}}$ and larger masses, leading --- barring some remedy --- to overproduction (see also \cite{Arvanitaki:2009fg}).  We will describe two very different 
remedies: one that is geometric and one that is cosmological.

The geometric solution is to find configurations with hierarchical decay constants, with $f_{\text{other}} \ll f_{\text{fuzzy}}$.  For certain compactification topologies, such hierarchies do arise, but only
in special regions of moduli space, i.e.~regions that are disfavored by the  Weil-Petersson  measure.  We review the mechanism in
Section \ref{sec:fiber} and give an explicit example in Section \ref{ex:fiber}. 
We should stress that it is not our goal here to make any claim about the relative prevalence of hierarchical versus non-hierarchical decay constants in a dynamically-populated landscape of CY compactifications of type IIB string theory.  
We lack a solution to the measure problem of eternal inflation, and we have not accounted for selection effects that may result from the dynamics of moduli stabilization, which is not understood in full generality.  Nonetheless, we find it useful to make `kinematic' statements about relative prevalence at the level of the natural (Weil-Petersson) measure on the moduli space, and it is in this very restricted sense that hierarchical decay constants are non-generic for $h^{1,1} \lesssim 10$.  For the remainder of this work, we will use the words `generic' and `non-generic' in the senses just defined.

For axion theories arising at generic points in moduli space, a cosmological remedy for overproduction is necessary.
To avoid overproduction of axions that are stable over cosmological timescales, one needs to reduce their initial misalignment angles \cite{Fox:2004kb,Arvanitaki:2009fg,Marsh:2015xka}, and/or dilute the resulting dark matter by  
making the reheating temperature sufficiently small together with appropriate engineering of the prior cosmology. However, stable axions with $m_a \lesssim 10^{-14}$ eV cannot be diluted in this way, because the reheating temperature would have to be below $5$ MeV, conflicting with Big Bang nucleosynthesis (BBN). To reduce the abundance of such axions, the only 
apparent
option is to reduce their initial angles.
These facts lead us to investigate the restrictions on the reheating temperature, and the level of misalignment angle tuning in fuzzy axion models, as explored in Section~\ref{sec:stats}.  

\begin{figure}
    \centering
    \includegraphics[width=0.9\linewidth]{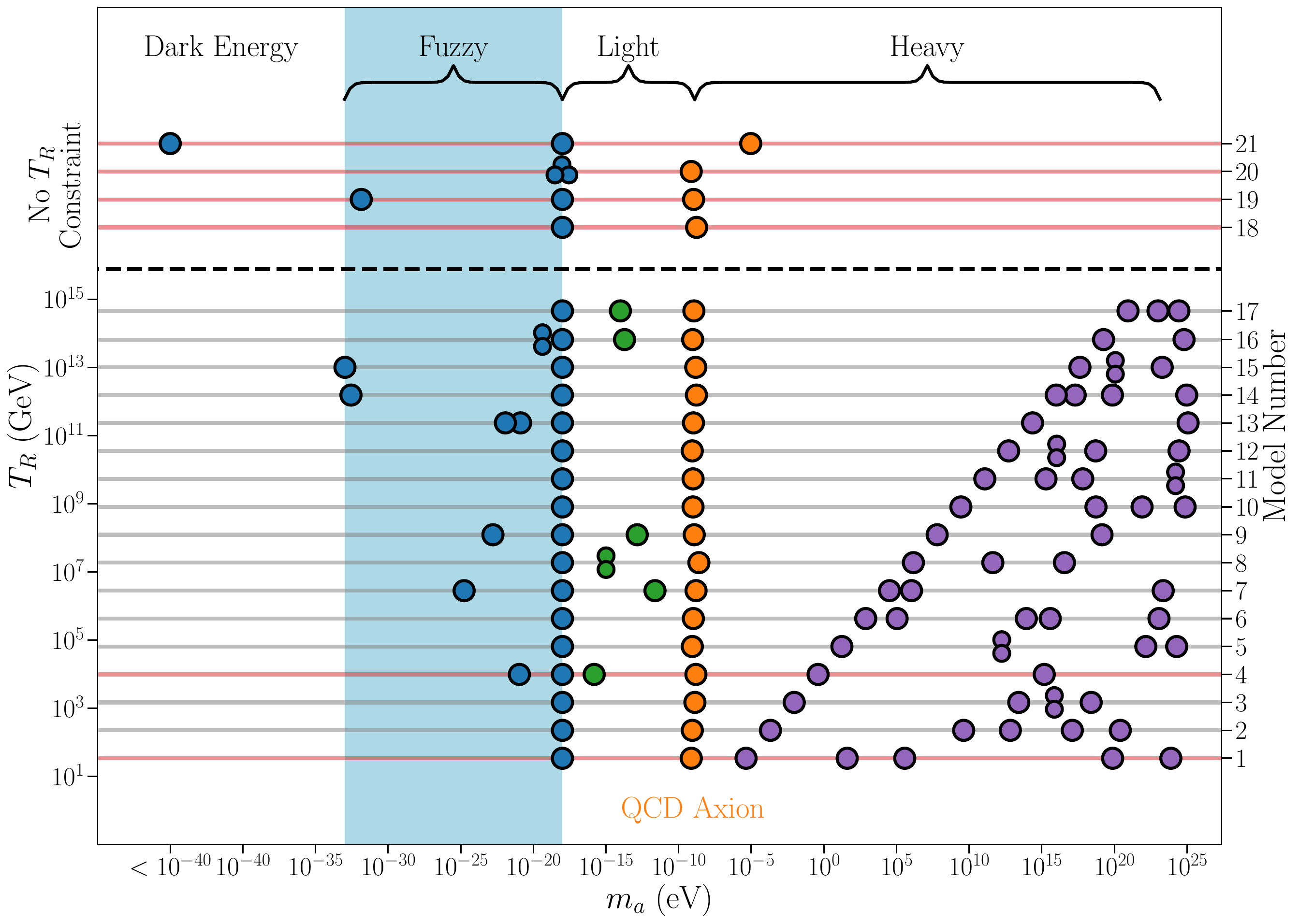} 
    \caption{Axion spectra for our six flagship example models (pink, discussed in detail in
    Section \ref{sec:examples}, K\"ahler parameters normalized here to have $m_\mathrm{fuzzy} = 10^{-18}$ eV) 
    along with prototypical large-abundance models from our ensemble (gray, see Section~\ref{sec:const_ensemble}  
    for details regarding the ensemble and 
    Section~\ref{sec:stats}  
    for discussion of the large-abundance subset). Cosmological parameters are chosen according to the prompt reheating scenario (see Section~\ref{se:choosecosmology}; see Figs.~\ref{fig:prompt_reheating} and \ref{fig:moduli_domination} for detailed statistics on the two reheating scenarios considered in this paper). The correlation $T_R \propto \sqrt{m_{a'}}$, for $m_{a'}$ the lightest heavy axion, emerges because $T_R$ is defined by $3H_R = m_{a'}$ and $H_R \propto T_R^2$.
    Model 21 (see Section~\ref{ex:fiber})
    has a different QCD axion mass than the rest of the ensemble since we used a special locus in K\"{a}hler moduli space to achieve hierarchical decay constants. 
    }
    \label{fig:golden_w=-1}
\end{figure}

A visual guide to how we classify the spectrum of axions in 21
different models is shown in Fig.~\ref{fig:golden_w=-1}. We distinguish the following classes of axions. \textit{Heavy} axions are those that are heavier than the QCD axion: these axions can have problematic abundances. The figure illustrates the reheating temperature needed to dilute all such heavy axions, and models were chosen to equally space this on a logarithmic scale.  \textit{Light} axions are those that are lighter than the QCD axion but heavier than the fuzzy axions. Since we require $T_R > 5 \text{ MeV}$ for successful BBN, these axions cannot be inflated away or diluted by entropy production, yet they can provide a too-large DM abundance. Thus, light axions necessarily imply tuning in the initial angles. Finally we classify axions that only contribute to the
dark energy density, i.e.~those with
$m_a<H_0\approx 10^{-33}\text{ eV}$.

This work explores the following features of the fuzzy axiverse:  
\begin{itemize}
    \item We compute the misalignment abundance of all axions in the spectrum and investigate the necessary reheating temperature, reheating equation of state, and tuning of initial misalignment angles to avoid DM overabundance.  
    \item We find models with multiple 
    fuzzy axions, and we simulate in full three dimensions a DM halo with three ultralight axions.
    \item From the statistics of the orientifold projection, we find that dark photon fields are generically present in the KS axiverse, and are more abundant in the theories with small numbers of axions --- the same class of theories that admit a significant relic abundance of fuzzy axion dark matter. 
    \item We examine in our ensemble the necessary conditions on the realization of the SM gauge sector for ultralight axions to  explain observational hints for cosmic birefringence~\cite{Minami:2020odp,Diego-Palazuelos:2022dsq}.
    \item We compute for the first time joint constraints on ultralight axions and dark radiation from a combination of cosmic microwave background and galaxy survey data.
\end{itemize}
 
Our findings about fuzzy DM in the string axiverse accord with observations originally made in 
\cite{Arvanitaki:2009fg}, but with some important qualifiers.
The axiverse envisioned in \cite{Arvanitaki:2009fg} had $N \gg 1$ axions with masses widely distributed on a logarithmic scale, but with decay constants $f_a$ clustered around a high scale $\langle f\rangle \sim 10^{16}\text{ GeV}$.
The first instantiation of a large axiverse from explicit string compactifications appeared in \cite{Demirtas:2018akl},
where it was shown that  
for compactifications in the geometric regime,
the mean decay constant $\langle f\rangle$ \emph{decreases} with $N = h^{1,1}$.  This trend is a consequence of the increasingly complex topology and geometry at large $h^{1,1}$, for which the condition that no cycle is small turns out to imply that some cycles are very large \cite{Demirtas:2018akl}.\footnote{This geometric result impacts the outlook for questions in axion physics including superradiance \cite{Mehta:2021pwf}, the strong CP problem \cite{Demirtas:2021gsq}, axion-photon couplings \cite{Gendler:2023kjt}, and QCD axion direct detection~\cite{Gendler:2024adn}.}
As a result,  
at large $h^{1,1}$ in the geometric regime of the KS axiverse, a typical model will have negligible fuzzy dark matter abundance (see Fig.~\ref{fig:h11_spread}), and we have accordingly restricted the present work to small $h^{1,1}$. 
We find that the 
small-$h^{1,1}$ corner of the KS axiverse manifests axion dark matter phenomenology largely as envisioned in \cite{Arvanitaki:2009fg}, albeit with few axions, whereas at large $h^{1,1}$ one finds smaller decay constants and hence very little relic abundance of ultralight axions.

Significant prior work on fuzzy DM in string compactifications appeared in~\cite{Cicoli:2021gss},
which examined the abundance of fuzzy DM in two canonical moduli stabilization scenarios.\footnote{See also the related work \cite{Reig:2021ipa,Kaloper:2024lrr}.} 
The present work differs from \cite{Cicoli:2021gss} in several ways: we work in explicit CY orientifolds, but do not incorporate moduli stabilization;
we impose the existence of a QCD axion; and we study the cosmological effects of the full spectrum of axions.

This paper is organized as follows. In Section~\ref{sec:cosmosetup} 
we lay out our cosmological assumptions and review the computation of axion relic densities. 
In Section~\ref{sec:stsetup} we review the ingredients and construction of axion models in type IIB compactifications. 
In Section~\ref{sec:bestiary} we report our main results on constructions of fuzzy axion models and their novel phenomenology: in particular, we present emblematic examples 
in Section~\ref{sec:examples}, and we present statistics characterizing fuzzy DM models in our ensemble in Section~\ref{sec:stats}.
In Section~\ref{sec:new_pheno} we provide new computations of two aspects of fuzzy axion phenomenology. After presenting some additional discussion in \cref{sec:discussion}, we conclude in Section~\ref{sec:conclusions}. 

The data needed to reproduce the examples discussed in Section \ref{sec:examples} can be found in supplementary materials, namely a dedicated \href{https://github.com/sheride/fuzzy_axions}{GitHub repository} which includes a data file and a Jupyter notebook demonstrating how to read and manipulate the data using \textsc{CYTools}~\cite{Demirtas:2022hqf}.

\section{Cosmological setup}
\label{sec:cosmosetup}

In this section, we outline our assumptions for possible cosmological histories and compute the abundance of axions 
resulting from 
the misalignment mechanism. 
In Section~\ref{sec:qcdab} we collect analytical estimates for the axion abundance based on the epoch in which a given axion starts to roll: namely, during matter domination, radiation domination, or reheating with a general equation of state $w_R$. 
Then, in Section~\ref{se:choosecosmology} we present the two reheating scenarios that we will consider.
Finally, in Section~\ref{sec:observables} we give an overview of our methodology for obtaining constraints on fuzzy axions.\\

We assume that in the early Universe there was a reheating phase with unknown equation of state $w_R$. Then, at a redshift around $z_{R}$, the Universe reheated to a thermal bath at temperature $T_{R}$, carrying the SM degrees of freedom. The usual cosmology unfolds after reheating.

Employing comoving entropy conservation within the SM thermodynamic bath for all $z \leq z_{R}$, we have 
\begin{align}
\label{eq:entropy_conservation}
    g_s(z)\,T^3(z) = g_{s,0}\,T_0^3\,(1+z)^3\,,
\end{align}
where $g_{s}(z)$ is the effective number of SM relativistic degrees of freedom that corresponds to the entropy, $g_{s,0} \approx 3.93$ today (including neutrinos), and $T_0 \approx 2.72$ K is the CMB temperature today. Based on our cosmological setup, the Friedmann equation can be approximated as
\begin{align}
\label{eq:Friedmann_main}
    3H^2\mpl^2 = \Bigl(\rho_{\rm r} + \rho_{\rm m} + \rho_{\Lambda}\Bigr)\Theta(z < z_{R}) + 3H^2_{R}\mpl^2\left(\frac{1+z}{1+z_{R}}\right)^{3(1+w_R)}\,\Theta(z \geq z_{R})\,,
\end{align}
where
\begin{align}
    \rho_{\rm r} &= \frac{\pi^2}{30}g_{\rho} (T)\,T^4 = \frac{\pi^2}{30}g_{\rho} (z)\left(\frac{g_{s,0}}{g_s(z)}\right)^{4/3}T_0^4\,(1+z)^4\,,\nonumber\\
    \rho_{\rm m} &= 3H_0^2\mpl^2\,\Omega_{\rm m}\,(1+z)^3\,,\nonumber\\
    \rho_{\Lambda} &= 3H_0^2\mpl^2\,\Omega_{\Lambda}\,,
\end{align}
are the energy densities of the SM radiation bath, matter, and cosmological constant respectively.  Here, $\Omega_{m}h^2 = \Omega_c h^2+\Omega_b h^2 \approx 0.14$ and $\Omega_{\Lambda} = 0.7$ are the total matter (including cold dark matter, $c$, and baryons, $b$) and cosmological constant densities today, and $H_0 \approx 2.14\,h\times10^{-33}\, {\rm eV} =: hM_H$ 
eV is the Hubble parameter today~\cite{Planck:2018vyg}. We denote by $g_{\rho}$ the effective number of SM relativistic degrees of freedom that corresponds to the energy density.\footnote{We use $g_{\rho}$ and $g_s$ as functions of temperature from Ref.~\cite{Saikawa:2018rcs}. Using entropy conservation as in \eqref{eq:entropy_conservation}, we get $g_{\rho}$ and $g_s$ as functions of redshift.} In $\rho_{\rm r}$ above, we used \eqref{eq:entropy_conservation} to get to the second equality. Since we have radiation domination at reheating, the corresponding $H_{R}$  (and the corresponding redshift $z_{R}$) is related to the reheating temperature $T_{R}$ through energy conservation: $\rho_{\rm r}(T_{R}) = 3H_{R}^2\mpl^2$.

A misaligned homogeneous axion field starts to roll when $3H(z_a) \simeq m_a$, and its energy density subsequently redshifts as $(1+z)^3/(1+z_a)^3$.
If the axion is stable on the timescale of the age of the Universe --- see Section \ref{subsec:stable_axions} --- then it eventually contributes towards $\Omega_m$ today.  In order to avoid overproduction of DM from axions with large $(m_a,f_a)$ combinations, one may need to tune their initial misalignment angles $\theta_a$,\footnote{If the Universe underwent an early epoch of inflation, the initial angle $\theta_a$ cannot be dialed down below the quantum dispersion 
value $\sim H_{I}/(2\pi f_a)$, where $H_{I}$ is the inflationary Hubble rate.} the reheating temperature $T_{R}$, and/or the equation of state $w_R$ of the reheating phase. 

For our main results we compute the relic abundance of axions numerically\footnote{To approximate the total abundance of a collection of axions, we add their separate relic abundances, rather than solving the coupled equations of motion directly. 
Investigation of resonant couplings as in \cite{Cyncynates:2021xzw,Murai:2023xjn} will be the
subject of future work.} based on $z_a$ and \eqref{eq:Friedmann_main}.
Even so, in order to more completely lay our cosmological assumptions, we now turn to an analytical treatment of relic abundance in our setting. 

\subsection{Axion abundance}

\subsubsection*{Stable axions}
\label{subsec:stable_axions}

The axion relic abundance today is only composed of axions that are stable on the timescale of the age of the Universe. Therefore, we begin by assessing the  stability of axions to decays. 
Axions could couple to the SM photons through the 
dimension-$5$ operator
\begin{align}
    \mathcal{L}_{\rm int} =\frac{1}{4}\,\frac{\alpha\,c_{a\gamma\gamma}}{2\pi f_a}\,a\,F\tilde{F}\,,
\end{align}
where $f_a$ is the axion decay constant, $\alpha$ is the fine structure constant, and $c_{a\gamma\gamma}$ is a model dependent constant.\footnote{Comparing to the notation of \cite{Gendler:2023kjt}, 
$C_\gamma^{\text{there}} = c_{a\gamma\gamma}^{\text{here}}$.}
 The resulting axion-photon decay rate is\footnote{We only consider the leading two photon perturbative decay mode. Due to the cosmological redshift, there is no significant resonant production (induced emission) of photons. See e.g. Refs.~\cite{Preskill:1982cy,Abbott:1982af,Alonso-Alvarez:2019ssa} for discussions of higher-order number-changing processes.}
\begin{align}
\label{eq:decayrate}
    \Gamma_{a\gamma\gamma} = c_{a\gamma\gamma}^2\left(\frac{10^{16}\,{\rm GeV}}{f_a}\right)^2\left(\frac{m_a}{4.3 \cdot 10^5\,{\rm GeV}}\right)^3\,6.3 \cdot 10^{-15}\,{\rm eV}\,.
\end{align}
Here we have pulled out an overall scale of $6.3 \cdot 10^{-15}$ eV, which is close to $H_{\rm BBN}$ (at a temperature of about $5$ MeV).

Axions with decay rates $\Gamma_{a\gamma\gamma}  \gtrsim H_{\rm BBN}$
would decay before BBN and leave the subsequent cosmology largely unchanged.
On the other hand, axions with decay rates comparable to or smaller than $H_{\rm BBN}$ would decay near or after BBN, and can possibly impact the subsequent cosmology~\cite{Cadamuro:2010cz,Cadamuro:2011fd,Millea:2015qra,Depta:2020wmr,Langhoff:2022bij,Balazs:2022tjl}. 
In particular, such axions can have observational consequences as decaying dark matter~\cite{Langhoff:2022bij}, but in this work we concern ourselves only with their relic abundance, which we will now compute.

\subsubsection*{Abundance of the QCD axion}\label{sec:qcdab}

Let us begin by reviewing the QCD axion~\cite{Peccei:1977hh,Weinberg:1977ma,Wilczek:1977pj} abundance due to misalignment. The QCD axion mass is temperature dependent, and can be approximated as  
\begin{align}
\label{eq:QCDaxion_mass}
    m_a(T) := \frac{\left(\chi_{\rm QCD}(T)\right)^{1/2}}{f_a} \approx 5.7 \cdot 10^{-10}\,{\rm eV}\left(\frac{10^{16}\,{\rm GeV}}{f_a}\right)\begin{cases}
        1 &\qquad T < T_c\\
        \left(T_c/T\right)^{4} &\qquad T > T_c
    \end{cases}\,.
\end{align} 
Here, $\chi_{\rm QCD} = (\partial^2{S[\theta_a]}/\partial\theta_a^2)_{\theta_a=0}$,
where $S[\theta_a]$ is the (temperature dependent) effective QCD action, is called the topological susceptibility and measures fluctuations of the topological charge in the QCD vacuum. $T_c$ is the QCD crossover temperature. In the above, the low $T$ (temperature independent) behavior can be captured from chiral perturbation theory~\cite{Weinberg:1977ma,GrillidiCortona:2015jxo}. The high $T$ behavior, on the other hand, can be obtained from the dilute instanton gas approximation~\cite{Gross:1980br}. Chiral perturbation theory~\cite{GrillidiCortona:2015jxo} gives $\chi_{\rm QCD} = (75.5\,{\rm MeV})^4$ at zero temperature, while lattice QCD results~\cite{Borsanyi:2016ksw} fix $T_c \approx 150$ MeV and corroborate the $\sim T^{-4}$ instanton calculation.

Depending upon the reheating temperature $T_R$, there are two main cases of interest. First, let us consider the case when the QCD axion starts to oscillate during radiation domination, i.e. $T_R > T_{\rm osc}$. Combining $3H(T_{\rm osc}) \approx m_a(T_{\rm osc})$ and \eqref{eq:Friedmann_main} (with radiation domination), it can be seen that 
$T_{\rm osc}$ is given by the solution of
\begin{align}
\label{eq:ToscQCD}
    T_{\rm osc} \times \Bigl(g_{\rho}(T_{\rm osc})\Bigr)^{1/12} \approx 298\,{\rm MeV}\,\Biggl(\frac{10^{16}\,{\rm GeV}}{f_a}\Biggr)^{1/6}\,,
\end{align}
where we have used the $m_a \propto (T_c/T)^4$ dependence in \eqref{eq:QCDaxion_mass}.\footnote{For very large $f_a\gtrsim 2 \cdot 10^{17}$, oscillations occur at $T_{\rm osc}<T_c$~\cite{Fox:2004kb}, but this does not occur anywhere in our model space.} Taking initial misalignment angle $\theta_a \lesssim 1$,\footnote{For larger angles, anharmonicity of the potential in the initial phase of rolling induces positive corrections to the abundance. In particular $\theta^2 \rightarrow \theta^2 f(\theta)$, where $f(\theta)$ can be approximated following Refs.~\cite{Lyth:1991ub,Visinelli:2009zm,Diez-Tejedor:2017ivd}.} the relic abundance can be approximated as
\begin{align}
\label{eq:Omega_aQCD_ini}
    \frac{\rho_{a,0}}{3 M_H^2 \mpl^2} \approx \frac{1}{3 M_H^2 \mpl^2}\frac{m_a(0)\,m_a(T_{\rm osc})\,f_a^2\theta^2_{a, \rm QCD}}{2(1+z_{\rm osc})^3} =: \Omega_{a,\rm QCD} h^2 \,,
\end{align}
where $z_{\rm osc}$ is the redshift at $T_{\rm osc}$. Putting everything together gives the familiar result\footnote{While a more robust computation of the QCD axion relic density using accurate $T$-dependence of its mass can be performed as in~\cite{Borsanyi:2016ksw} (also see~\cite{Wantz:2009it}), we shall resort to \eqref{eq:QCDaxion_mass} in this paper.} 
\begin{align}
\label{eq:Omega_aQCD}
    \Omega_{a,\rm QCD}h^2 \approx 0.12\left(\frac{\theta_{a,\rm QCD}}{4.7 \cdot 10^{-3}}\right)^2\left(\frac{f_a}{10^{16}\,{\rm GeV}}\right)^{7/6}\,.
\end{align}
Here, for the purposes of illustration, we have used $g_{\rho} \approx 44$ and $g_{s} \approx 41$, which are their respective values around temperatures of $T_{\rm osc} \approx 218$ MeV (for $f_a \approx 10^{16}$ GeV).

On the other hand, if $T_R < T_{\rm osc}$, then we need to specify the cosmology during the reheating phase, since the QCD axion would start to roll in this phase. Assuming that the SM bath is not in equilibrium during this phase and therefore there is no notion of a temperature, only $m_a(0)$ could be the relevant mass factor. This reasoning would apply as-is to the $T_R < T_c$ regime. For the $T_c < T_R < T_{\rm osc}$ regime on the other hand, the QCD axion mass will have $m_a \sim T^{-4}$ behavior upon reheating. However, the mass will always remain larger than the Hubble rate. Therefore, the corresponding abundance may be roughly approximated as in \eqref{eq:Omega_aQCD_ini} with $m_a(T) \rightarrow m_a(0)$, and where $z_{\rm osc}$ is the redshift when $3H \approx m_a(0)$ in the unknown reheating phase, with equation of state $w_R$ (see \eqref{eq:z_aforw} below). 
While we shall do this for the purposes of the present paper, we note that this is subject to uncertainties since we do not have any reasonable information on the behavior of the QCD axion mass in a non-equilibrated QCD bath. 
We shall indicate cases where this uncertainty might be relevant.

\subsubsection*{Abundance of axion-like particles}
\label{eq:alpab}

Having addressed the QCD axion, we turn to considering all 
other axions. For an axion that starts to oscillate with some initial misalignment angle $\theta_{a} \lesssim 1$ and thus initial energy density $\rho_a \approx m_a^2f_a^2\theta_a^2/2$, the relic abundance can be estimated as
\begin{align}
\label{eq:Omega_a_general}
    \Omega_a h^2 \equiv \frac{\rho_{a,0}}{3 M_H^2 M_{pl}^2} := \frac{1}{3 M_H^2 M_{pl}^2}\frac{m_a^2f_a^2\theta^2_{a}}{2(1+z_{a})^3}\,.
\end{align}
Here $z_a$ corresponds to the redshift of the onset of axion oscillations, $3H(z_a) \approx m_a$. For simplicity, we break the computation into three broad categories of interest.

The first category corresponds to axions that start to roll during matter domination, i.e., when $\rho_{\rm m}$ in \eqref{eq:Friedmann_main} constitutes almost all of the energy budget of the Universe, and $3H_0 \lesssim m_a \lesssim 3H_{\rm eq} \approx 6.5 \cdot 10^{-28}$ eV. One gets  
\begin{equation}
    (1+z_a)^3 \approx \dfrac{(m_a/3H_0)^{2}}{\Omega_{\rm m}}\,,
\end{equation}
giving (see e.g. \cite{Marsh:2015xka}):
\begin{align}
\label{eq:Omega_a_matterdominated}
    \Omega_a\,h^2\Bigr|_{\rm m} \approx 7.6\times10^{-6}\,\theta_a^2\left(\frac{f_a}{10^{16}\,{\rm GeV}}\right)^2\,.
\end{align}
The relic density is independent of the axion mass, since $z_a^3 \propto H_a^2 \propto m_a^2$ in a matter dominated phase. All such axions can only comprise a small fraction of the total dark matter (even if $f_a \sim \mpl$). 
Moreover,  
all axions with $m_a \lesssim 3H_0$ are still in the slow-roll phase of their evolution, and can only contribute towards the effective cosmological constant: we neglect such axions.

The second category corresponds to axions that start to roll during the radiation dominated phase, i.e., when $\rho_r$ in \eqref{eq:Friedmann_main} constitutes almost all of the energy budget of the Universe, and $3H_{\rm eq} \lesssim m_a \lesssim 3H_{R}$. Using $3H(z_a) \approx m_a$, one finds that $z_a$ is the solution of
\begin{align}
\label{eq:z_a2}
   \frac{\Bigl(g_{\rho}(z_a)\Bigr)^{1/4}}{\Bigl(g_s(z_a)\Bigr)^{1/3}}\,(1+z_a) \approx \left(\frac{10}{\pi^2}\right)^{1/4}\frac{\sqrt{m_a\mpl}}{g_{s,0}^{1/3}\,T_0}\,.
\end{align}
Upon using this in \eqref{eq:Omega_a_general}, we get the familiar result $\Omega_a \propto \theta^2f_a^2\sqrt{m_a}$:
\begin{align}
\label{eq:Omega_a_raddominated}
    \Omega_{a}h^2\Bigr|_{\rm r} \approx 0.12\,\theta_a^2\left(\frac{f_a}{10^{16}\,{\rm GeV}}\right)^2\left(\frac{m_a}{4.4 \cdot 10^{-19}\,{\rm eV}}\right)^{1/2}\,,
\end{align}
where for illustration purposes we have neglected the dependence on $g_{\rho}(z_a)$ and $g_{s}(z_a)$.

Finally, the third category of interest corresponds to axions with $m_a \gtrsim 3H_{R}$. These start to oscillate during the reheating phase with equation of state $w_R$ when $z > z_{R}$. Using $3H(z_a) = m_a$ in \eqref{eq:Friedmann_main} gives
\begin{align}
\label{eq:z_aforw}
    1+z_a \approx (1+z_{R})\left(\frac{m_a}{3H_{R}}\right)^{\frac{2}{3(1+w_R)}}\,,
\end{align}
leading to the following relic abundance:
\begin{align}
\label{eq:Omega_a_w}
    \Omega_ah^2\Bigr|_{w_R} \approx 0.12\left(\frac{\eta}{0.4}\right)\left(\frac{\theta_{a}}{4.4 \cdot 10^{-3}}\right)^2\left(\frac{3H_{R}}{5.7 \cdot 10^{-10}\,{\rm eV}}\right)^{1/2}\left(\frac{f_a}{10^{16}\,{\rm GeV}}\right)^2\left(\frac{m_a}{3H_{R}}\right)^{\frac{2w_R}{1+w_R}}\,.
\end{align}
In the above, we used energy conservation at $z=z_{R}$ to get $z_{R}$ in terms of $H_{R}$,
resulting in the factor of 
\begin{align}
\eta := \frac{(g_{\rho}(z_{R}))^{3/4}}{g_{s}(z_{R})}\,,
\end{align}
which has a weak dependence on $T_R$.\footnote{For $T_{R} = \mathcal{O}(220)$ MeV,
$\eta \approx 0.42$; while for $T_{R} = \mathcal{O}(100)$ GeV and above, it saturates to about $\eta \approx 0.31$.} 
 
Combining \eqref{eq:Omega_aQCD}, \eqref{eq:Omega_a_matterdominated}, \eqref{eq:Omega_a_raddominated}, and \eqref{eq:Omega_a_w}, the total abundance due to all axions of interest is given by 
\begin{align}
\label{eq:Omega_total}
    \Omega_{\rm total}h^2 = \Omega_{a,\rm QCD}h^2 + \sum_{a}\Omega_a h^2\Bigr|_{\rm m} + \sum_{a}\Omega_a h^2\Bigr|_{\rm r} + \sum_{a}\Omega_a h^2\Bigr|_{w_R} \leq \Omega_{\rm DM}h^2 \approx 0.12\,.
\end{align}

\subsection{Reheating}
\label{se:choosecosmology}

The total abundance \eqref{eq:Omega_total} 
sets up a relationship between 
the reheating  
temperature 
of the Universe, reheating equation of state, the
axion masses, decay constants, and initial angles.  In this section we explain this relationship and present the two reheating scenarios that will be considered in this paper.

Given a spectrum of $m_a \gtrsim 3H_{R}$ axions along with their decay constants, 
and recalling that we are assuming the usual cosmology post-reheating, \eqref{eq:Omega_a_w} dictates different possible choices for the reheating epoch.
Specifically,
taking $w_R=0$, the abundance is $\Omega_a h^2\Bigr|_{w_R}=0.12$ when the initial angle takes the critical value
\begin{align}
\label{eq:theta_crit}
    \theta_{\rm crit} \approx 5.6 \cdot 10^{-3}\left(\frac{4.45 \cdot 10^{-10}\,{\rm eV}}{3H_{R}}\right)^{1/4}\left(\frac{10^{16}\,{\rm GeV}}{f_a}\right)\,,
\end{align}
where we have taken $\eta \approx 0.31$.  Moreover, for any $w_R>0$, the abundance is larger than for $w_R=0$.  
Thus, in the \textit{absence} of tuning of the initial angle $\theta_a$ below $\theta_{\rm crit}$,
a \textit{necessary} requirement to keep such an axion from giving rise to too much DM is to have $w_R < 0$.

\begin{figure}
    \centering
    \includegraphics[width=0.99\linewidth]{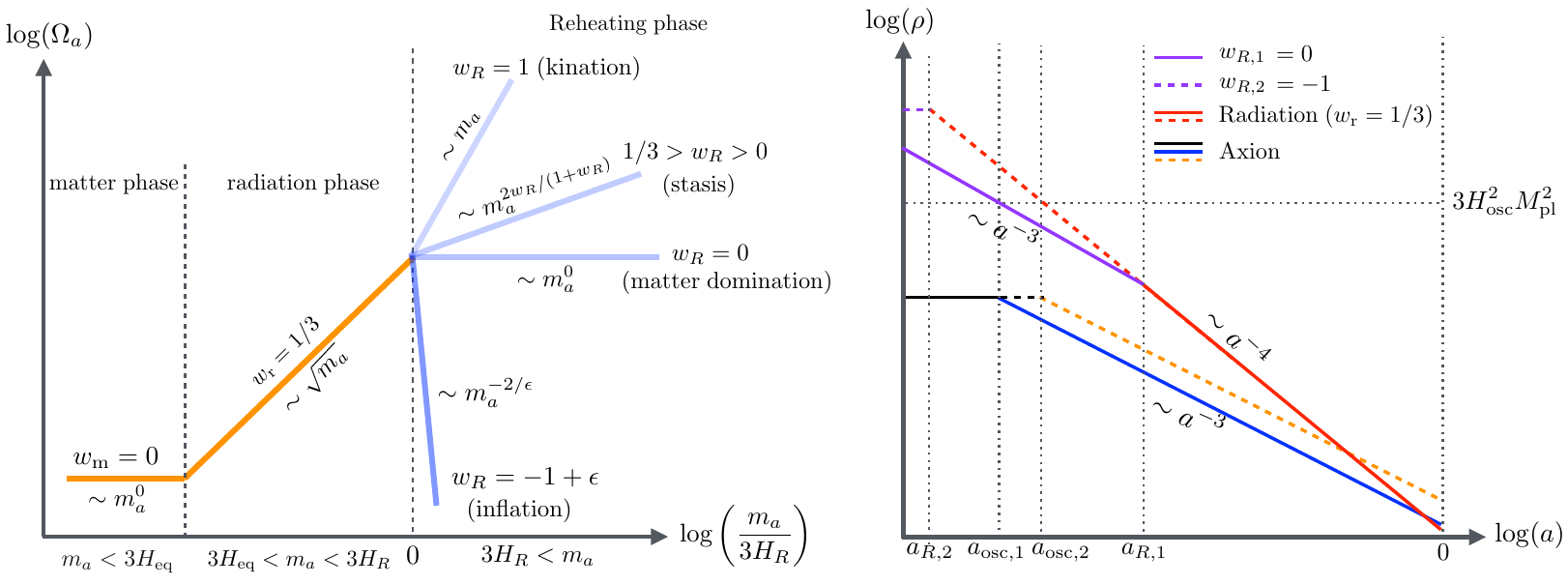}
    \caption{\emph{Left:} Relic abundance of a stable axion as a function of its mass with fixed $\theta_a$ and $f_a$, for different equations of state  $w_R$ during reheating. In general, lower $w_R$ means lower final abundance for a given axion. 
    For any $w_R \geq 0$, all $m_a > 3H_{R}$ axions are more abundant than all $m_{a} < 3H_{R}$ axions with the same
    $\theta_a$ and $f_a$. 
    \emph{Right:} Energy density of a stable axion 
    as a function of the scale factor $a$, demonstrating how a $w_R=0$ reheating epoch (subscript $1$) leads to a lower abundance than the standard $w_{\rm r}=1/3$ radiation dominated early Universe (subscript 2). The curves associated to the two cosmologies are depicted by solid and dashed curves respectively. The $a_{\rm osc, 1}$, $a_{\rm osc, 2}$ correspond to times when this axion starts to oscillate for these two cosmologies, while $a_{R, 1}$, $a_{R, 2}$ gives the time of reheating.  
    } \label{fig:xi_cartoonplot.pdf}
    \end{figure}

The effect of the reheating phase with equation of state $w_R$ on the axion abundance is represented in Fig.~\ref{fig:xi_cartoonplot.pdf}.
The left panel shows how changing $w_R$ affects the abundance of axions of different masses, holding all other parameters fixed. For any $w_R \geq 0$, all $m_a > 3H_R$ axions are more abundant than all $m_{a} < 3H_R$ axions, i.e., only $w_R<0$ suppresses abundances of heavy compared to light axions with the same $f_a$ and $\theta_a$ tuning. In the right panel, we show how the relic energy density of an axion, with a given $(m_a, f_a, \theta_a)$, is lowered when it starts to oscillate in the $w_R$ = 0 epoch compared to the standard scenario when it starts to roll in the radiation dominated $w_{\rm r} = 1/3$ epoch.

For the purposes of this paper, the key lesson of \eqref{eq:theta_crit}
and Fig.~\ref{fig:xi_cartoonplot.pdf} is that axions with $m_a \gtrsim 3H_R$ and $f_a \gtrsim 10^{15}\,{\rm GeV}$ (obtained by setting reheating Hubble at its minimum, $H_R = H_{\rm BBN}$) are dangerous: they will be overproduced unless either
\begin{enumerate}
    \item $w_R \le 0$\,, \text{or}
    \item \text{the initial angle }$\theta_a$\text{ is tuned to be sufficiently small.}
\end{enumerate}
In our ensemble of axion effective theories, we will find that axions that are dangerous in the above sense are very common.  We therefore turn to considering reheating scenarios that can mitigate the risk of overproduction.

We work with two different scenarios in this paper: 
\begin{enumerate}
    \item Inflation followed by prompt reheating ($w_R \rightarrow -1^{+}$).
    \item Modulus (matter) domination ($w_R \rightarrow 0^{-}$). 
\end{enumerate}
Reheating with $w_R>0$ would require an extreme degree of tuning of the initial angles $\theta_a$, and will not be considered here.

In each scenario, there are two important free cosmological model parameters: the reheating temperature $T_R$ and the total initial misalignment tuning,
which we define as the product $\delta$ of all initial angles,\footnote{We note that fixing the product of initial misalignment angles $\delta$ and enforcing that the total abundance is minimized uniquely determines the individual angles $\theta_a$: in particular, they are fixed such that the abundances $\Omega_a$ all coincide. It is for this reason that we can parameterize all of the misalignment angles with the single variable $\delta$.}
\begin{align}
   \delta \coloneqq \prod_a\theta_a\,,
\end{align}
where the product runs over all axions whose initial angles require tuning. 
  
In scenario (1), we let $H_{R}$ be as high as possible and do not allow any tuning of angles for axions heavier than the QCD axion. The idea is to dilute the abundance of all $m_a \gtrsim 3H_{R}$ axions by choosing $w_R$ sufficiently smaller than $0$ and close to $-1$ 
(thus exponentially diluting them), allowing us to place the reheating scale $H_{R}$  
right below the mass of the first axion heavier than the QCD axion, say $a'$. We set
\begin{align}
        \rho_{\rm r}(T_{R}) = 3H_{R}^2\mpl^2 = \frac{1}{3}m_{a'}^2\mpl^2\,,
\end{align}
fetching $T_R$ as a function of $m_{a'}$ (specifically, $T_R \sim \sqrt{m_{a'}}$, modulo corrections due to $g_{\rho}(T_R)$).
We then let $\delta$ take its maximum value subject to the constraint that the total axion abundance should agree with the observed value, tuning initial misalignment angles of all axions at least as light as the QCD axion. 
See Fig.~\ref{fig:golden_w=-1} for a summary plot in this scenario. Also see Fig.~\ref{fig:prompt_reheating} for statistics of the misalignment angle tunings, and reheating temperatures, depending upon the number of light and heavy axions respectively.

Scenario (2) is the `borderline' case since $\Omega_a$ becomes independent of $m_a$, and the problem of overabundance does not increase for successively heavier axions. 
Here, we shall impose a prior constraint on how much tuning is allowed, which in turn would enforce a particular $H_{R}$ (equivalently $T_{R}$). 
Explicitly, we consider fixed values of $\delta$ and select $T_R$ such that the optimal assignment of initial misalignment angles with total product $\delta$ yields a relic abundance of the observed value.     
We expect that such a tuning measure will pull $H_{R}$  towards $H_{\rm BBN}$ depending upon how small $\delta$ is, and the number and masses of heavier axions. 
The corresponding reheating temperature is obtained as usual from $\rho_{\rm r}(T_{R}) \approx 3H_{R}^2\mpl^2$: see Fig.~\ref{fig:moduli_domination}.
We remark that for reheating scenario (2), some of the models we will obtain in Section \ref{sec:bestiary} will be rejected because $T_R$ falls below $5~\rm{MeV}$, conflicting with BBN.

The two reheating scenarios that we have discussed here will be used to provide concrete results for axion DM abundance in the string theory models obtained in Section \ref{sec:bestiary}.\footnote{For our cosmological considerations, we always discard axions with masses $m_a > 3H_{I,\rm max} \approx 1.4 \cdot 10^{14}$ GeV, where $H_{I, \rm max}$ is the observational bound due to CMB measurements.} 

One may assume that reheating is determined by perturbative decay of a heavy particle of mass $m$, with decay rate $\Gamma_{R}$ which fixes $H(T_{R}) \approx \Gamma_{R}$. If this decay is mediated by a dimension-$5$ operator that couples the heavy particle to photons, suppressed by a scale $\Lambda$, then $\Gamma$ is given by\footnote{To draw a comparison with \eqref{eq:decayrate}, here we have absorbed the anomaly coefficient in the UV (decay constant) scale, and have called it $\Lambda$.}
\begin{align}
\label{eq:Gamma_R}
    \Gamma_{R} = \frac{m^3}{64\pi\Lambda^2}\,.
\end{align}
If this particle is a modulus with mass comparable to the supersymmetry (SUSY) breaking scale $M_{\rm SUSY}$, and if $\Lambda \lesssim \mpl$, then a definite relationship can be set up between SUSY breaking scale and reheating temperature~\cite{Coughlan:1983ci,Acharya:2008bk,Iliesiu:2013rqa}. With $M_{\rm SUSY} \gtrsim 100$ TeV, this leads to $T_{R}~\gtrsim~T_{\rm BBN}$. Later in Section~\ref{sec:prefactor}, we discuss how $m_{\rm SUSY}$  
can be connected to the flux superpotential $W_0$.

\subsection{Observables and constraints}\label{sec:observables}
\begin{figure}
    \centering
    \includegraphics[width=1\textwidth]{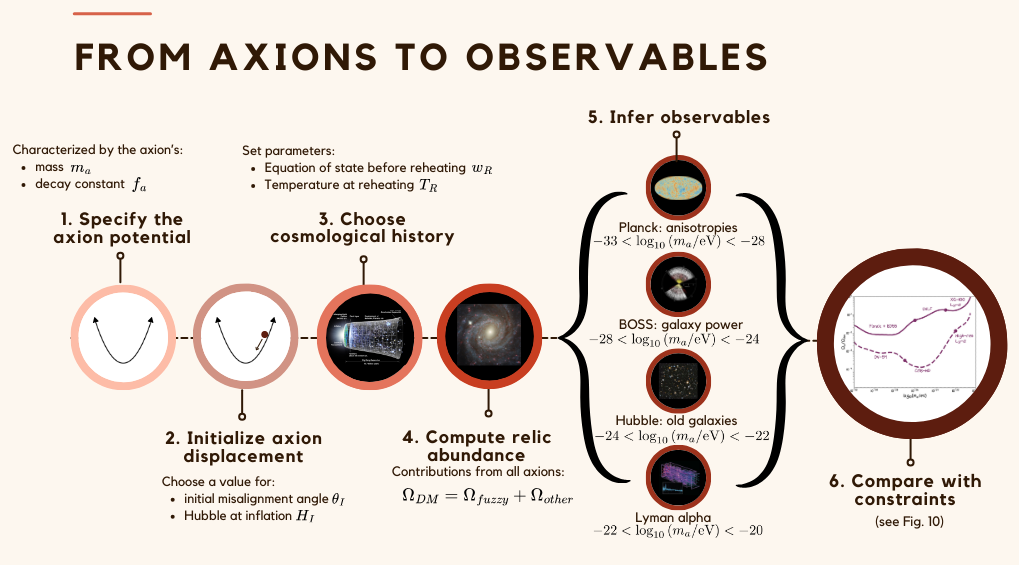}
\caption{A representation of the pipeline that takes as input axion characteristics and choices about the cosmology, and outputs constraints. The observables in region 5 are arranged from top to bottom according to scale, with large, linear scales at the top, and small, non-linear scales at the bottom. The CMB anisotropies are constrained via $C_\ell$ two point temperature and polarization  
correlations, with axions leading to effects dominantly via the expansion rate. The other observables are sensitive to the matter power spectrum, $P(k)$, where axions manifest via their large de Broglie wavelength suppressing structure formation relative to CDM. Probes of $P(k)$ on smaller scales involve increasing forward model complexity. (Images: NASA, ESA/Hubble, Planck, BOSS, LBL.)} 
    \label{fig:cartoon_cosmo}
\end{figure}


In this work we focus on obtaining a large relic density of axions in the fuzzy window, because such axions can be searched for via their effects on cosmological and astrophysical observables~\cite{Hu:2000ke,Schive:2014dra,Schive:2015kza,LinaresCedeno:2020dte,Cedeno:2017sou,Urena-Lopez:2015gur,Marsh:2018zyw,Dalal:2022rmp,Amendola:2005ad,Arvanitaki:2009fg,Hlozek:2014lca,Hlozek:2017zzf,Dentler:2021zij,Hotinli:2021vxg,Farren:2021jcd,Winch:2024mrt,Rogers:2023ezo,Lague:2021frh,Rogers:2020ltq,Rogers:2023upm,Poulin:2018dzj,Hui:2016ltb,Lazare:2024uvj,Shevchuk:2023ccb,Flitter:2022pzf,Blum:2024igb,Zimmermann:2024xvd,Arvanitaki:2010sy,Hoof:2024quk,Stott:2018opm,Marsh:2015daa,Marsh:2015xka,Winch:2023qzl,Leong:2018opi,Schive:2017biq,Zhang:2017dpp,Khlopov:1985fch,Hwang:2009js,Park:2012ru,Marsh:2015wka,Gonzalez-Morales:2016yaf,Bar:2018acw,Bar-Or:2018pxz,Bozek:2014uqa,Corasaniti:2016epp,Kobayashi:2017jcf}. Ultralight axions distinguish themselves from CDM via a variety of physical effects. At early times, the axion field is frozen by Hubble friction and acts like dark energy, rather than the usual matter-like 
behavior of CDM. This difference leads to a modified expansion rate $H(z)$ of the Universe, which impacts the CMB anisotropies via Silk damping and the Sachs-Wolfe effect, as well as other measures of the expansion rate at late times such as baryon acoustic oscillations (BAO) 
or supernovae. 
Ultralight axions also possess a large de Broglie wavelength. In cosmological perturbation theory, this manifests via the gradient terms in the Klein-Gordon equation, leading to an effective sound speed and Jeans scale, and ultimately a suppressed matter power spectrum, $P(k)$, compared to pure CDM. On the other hand, if the axion initial misalignment angle is large, then the convex potential leads to an instability that can enhance $P(k)$ in a narrow range of scales. These constraints are purely cosmological.

There are also a number of astrophysical probes of fuzzy axion dark matter.
In the non-linear regime inside DM halos, the wave nature of ultralight axions leads to time-varying interference effects, which can scatter stars, leading to heating or cooling of stellar distributions that is absent in CDM. Deep in the center of DM halos, ultralight axions form coherent solitons known as axion stars, the presence of which modifies the CDM Navarro-Frenk-White density profile, affecting the rotation curves and stellar velocity dispersions of galaxies. Lastly, an instability of the Klein-Gordon equation on Kerr spacetime causes the build up of axions near astrophysical black holes, which extracts spin from the black hole, and if efficient enough this can alter black hole evolution in an observable way. These constraints are non-cosmological, and rely on varying degrees of astrophysical modeling, with systematics that are difficult to control due to small sample sizes.

In this paper we will only examine cosmological constraints. The road from model parameters $(m_a,f_a,\Omega_a h^2)$ to constraints 
is sketched in Fig.~\ref{fig:cartoon_cosmo}. Cosmological constraints are derived at their core from linear perturbation theory, giving strong theoretical control. In the non-linear regime the models are well calibrated by simulations. Furthermore, the datasets are large, and the statistics are thus also under control, leading to Bayesian upper limits on $\Omega_a h^2(m_a)$, marginalized over many additional variables. We consider only those data leading to the strongest 95\% C.L. upper limits at a given axion mass (we do not recompute observables and find the posterior probability for a given string model: this task would be significantly more computationally intensive and we leave it for a future work). In most cases, these limits are arrived at in individual mass bins, and we spline interpolate to give a smooth upper limit 
curve.\footnote{Note that this upper limit curve is not the same quantity one would obtain from a consistent global Bayesian posterior on $(m_a,\Omega_a h^2)$. See e.g. the discussion in Refs.~\cite{Hlozek:2014lca,Hlozek:2017zzf}.} The most powerful current limits on $\Omega_a h^2$ are as follows: \emph{Planck} CMB combined with BOSS galaxy power spectrum shape~\cite{Rogers:2023ezo}, which dominates at the lowest masses; the ultraviolet luminosity function (UVLF) from the Hubble space telescope~\cite{Winch:2024mrt}, which dominates at intermediate masses; the XQ-100 Lyman-$\alpha$ forest flux power spectrum~\cite{Kobayashi:2017jcf}, which dominates at the highest masses.
\begin{figure}
    \centering
    \includegraphics[width=0.9\textwidth]{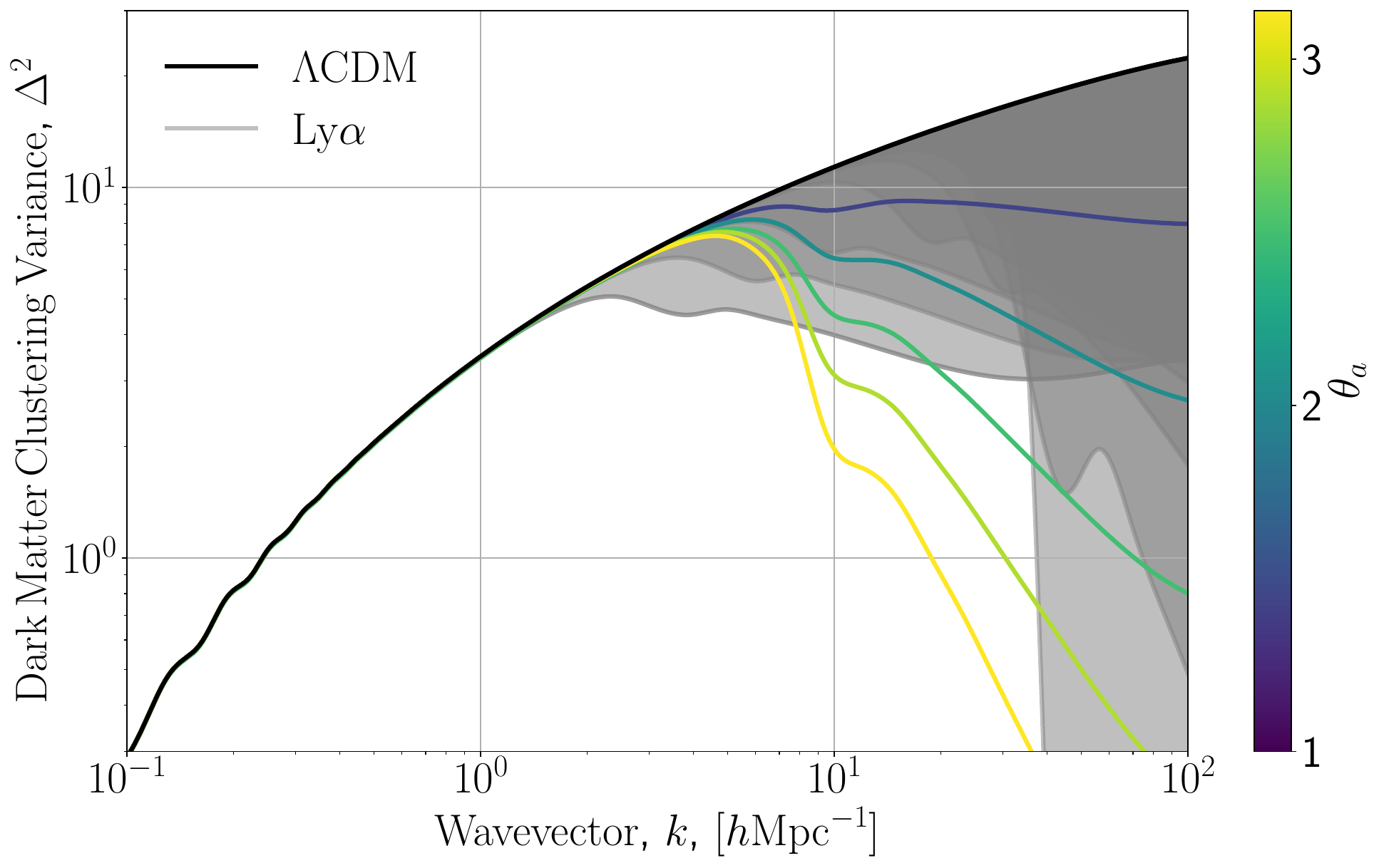}
    \caption{Linear matter power spectrum $\Delta^2(k)$ at the present day, computed with \textsc{axionCAMB}~\cite{Hlozek:2014lca} for $\Lambda$CDM compared to fuzzy DM models with $m_a=10^{-22}\text{ eV}$, for a range of DM fractions that are consistent with our optimized model with the largest abundance 
    (see Section~\ref{sec:best_abundance}),
    which has $f_a\approx 2\times 10^{16}\text{ GeV}$ at this mass.  The gray shaded region is constructed from the 95\% C.L. constraints on fuzzy DM from the Lyman alpha forest flux power spectrum (Ly$\alpha$)~\cite{Kobayashi:2017jcf}. Models outside of the gray region are excluded, which in this case is determined by the value of the initial misalignment angle, $\theta_a$. This result was generated assuming a quadratic axion potential.}
    
    \label{fig:pk_plot}
\end{figure}

Future cosmological observations will improve on these limits. Forecasts are typically made using a Bayesian Fisher matrix formalism, marginalizing 
over nuisance parameters. This is again with fixed mass bins, which we interpolate 
to produce a smooth forecast limit curve. The most powerful future limits on $\Omega_a h^2$ are as follows: the Ostriker-Vishniac (OV) effect at high multipoles, measured with precision equivalent to CMB-S4~\cite{Dvorkin:2022bsc}; the `high definition' (HD) CMB measurement of the same effect~\cite{Farren:2021jcd}; and a future high resolution Lyman-$\alpha$ measurement by e.g. the Dark Energy Spectroscopic Instrument~\cite{Grin:2019mub}.\footnote{A forecast for the OV effect at low mass, $m_a< 10^{-27}\text{ eV}$, is only available for CMB-S4, which may not continue in the form these forecasts assumed. However, since the constraints are driven by high multipole CMB measurements, we expect the forecast limit to be the same or better if the CMB-HD forecast were extended to low mass.}
 
Let us illustrate our approach to inferring constraints on fuzzy axions from $(m_a,\Omega_a h^2)$ with a more detailed example based on the matter power spectrum $\Delta^2(k)=k^3P(k)/2\pi^2$. Fuzzy DM suppresses the linear matter power spectrum relative to CDM in the early Universe, which correspondingly leads to suppression of the Ly-$\alpha$ forest flux power spectrum at later times, and thus Ly-$\alpha$ constrains the fuzzy DM mass and density~\cite{Amendola:2005ad,Rogers:2020ltq,Kobayashi:2017jcf}. On the other hand, it is not straightforward to interpret the Ly-$\alpha$ measurements as direct constraints on $\Delta^2(k)$~\cite{Chabanier:2019eai}. In order to provide an intuitive way to illustrate the constraints, we take the 95\% C.L. exclusions on $(m_a,\Omega_a h^2)$ of Ref.~\cite{Kobayashi:2017jcf} and construct the envelope of allowed fuzzy DM models they define on $\Delta^2(k)$. We then compare this envelope to $\Delta^2(k)$ as predicted by one of our string models (see Section~\ref{sec:best_abundance}) at $m_a=10^{-22}\text{ eV}$ for different $\theta_a$. The result is shown in Fig.~\ref{fig:pk_plot}, where we see that the suppression in $\Delta^2(k)$ caused by the fuzzy DM Jeans scale near $k=5\, h\,\text{Mpc}^{-1}$ causes the predicted linear power spectrum to fall outside the allowed envelope for $\theta_a\gtrsim 2$. Fig.~\ref{fig:pk_plot} was generated assuming a quadratic axion potential. For the largest $\theta_a\approx 3$ (yellow line), this approximation is expected to break down, correcting the relic abundance and also the power spectrum itself, which may lead to a loosening of constraints for large, fine-tuned $\theta_a$, as discussed in e.g. Refs.~\cite{Schive:2017biq,Winch:2023qzl}.

An important caveat to the process of comparing string models with multiple axions to cosmological constraints is that all current cosmological constraints are derived assuming a single ultralight axion plus an additional CDM component. In our models the CDM component is naturally provided by the QCD axion, which is too heavy to affect any of the considered observables in a non-trivial way. However, we will find some models with multiple fuzzy axions. For these models it is no longer strictly consistent to compare them to constraints derived assuming a single fuzzy axion: the multi-field constraint on $\Omega_a h^2(m_a)$ is likely given by the sum total abundance in relatively wide mass bins. At present, however, it is impossible to compute multi-axion cosmological constraints due to the lack of a Boltzmann code capable of handling multiple species (although see Ref.~\cite{Chen:2023unc}).

All of the above-mentioned effects distinguishing fuzzy axions from CDM are purely gravitational, and so do not depend on the couplings of axions to the SM. We will consider only one effect caused by such a coupling of ultralight axions: the uniform rotation of the CMB polarization 
angle due to the axion-photon coupling for $10^{-33}\text{ eV}\lesssim\,m_a\lesssim 10^{-28}\text{ eV}$, 
known as cosmic birefringence. In Section~\ref{sec:birefringence_model} we assess the ability of string models to explain the observational hint for such an effect found by Refs.~\cite{Minami:2020odp,Diego-Palazuelos:2022dsq}.

\section{String theory setup}
\label{sec:stsetup}
 
String theory  famously gives rise to  
many
axions in four dimensions through the compactification of the extra dimensions.
In our setting, the relevant Lagrangian for the axions $\theta^i$ (where the index $i$ runs over the number of axions) takes the form\footnote{Indices on $\theta_i$ are raised and lowered with the identity matrix.}
\begin{equation}\label{eq:axionL}
    \mathcal{L}\supset -\frac{1}{2}\mpl^2K_{ij}\partial_{\mu}\theta^i\partial^{\mu}\theta^j -V_{\mathrm{stringy}}(\theta_i)-V_{\mathrm{QCD}}(\theta_{\mathrm{QCD}})+\mathcal{L}_{\mathrm{SM}}\,,
\end{equation}
where $K_{ij}$ is the K\"{a}hler metric and
\begin{align}
    V_{\mathrm{stringy}} = \sum_{A}\Lambda^4_{A}\left(1-\cos(2\pi Q_{A}^i\theta_i)\right)\,,
    \label{eq:Vstringy}
\end{align}
with $\Lambda_{A}^4$ the symmetry-breaking scales and $Q_{A}^i$ the entries of the instanton charge matrix; these quantities will be defined in detail in Section \ref{sec:effpotential}. The index $A$ runs over all instanton contributions. The $\theta_i$ are dimensionless angular variables, and the Lagrangian is not in its canonical form. We can canonically normalize the fields by constructing a change-of-basis matrix that brings the original charges $Q_{A}^i$ to the kinetic and approximate mass eigenbasis, as detailed in \cite{Gendler:2023kjt}. In this basis, the QCD axion $\theta_{\mathrm{QCD}}$ is a particular linear combination of the $\theta_i$ and receives an additional contribution to its mass, given at low temperature by low-energy non-perturbative QCD effects:
\begin{align}\label{eq:QCDpotential}
    V_{\mathrm{QCD}}(\theta_{\mathrm{QCD}}) \approx \frac{1}{2}\chi_{\rm QCD} \theta_{\mathrm{QCD}}^2\,.
\end{align}
Note that \eqref{eq:Vstringy} contains, in general, terms that also depend on $\theta_{\mathrm{QCD}}$. Such terms that are large enough can spoil the QCD axion solution to the strong CP problem, as studied in \cite{Demirtas:2021gsq}. See Section~\ref{sec:PQq} for a discussion of the interplay between Peccei-Quinn quality and fuzzy axion dark matter.

In the following sections, we show how to specify the data in  \cref{eq:axionL} from string theory. We compactify the 10D theory on explicit CY manifolds (see Section \ref{sec:CYandOrientifold}) and compute the relevant terms in the 4D supergravity Lagrangian (see Section \ref{sec:effpotential}). The full procedure is visually summarized in \cref{fig:cartoon}.

\subsection{Choosing Calabi-Yau   orientifolds}\label{sec:CYandOrientifold}

\noindent{\bf Choosing a CY:} Type IIB superstring theory enjoys maximal supersymmetry in $\mathbb{R}^{1,9}$.
Given that we are interested in four-dimensional phenomena, we instead consider the ten-dimensional spacetime $M = \mathbb{R}^{1,3} \times X$, with $X$ a compact six-dimensional Euclidean manifold. 
Kaluza-Klein theory then indicates that at energies smaller than the inverse radius of $X$, the low-energy EFT is effectively four-dimensional, but depends strongly on the topology and geometry of $X$.

To find a solution to the 
ten-dimensional vacuum Einstein equations, 
the extra dimensions must be described by a Ricci-flat manifold. A non-trivial class of such manifolds consists of \textit{Calabi-Yau spaces}, i.e., compact complex manifolds that are K\"ahler and have trivial canonical bundle. These properties imply the preservation of some supersymmetry in the 4D effective field theory. We will construct CY threefolds ($3=\mathrm{dim}_\mathbb{C} = \frac{1}{2} \; \mathrm{dim}_\mathbb{R}$) and compute the topological and geometric data necessary to understand the resulting physics in $\mathbb{R}^{1,3}$.

Differential geometry, though the natural place to start, is unwieldy for this purpose.
To take a step toward tractability, we move to algebraic geometry, where the spaces, known as algebraic varieties,  come from sewing together vanishing sets of systems of polynomials. In particular, these algebraic varieties are fully determined by the polynomial functions defined on them, meaning that topological and geometric calculations can be turned into commutative algebra, where computational algorithms abound (e.g., \textsc{Macaulay2} \cite{M2}). 

In fact, we typically go one step further: the polynomials defined on varieties containing a torus as a dense subset turn out to be particularly simple, being generated by the lattice points inside polyhedral cones.
Sewing these varieties together amounts to gluing their cones together along their faces, yielding so-called \textit{toric varieties} parameterized by \textit{fans}, namely collections of cones closed under taking faces and intersections. 
Topological and geometric data of toric varieties can then be computed from the combinatorial data of the cones making up the fan, including 
the structure of their faces and how they intersect \cite{cox2011toric, hori2003mirror}.

Several constructions of CY spaces are present in the literature. In this paper, we focus on hypersurfaces in compact toric varieties associated to triangulations of polytopes (higher-dimensional polygons). 
Batyrev established \cite{batyrev1993dual}
that toric fans corresponding to suitable triangulations of four-dimensional reflexive polytopes
define toric varieties in which 
one can always construct a smooth CY threefold hypersurface.
Kreuzer and Skarke then classified all
four-dimensional reflexive  polytopes
\cite{Kreuzer:2000xy}, assembling the \emph{KS database} of $473{,}800{,}776$ polytopes.

The KS landscape is vast: the
number of 
triangulations of four-dimensional reflexive polytopes giving rise to smooth, homotopy inequivalent CY hypersurfaces 
is bounded from above by $10^{428}$ \cite{Demirtas:2020dbm}.
Most of this potential wealth of geometries occurs at large $h^{1,1}$, and more specifically at $h^{1,1} = 491$.  We remark that it is not known whether CY threefold hypersurfaces in toric varieties obtained from the KS list are a representative sample of all CY threefolds; however, they are the largest currently-established ensemble, and are moreover very conveniently accessible via combinatorial computations.
The KS database is publicly accessible, and \textsc{CYTools} \cite{Demirtas:2022hqf} is an open-source Python library allowing users to triangulate polytopes  and study the topology of the induced CY hypersurfaces.

As explained in the Introduction, 
in this work we consider $2 \le h^{1,1} \le 7$.
The lower bound is chosen in order to have both a fuzzy axion and a QCD axion.  The upper bound is dictated both by physics --- for $h^{1,1} \gg 10$, fuzzy axions are rare --- and by computational limitations.  The analyses presented in the remainder of this paper could be carried out for any desired CY, even one with $h^{1,1}=491$, but a study of all polytopes, or even of all CYs resulting from triangulations of one polytope, is out of reach for $h^{1,1} \gg 10$.
Here we will give an exhaustive treatment for $2 \le h^{1,1} \le 7$: in particular, we will construct \emph{all} CY threefold hypersurfaces 
that result from 
fine, regular, star triangulations (FRSTs --- see e.g. \cite{Demirtas:2018akl}) 
of favorable polytopes\footnote{Favorable polytopes are those for which $h^{1,1}(X)$ agrees with the rank of the Picard group of the ambient 4D toric variety: in such cases many topological calculations associated with the CY are simplified.} in this range of $h^{1,1}$. 
We emphasize that while even in this range of Hodge numbers the number of triangulations is quite large, many CY hypersurfaces are topologically redundant, allowing us to be comprehensive by considering only a subset of the total triangulations. We briefly review this fact now.

By Wall's theorem \cite{wall}, the diffeomorphism class of a CY threefold is determined by its intersection numbers and second Chern class, and for CYs constructed from FRSTs, these data are determined by the restriction of the FRST to two-faces of the reflexive polytope (i.e., the two-skeleton of the triangulation). Thus, if the two-face triangulations of two such CYs agree up to an automorphism of the polytope, 
the CYs are trivially topologically equivalent: following \cite{gendler2023counting}, we say these CYs belong to the same FRST class. 
 
The methods developed in \cite{MacFadden:2023cyf} then describe how one can directly enumerate  equivalence classes of triangulations,
saving considerable time and memory.\\

\noindent{\bf Choosing an orientifold:}
The effective action of type IIB string theory compactified on a CY $X$ enjoys $\mathcal{N}=2$ supersymmetry in four dimensions. 
This amount of supersymmetry
ensures the existence of an exact moduli space, and in particular the axions that are the subject of this paper remain exactly massless if $\mathcal{N}=2$ supersymmetry is preserved. 
In order to make a more realistic phenomenological model, we consider an orientifold of the CY, which breaks the supersymmetry to $\mathcal{N}=1$. Specifically, we gauge a discrete symmetry of the form $(-1)^{F_L}\Omega_p\sigma$ where $\Omega_p$ is the parity on the worldsheet, $F_L$ is the left-moving fermion number, and $\sigma: X\to X$ is an isometric and holomorphic involution of the CY space $X$. The fixed loci for this choice of orientifold are of codimension two or six in $X$, and therefore correspond to O7-planes and O3-planes, respectively.  

Let $\sigma^*$ be the pullback of $\sigma$. Harmonic $(p,q)$-forms are either even or odd eigenstates of $\sigma^*$, and thus the cohomology splits as
\begin{equation}
H^{p,q}=H^{p,q}_+\oplus H^{p,q}_-\,,\;\; h^{p,q}_{\pm} = \mathrm{dim}(H^{p,q}_{\pm})\, .
\end{equation}
The number and type of fields in the 4D EFT depend on the values of the  Hodge numbers  $h^{1,1}_{\pm},\ h^{2,1}_{\pm}$. Orientifolds of CY manifolds in the KS database, with involutions $\sigma$ of the CY inherited from involutions of the ambient space, can be constructed systematically using the methods of \cite{Moritz:2023jdb}. 
For simplicity, we restrict to involutions with $h^{1,1}_-=0$,\footnote{We further restrict to orientifolds for which there are no frozen conifold singularities on top of O$7$-planes, in the sense of \cite{carta2020landscape}.} which ensures that all of the potential $C_4$ axions featuring in our models will survive orientifolding. 

The scalar fields of the resulting 4D $\mathcal{N}=1$ EFT are contained in $h^{1,1}_++h^{2,1}_-+1$ chiral multiplets: $h^{1,1}_+$ K\"ahler moduli and their axion partners, $h^{2,1}_-$ complex structure moduli, and the axio-dilaton. 
These moduli need to be stabilized to avoid fifth force constraints. The complex structure moduli and the axio-dilaton can be stabilized at energies higher than those for the $h^{1,1}_+$ 
K\"ahler moduli, and can be integrated out. Thus, they effectively contribute to our theory as constants: the vacuum expectation value of the dilaton parametrizes the string coupling, while with $W_0$ we denote the complex-valued Gukov-Vafa-Witten superpotential \cite{Gukov:1999ya} generated by stabilization, via fluxes of the complex structure moduli \cite{Giddings:2001yu}. Our assumptions about the stabilization of the $h^{1,1}_+$ K\"ahler moduli will be discussed in Section \ref{sec:KC}.

Let us also note that the 4D theory contains $h^{2,1}_+$ U$(1)$ vector multiplets, which potentially imply the presence of dark photons in the effective theory. 
The associated vector fields $V^a$, $a=1,\dots,h^{2,1}_+$, arise from the expansion of the four-form potential $C_4$ in terms of harmonic forms, and its dimensional reduction to 4D \cite{Grimm:2004uq}. In more detail, let $(\alpha_a, \beta^a)$ be a real, symplectic basis for $H^3_+=H^{1,2}_+\oplus H^{2,1}_+$, and let $w_2^i$ (respectively its dual $w_4^i$) be a basis of harmonic forms for $H^{1,1}_{+}$ (respectively $H^{2,2}_{+}$). Then the four-form potential expands as 
\begin{equation}\label{eq:C4expansion}
    C_4=\theta_i\, \omega_4^i + V^a\wedge \alpha_a - \tilde{V}_a\wedge \beta^a +Q_i\wedge \omega_2^i \, .
\end{equation}
Imposing the self-duality condition of $\tilde{F}_5=F_5-\frac{1}{2}C_2\wedge H_3+\frac{1}{2}B_2\wedge F_3$, half of the degrees of freedom of $C_4$ are projected out. One conventionally chooses to eliminate the two-forms $Q_i$ and the U$(1)$ vectors $\tilde{V}_a$ in favor of the axions $\theta_i$ and the U$(1)$ vectors $V_a$. 
We comment on the prevalence of dark photons in Section \ref{sec:const_ensemble}.

\subsection{Effective theory for $C_4$ axions}

\label{sec:effpotential}

We now derive \cref{eq:axionL} for axions arising from the 10-dimensional gauge potential $C_4$ integrated over   
a basis of 4-cycles $D_i$, with $i=1,\dots,h^{1,1}_+$. The low energy effective theory inherits $h^{1,1}_+$ axion fields as
\begin{equation}\label{eq:defaxion}
    \theta_i \coloneqq  \int_{D_i} C_4 \,,
\end{equation}
and their periodicity descends from the 10D gauge symmetry 
of $C_4$.  
The K\"ahler coordinates take the simple form\footnote{For background on the effective theories in flux compactifications, see e.g.~\cite{McAllister:2023vgy}.}
\begin{equation}
    T_i=\int_{D_i}\frac{1}{2}J\wedge J \;+\; i \int_{D_i}C_4\equiv\tau_i+i\theta_i\, ,\quad i=1,\dots,h^{1,1}_+\, ,
\end{equation}
where $J$ is the K\"ahler form and $\tau_i = \text{Vol}(D_i)$ is a modulus parametrizing the volume of $D_i$. 

As defined in \cref{eq:defaxion}, $\theta_i$ enjoys a continuous shift symmetry at the perturbative level, which descends from the gauge transformation of the gauge potential. However, introducing non-perturbative effects breaks this symmetry to a discrete one. 
In what follows we focus on non-perturbative contributions to the superpotential\footnote{Euclidean D3-brane contributions to the K\"ahler potential can be important if the scale of supersymmetry breaking is high, but are almost entirely unknown, and will not be incorporated in this work: see the discussions in \cite{Demirtas:2019lfi,Demirtas:2021gsq}.}
from Euclidean D3-branes wrapping 4-cycles.
In general there are infinitely many 4-cycles $\Sigma_A$ that support such contributions: we can write
\begin{equation}
    \Sigma_A = Q_A^i D_i\, 
\end{equation} for $A=1,\ldots,\infty$ and with $Q_A^i$ a set of coefficients.
However, we will consider only Euclidean D3-branes wrapping 
\textit{prime toric divisors}, i.e.~divisors that result from intersecting the 
CY hypersurface with the zero loci of the $h^{1,1}+4$ toric coordinates.
Prime toric divisors generate (over $\mathbb{Z}$, but not in general over $\mathbb{Z}_+$) the cone of effective divisors, and they are irreducible, smooth, and very often rigid, and so they typically support  leading contributions to the superpotential.\footnote{We will also suppose that QCD is supported on a prime toric divisor, for a similar reason: 
D7-branes wrapping a rigid irreducible cycle have no adjoint matter.}
We write the charges of the prime toric divisors as $Q_\alpha^i$, 
and we note that $Q_\alpha^i$ agrees with the GLSM (gauged linear sigma model) charge matrix of the ambient toric variety, and is determined by the polytope. 
We also refer to $Q_\alpha^i$ as the instanton charge matrix.

The Euclidean D3-brane corrections to the superpotential  $W$  depend on Vol$(\Sigma_\alpha)\equiv Q_\alpha^i \tau_i$ through the instanton actions
\begin{equation}
    S_\alpha=2\pi Q_\alpha^i T_i =  2\pi \text{Vol}(\Sigma_\alpha)+2\pi i Q_\alpha^i\theta_i \,.
    \label{eq:volume}
\end{equation}
The 4D $\mathcal{N}=1$ theory can be characterized in terms of the K\"ahler potential, the superpotential and gauge kinetic functions. The tree-level K\"ahler potential for the K\"ahler moduli is
\begin{align}
    K = -2\log(\mathcal{V})\; , \quad \mathcal{V} = \dfrac{1}{6}\int_{X} J\wedge J\wedge J= \dfrac{1}{6}\kappa_{ijk} t^i t^j t^k\,,
    \label{eq:kahlerpot}
\end{align}
where $\mathcal{V}$ denotes the volume of the CY $X$, and $\kappa_{ijk}$ are the triple intersection numbers of $X$. 
In \cref{eq:axionL}, the kinetic terms for the axions are expressed in terms of the K\"ahler metric, given by
\begin{align}\label{eq:KM}
    K_{ij} = 2 \frac{\partial}{\partial T_i} \frac{\partial}{\partial \overline{T}_j} K =  \dfrac{1}{2}\frac{\partial}{\partial \tau_i} \frac{\partial}{\partial \tau_j} K\,.
\end{align}
Taking into account the instanton corrections described above, the total superpotential reads
\begin{equation}
 W=W_0+\sum_{\alpha}\, A_\alpha\,  \mathrm{e}^{-S_\alpha}\,.
 \label{eq:superpot}
\end{equation}
The sum of exponentials contains the contributions of Euclidean D3-branes. 
In the rest of the paper we will set $A_{\alpha}\,=1$.  

The $F$-term scalar potential $V_F$ in the K\"ahler moduli sector is
\begin{equation}
V_F=e^{K_{\rm{tot}}}\left(K^{i\bar{\jmath}}D_iW\, D_{\bar{\jmath}}\overline{W}-3 |W|^2\right)\,,\quad D_iW = \partial_{T_i}W + (\partial_{T_i}K)\, W\, ,
\label{eq:fterm}
\end{equation}
where  $K^{i\bar{\jmath}}=2K^{ij}$ and $K^{ij}$ is the inverse of the K\"ahler metric \eqref{eq:KM}.
We have introduced the total K\"ahler potential,
\begin{equation}
K_{\rm{tot}} = K + K_{\rm{other}}\,,
\end{equation}
where $K$ is the K\"ahler potential for the K\"ahler moduli, as given in \eqref{eq:kahlerpot}, and 
$K_{\rm{other}}$ is the 
K\"ahler potential for the remaining moduli, i.e.~the complex structure moduli and the axio-dilaton. As these remaining moduli are stabilized at a high mass scale, their effect on \eqref{eq:fterm} occurs via a  constant  prefactor,
\begin{equation}\label{eq:pdef}
\mathcal{P} \coloneqq e^{K_{\rm{other}}}\,.
\end{equation}
The factor $\mathcal{P}$ varies depending on the string coupling and the stabilized vevs of the complex structure moduli, as we explain further in Section \ref{sec:prefactor}.

Plugging in the K\"ahler potential \eqref{eq:kahlerpot} and superpotential \eqref{eq:superpot} into \eqref{eq:fterm} and expanding, we arrive at a scalar potential of the following schematic form
\begin{equation}\label{eq:fullpot}
    V_F = V_{\mathrm{mod}}(\tau_i) + V_{\mathrm{stringy}}(\tau_i,\theta_i)\, .
\end{equation} 
Here, $V_{\mathrm{mod}}(\tau_i)$ depends only on the saxions $\tau_i = \mathrm{Re}(T_i)$, which we discuss in Section~\ref{sec:KC}. The stringy  potential for the axions (for a fixed choice of $\langle \tau_i \rangle$) can be written as 
\begin{equation}\label{eq:axionpot}
    V_{\mathrm{stringy}}(\theta_i)  =  \sum_{\alpha} \Lambda_{\alpha}^4 \left[1 - \cos(2\pi Q_\alpha^i \, \theta_i ) \right]\,,
\end{equation} 
where in the approximation that single-instanton contributions dominate, the instanton scales take the form   
\begin{align}
\label{eq:Lambda_and_m3/2}
    \Lambda_{\alpha}^4 = 8\pi \sqrt{\mathcal{P}} \, m_{3/2} \frac{Q_\alpha^i \, \tau_i}{\mathcal{V}} e^{-2\pi Q_\alpha^i \, \tau_i}\, , \quad m_{3/2} = \sqrt{\mathcal{P}}e^{K/2}|W|\,,
\end{align}
where $m_{3/2}$ is the gravitino mass and sets the scale of supersymmetry breaking, $M_{\rm{SUSY}}$.  

The primary quantities that we extract from \cref{eq:axionL} are the masses and decay constants of the axions, which we compute using the algorithm laid out in~\cite{Gendler:2023kjt}. The axion decay constants and masses are computed for the canonically normalized fields $\phi_i = M_i\,^j \,\theta_j$ as
\begin{equation}\label{eq:axion_fm}
    f_{i}= \frac{\mpl}{2\pi}\left[Q_{ij}(M^{-1})^j\,_i\right]^{-1} \, , \quad  m_{i}^2 = \frac{\Lambda_i^4}{f_{i}^2} \, .
\end{equation}
The matrix $Q_{ij}$ denotes the reduced charge matrix for the $h^{1,1}$ leading order contributions to \eqref{eq:axionpot}.

A key assumption used in the construction of our ensemble is that the dominant instanton scales, $\Lambda_i$, appearing in the axion potential \eqref{eq:axionL} are hierarchically separated. That is,   
\begin{equation}\label{eq:lambdahier}
\epsilon_{i}\coloneqq\dfrac{\Lambda_{i+1}}{\Lambda_{i}} \ll 1, \ \forall i \,,
\end{equation}
where without loss of generality, we have sorted the instanton scales such that $\Lambda_{i+1}\leq \Lambda_i$ $\forall \, i$.  In particular, the formula for the mass in \eqref{eq:axion_fm} is valid only when \eqref{eq:lambdahier} holds, and so the exact values of masses in resonant blocks in our statistics are subject to small errors. In one of our explicit models, given in Section \ref{ex:degen}, we showcase an example where some of the $\Lambda_\alpha$ are degenerate, and a computation based on the assumption of hierarchies is untenable. For this case, we computed masses and decay constants by explicitly diagonalizing the Hessian matrix using arbitrary precision.\footnote{We thank 
Sebastian Vander Ploeg Fallon for useful discussions regarding this calculation, and for sharing relevant code.}

Axions also couple to SM fields, which we have suppressed in \eqref{eq:axionL}. The only such couplings that will be relevant for us are the couplings of axions to photons, which are obtained from \cite{Gendler:2023kjt} and will be used in Section \ref{sec:birefringence_model} to calculate the CMB birefringence angle.
The dimensionless parameter 
$c_{a\gamma\gamma}$
appearing in the axion-photon coupling \eqref{eq:decayrate} 
is of order unity  
only for a subset of axions \cite{Gendler:2023kjt}: 
in particular,
axions 
lighter than the mass scale generated by Euclidean D3-branes on the same cycle as the D7-branes hosting electromagnetism
have strongly suppressed couplings to photons.

A complete computation of the couplings of axions to SM fields requires an explicit realization of the SM, as well as an understanding of corrections to $K$. We leave this difficult task to future work.

\subsection{The K\"ahler cone and the visible sector}\label{sec:KC}

The data characterizing the effective Lagrangian \eqref{eq:axionL} can be computed in terms of the geometric and topological data of the compactification. 
In this section, we will describe how we fix the parameters of $X$ in order to calculate axion masses and decay constants.

As explained before, EFTs obtained from compactifications of type IIB string theory on a CY orientifold contain a number of massless fields. In particular, the saxions (the real parts of the K\"ahler moduli) must be stabilized, or else the theory would be ruled out by fifth-force constraints. 

The values that the saxions can assume are given by the \textit{K\"ahler cone}, $K_X$, which is defined as the set of K\"ahler forms $J$ for which all holomorphic curves $C$ in $X$ have non-negative volumes
\begin{align}
    K_X \coloneqq  \biggl\{ J \ \biggl | \  \int_C J \geq 0 \  \forall \  C \in X \biggr\}\, .
\end{align}
Fixing a point in $K_X$ fixes the volumes of all holomorphic 
cycles in $X$, and correspondingly determines the semi-classical actions of Euclidean D3-brane contributions to the superpotential. When the scale of SUSY breaking is sufficiently far below the compactification scale, such superpotential terms determine the leading contributions to the axion masses.

We will not explicitly stabilize the K\"ahler moduli in this work.
That is, we will not construct the potential $V_{\mathrm{mod}}(\tau_i)$ for the saxions in \eqref{eq:fullpot} and find the points in $K_X$ where $V_{\mathrm{mod}}(\tau_i)$ has local minima. Instead, we will study axion theories across the entire domain defined by $K_X$, and will suppose\footnote{We furthermore require that the saxions be stabilized with masses sufficiently high such that the Lagrangian \eqref{eq:axionL} involves no dynamical mixings between the axions and saxions, and we are free to compute axion masses and decay constants treating $K_{i\bar{\jmath}}$, $\tau_i$, and $\mathcal{V}$ as (matrices and vectors of) numbers.}
that the saxions could in principle be stabilized anywhere in $K_X$.\footnote{Perturbative corrections to the K\"ahler potential are a plausible mechanism for stabilization relatively near to the walls of $K_X$, but such corrections are not currently known in enough detail to identify explicit minima (however, see  \cite{Cicoli:2024bwq}).  Other well-studied approaches such as~\cite{kklt,Balasubramanian:2005zx} rely on balancing classical effects against non-perturbative quantum corrections to the superpotential, and correspondingly lead to stabilization in special regions of $K_X$; for the phenomenology of fuzzy DM in such regions, see \cite{Cicoli:2021gss}.}
For this reason, our work does not establish the existence of fuzzy DM in the compactifications that we study; instead, it reveals the possible scope of fuzzy DM models in this setting.

Another key condition in our analysis is the imposition of computational control. Since \eqref{eq:axionL} is a sum over exponentially-suppressed corrections, if the divisor volumes $\tau_i$ are too small, then the sum will not converge. Knowledge of the entire, potentially infinite, series of corrections will then be necessary to compute the axion potential accurately. To avoid this, we require that all divisor volumes are sufficiently large that the series converges. Specifically, we ensure that $\tau_{\alpha} \geq 1$, $\alpha=1,\ldots,h^{1,1}+4$.
As a consequence of this choice, the results of this paper should be understood as applying only to CY hypersurface compactifications \emph{in the geometric regime}, not to all four-dimensional solutions of type IIB string theory.

Although we are not engineering an explicit construction of the SM in this work, an important part of fixing the moduli in $K_X$ is imposing that the SM gauge couplings are in the right range. The gauge coupling in the UV for a stack of D7-branes wrapped on a divisor $\Sigma$ is given by 
\begin{align}\label{eq:gcoupling}
    g_{\mathrm{UV}}^2 \propto \frac{1}{\text{vol}(\Sigma)}\,.
\end{align} 
Assuming no vector-like matter  coupled to QCD and a supersymmetric completion of the SM at some scale $M_{\mathrm{SUSY}}$, the UV gauge coupling can be matched to the IR gauge coupling as (see \cite{Demirtas:2021gsq}) 
\begin{align}
    e^{-8\pi/g_{\mathrm{UV}}^2} = e^{-8\pi/g_{\mathrm{Z}}^2} \left(\frac{M_{\mathrm{UV}}}{M_{\mathrm{SUSY}}} \right)^7 \left(\frac{M_{\mathrm{SUSY}}}{M_Z} \right)^3\,,
\end{align}
where $M_Z$ is the mass of the Z-boson. With $M_{\mathrm{SUSY}} \approx M_{\mathrm{UV}}$, a D7-brane stack on a divisor $\Sigma$ with $\text{vol}(\Sigma) \approx 40$ would result in the observed QCD gauge coupling, while with a SUSY-breaking scale in the TeV range, $\text{vol}(\Sigma) \approx 25$ would give the observed value. For this reason, we ensure that at least one divisor in $X$ has a volume between $25$ and $40$.

Our goal is to find examples of compactifications with large abundances of fuzzy dark matter. To this end, we furthermore impose that at least one of the axions in a given compactification has a mass of $10^{-18}$ eV.\footnote{In the small set of explicit examples presented in Section \ref{sec:examples} we also explore nearby masses, but in the ensemble constructed in Section \ref{sec:const_ensemble} the fuzzy axion mass is fixed to $10^{-18}$ eV.}

To summarize, we seek a point in the interior of the K\"ahler cone for which the following conditions are satisfied:
\begin{enumerate}
    \item   $\tau_{\alpha} \geq 1$, $\alpha=1,\ldots,h^{1,1}+4$, 
    \item  $\tau_{a}\in [25,40]$ for some $ a \in \{1,\ldots,h^{1,1} \}$, and 
    \item $m_a = 10^{-18}$  eV for at least one $ a \in \{1,\ldots,h^{1,1} \}$.
\end{enumerate}
One way to search for such a point is to begin at the tip of the stretched K\"ahler cone (see \cite{Demirtas:2018akl}), and then perform a homogeneous dilatation $J \rightarrow \lambda J$ until all three requirements are met. We provide further details on the construction of our ensemble in Section \ref{sec:const_ensemble}.

\subsection{The scale of the potential} \label{sec:prefactor}

In \eqref{eq:pdef} we introduced an overall prefactor $\mathcal{P}$ appearing in the F-term potential \eqref{eq:fterm}, and entering the axion masses as 
\begin{equation} \label{eq:massscale}
    m_a^2 \propto  \mathcal{P} \,\frac{(W_0\tau_a/\mathcal{V}^2)\exp(-2\pi\tau_a)}{f^2}\,,
\end{equation}
where $W_0$ is the vev of the flux superpotential.
We will now show that
the numerical value of $  \mathcal{P} \cdot W_0$
has very little impact on our analysis of fuzzy axion DM abundance.
Because the constants $W_0$ and $\mathcal{P}$ enter on the same footing via their product, we can without loss of generality discuss only $W_0$ below.
 
Under a dilation of the K\"ahler parameters 
\begin{equation}
 t \mapsto \lambda t\,,   
\end{equation} 
we have $\tau \mapsto \lambda^2\tau$, $\mathcal{V}\mapsto \lambda^3\mathcal{V}$, and $f \mapsto f/\lambda^2$.
Thus, the mass \eqref{eq:massscale} depends on 
$\lambda$ only through the exponential, 
i.e.~$m_a^2 \propto \exp(-2\pi\lambda^2\tau_a)$. 
Now consider changing $W_0 \mapsto cW_0$.
For any single desired axion $a$ --- for our purposes, this is usually the fuzzy axion --- we can keep the mass $m_a$ fixed by compensating  
the change in $W_0$ by a suitable dilation $t \mapsto \lambda(c) t$. 
Specifically, $m_a$ is fixed under
\begin{equation}\label{eq:comp}
    \Bigl(W_0, t\Bigr) \mapsto \left(cW_0, \Bigl(1 + \ln(c)/2\pi\tau_a\Bigr)^{1/2}t\right)\,.
\end{equation}
Thus, the net effect of changing $W_0$ and performing a compensating dilation to keep $m_a$ fixed is a logarithmic correction to $f$: 
\begin{equation}\label{eq:fchange}
   \delta f = \Bigl(1 + \ln(c)/2\pi\tau_a\Bigr)^{-1}f\,.
\end{equation}
For the fuzzy axion, a typical divisor volume will be $\tau_a \approx 35$.  Thus, unless the change in $c$ is extremely large, the impact on the fuzzy axion decay constant is negligible.
The most extreme change in $W_0$ we could see is about $c \approx 10^{-9}$ (for $W_0 \lesssim 10^{-9}$, $T_R$ falls below $5$ MeV --- see \eqref{eq:TR_and_W0}), for  
which the decay constant $f$ of a fuzzy axion would change by a factor of $\approx 1.1$.

We conclude that changing $W_0$, $\mathcal{P}$, or both in such a way that $ \mathcal{P} \cdot W_0$ changes by a factor $c$, and then performing a compensating rescaling of the CY to keep the fuzzy axion mass fixed, has the net effect of a small correction to the fuzzy axion decay constant, unless $c$ is exponentially large.   On the other hand, the masses of \emph{other} axions will change under \eqref{eq:comp}.

We can estimate the value of the prefactor $\mathcal{P}=e^{K_{\rm{other}}}$ in \eqref{eq:fterm}.
In the conventions of \cite{McAllister:2024lnt}, 
one finds
\begin{equation}
e^{K_{\rm{other}}}  = \frac{g_s}{128\widetilde{V}}\,,
\end{equation}
where $\widetilde{V}$ is the volume of the threefold $\widetilde{X}$ that is the mirror of $X$.
To model $\widetilde{V}$, we suppose for concreteness that $W_0 \ll 1$ arises from a perturbatively flat vacuum as in \cite{Demirtas:2019sip}.
Then one finds \cite{McAllister:2024lnt} 
\begin{equation}
\label{eq:P_set}
    \widetilde{V} \sim g_s^{-3} \Rightarrow \mathcal{P} \approx g_s^4/128\,.
\end{equation}
For the reasonable reference value $g_s=0.5$, we have $\mathcal{P} \sim 5\cdot 10^{-4}$, and this is the value we have used in our main analysis.

Furthermore, one may now also establish a connection between the reheating temperature $T_R$, and the flux superpotential $W_0$. Here, we assume that reheating is driven by the lightest massive modulus of mass $m \sim M_{\rm{SUSY}}$. Using Eqs. \eqref{eq:Gamma_R}, \eqref{eq:Lambda_and_m3/2}, and \eqref{eq:P_set}, along with $H(T_R) = \Gamma_R$, 
and
$K = -2\log(\mathcal{V})$, we have
\begin{align}
\label{eq:TR_and_W0}
    H_{R}(T_R)= \frac{g_s^{6}\,W^3_0\,\mpl^3}{2^{33/2}\pi \Lambda^2\mathcal{V}^{3}}\,.
\end{align}
In the above, we have also restored the factors of $\mpl$. Then, for typical values of $g_s=0.5$, $W_0=10^{-5}$, $\mathcal{V}=10^3$, and $\Lambda \sim \mpl$, we have $H_R(T_R) \sim 10^{-4}$ eV, or 
equivalently $T_R\sim 300$ GeV.

Under these assumptions, a choice of $W_0$ determines $T_R$. In the examples in Section~\ref{sec:examples}, we will use the value of $T_R$ (following the prescription of Section~\ref{se:choosecosmology}) to self-consistently set $W_0$. On the other hand, in the larger ensemble of models that we consider in Section~\ref{sec:stats}, we will make a different assumption: we will set the value of $W_0$ and $T_R$ independently, implicitly making the assumption that the scale of reheating is not set by the gravitino mass.

\subsection{Hierarchical axion decay constants}\label{sec:fiber}

As discussed in the Introduction, for generic points in the moduli space of CY manifolds, as measured by the Weil-Petersson measure on moduli space, the existence of fuzzy axion dark matter is accompanied by an overabundance of heavier axions, absent any other tuning. In this section, we will describe a geometric structure that, if present, can facilitate a \textit{geometric} tuning of the relative abundance of heavy axions, including the QCD axion. 

Consider a CY threefold whose volume \eqref{eq:kahlerpot} takes the following form\footnote{We thank Michele Cicoli for pointing us to geometries with this structure to achieve hierarchies.}
\begin{align}
    \mathcal{V} = \sqrt{\tau_{1}} \tau_2 - \sum_i \tau_i^{3/2}\,,
    \label{eq:fibervolume}
\end{align}
where $i=3,\ldots,h^{1,1}$ and $\tau_i \coloneqq \text{vol}(D_i)$ for all divisors $D_i$ in the basis. Such structures can arise in geometries that exhibit K3 fibrations, where $\tau_1$ is the volume of the K3 fiber. This structure will be utilized in Section~\ref{ex:fiber} to find a suitable example with hierarchical axion decay constants. In this work, we will not address the relationship between the existence of a K3 fiber and a volume of the form \eqref{eq:fibervolume}---instead, we will simply use the structure of \eqref{eq:fibervolume} to engineer hierarchies between axion decay constants in CY threefolds.

In this geometry, we would like to engineer a limit in which:
\begin{itemize}
    \item One divisor, $\tau_{\mathrm{fuzzy}}$, can host the fuzzy axion, i.e. has volume $\mathcal{O}(10)$.\footnote{Note that in this section, we are assuming, for simplicity, that the fuzzy axion is a mass eigenstate, such that its mass is dominantly set by the contribution of a Euclidean D3-brane on $D_{\mathrm{fuzzy}}$.}
    \item One divisor, $\tau_{\mathrm{QCD}}$ can host QCD, i.e. has volume $\mathcal{O}(25-40)$.
    \item The fuzzy axion has a large decay constant, in order to produce a significant fraction of dark matter.
    \item The QCD axion has a small decay constant to avoid overproduction of dark matter without the need to tune $\theta_a$.
\end{itemize}

In Section~\ref{ex:fiber}, we will give an explicit example that exhibits these features, but in this section we will outline the mechanism that allows one to engineer such a limit in geometries with volume forms given by \eqref{eq:fibervolume}.

For clarity, we take $h^{1,1} = 3$. For simplicity, we will parameterize the divisor volumes as 
\begin{align}
    \tau_1 \equiv \tau_{\mathrm{fuzzy}} = \lambda, \ \ \tau_2 \equiv \frac{\lambda}{\epsilon}, \ \ \tau_3 \equiv \tau_{\mathrm{QCD}} = \lambda\,.
    \label{eq:tauparams}
\end{align}
To illustrate the mechanism, we will consider the limit where $\lambda$ is fixed, and $\epsilon \rightarrow 0$.
 
The K\"ahler metric \eqref{eq:KM} is given by
\begin{align}
   K_{ij} = \frac{1}{2\mathcal{V}^2} 
   \begin{pmatrix}
    \frac{\tau_2 (2 \sqrt{\tau_1} \tau_2 - \tau_3^{3/2})}{2\tau_1^{3/2}}  & \frac{\tau_3^{3/2}}{\sqrt{\tau_1}} & -\frac{3 \tau_2 \sqrt{\tau_3}}{2\sqrt{\tau_1}}  \\[1em]
    \frac{\tau_3^{3/2}}{\sqrt{\tau_1}} & 2 \tau_1 & -3 \sqrt{\tau_1 \tau_3} \\[1em]
    -\frac{3 \tau_2 \sqrt{\tau_3}}{2\sqrt{\tau_1}} & -3 \sqrt{\tau_1 \tau_3} & \frac{3 (\sqrt{\tau_1} \tau_2 + 2 \tau_3^{3/2})}{2\sqrt{\tau_3}} 
    \end{pmatrix}\,.
\end{align}

We now plug in the parameterization \eqref{eq:tauparams}, and take the limit $\epsilon \rightarrow 0$. The resulting K\"ahler metric is approximately  
\begin{align}
  K_{ij} \approx  \frac{1}{2 \lambda^2} \begin{pmatrix}
        1 & \epsilon^2 & -\frac{3}{2} \epsilon \\[1em]
        \epsilon^2 & 2 \epsilon^2 & -3 \epsilon^2 \\[1em]
        -\frac{3}{2} \epsilon & -3 \epsilon^2 & \frac{3}{2} \epsilon
    \end{pmatrix}\,.
\end{align}
Assuming the square roots of the eigenvalues of $K_{ij}$ provide a good approximation of the true axion decay constants (which we will compute explicitly in Section~\ref{ex:fiber}), we can read off the scalings:
\begin{align}
    f_1 &\sim \mathcal{O}(1)\\
    f_2 &\sim \epsilon\\
    f_3 &\sim \sqrt{\epsilon}\,.
\end{align}
Thus, by taking $\tau_2 \gg 1$, we can achieve a parametric hierarchy between $f_1$ and $f_3$, as well as between $f_1$ and $f_2$. In particular, since $f_3$ represents the decay constant of the QCD axion, one can see that if an extreme enough limit is taken in $\tau_2$, the QCD axion decay constant can be suppressed enough to evade any limits on dark matter abundance.

The simple structure showcased in this section can become complicated in realistic examples. First of all, it is not clear that a limit of the type described here can be taken in general high-dimensional moduli spaces. In addition, a key feature of this toy example was that the field playing the role of the fuzzy axion received its mass dominantly from Euclidean D3-branes wrapping the \emph{same} cycle as the one that gives rise to the axion in 4D. In some examples, going to such extreme limits in moduli space induces more important contributions to the mass of this axion from Euclidean D3-branes on other cycles, rendering it no longer an ultralight field. It is also not clear that this mechanism can always provide a tuning such that all heavy axions have small enough decay constants to not overproduce dark matter. Finally, the utility of this mechanism depends on the prevalence of geometries with a volume form given by \eqref{eq:fibervolume}, see e.g. \cite{Avram:1996pj,Cicoli:2011it,Crino:2022zjk,Carta:2022web,Shukla:2022dhz} for progress in this direction. This requires further investigation.

\section{Results}\label{sec:bestiary}

In this section, we summarize our results for stringy models with fuzzy and QCD axions. We begin by constructing a database of CY orientifolds in the range 
$2 \le h^{1,1} \le 7$, and we explain how we generated
an ensemble of models by sampling points in K\"ahler moduli space.
We then present the full details of six examples in order to illustrate key phenomena.
Finally, we study the statistics of cosmological observables in our ensemble.

\begin{table}[t!]
\centering
\resizebox{\columnwidth}{!}{
\begin{tabular}{|c|c|c|c|c|c|c||c|}
\hline 
&  &  &  &  &  &  & \\ [-0.8em]
$h^{1,1}$ & 2 & 3 & 4 & 5 & 6 & 7  & total \\ [0.3em]
\hline 
\hline 
 &  &  &  &  &  &  & \\ [-0.8em]
favorable polytopes & 36 & 243 & 1,185 & 4,897 & 16,608 & 48,221 &  71,190 \\ [0.3em]
\hline 
 &  &  &  &  &  &  & \\ [-0.8em]
\makecell{FRST class} & 36 & 274 & 1,760 & 11,713 & 74,503 & 467,283 & 555,569 \\ [0.3em]
\hline 
\hline 
 &  &  &  &  &  &  & \\ [-0.8em]
\makecell{CYs with \\ inherited involutions} & 32 & 253 & 1,559 & 9,530 & 54,274 & 292,158 & 357,806 \\ [0.8em]
\hline 
 &  &  &  &  &  &  & \\ [-0.8em]
\makecell{CYs with inherited \\ $h^{1,1}_- = 0$ involutions} & 32 & 253 & 1,554 & 9,459 & 53,810 & 289,684 & 354,792 \\ [0.8em]
\hline  
 &  &  &  &  &  &  & \\ [-0.8em]
\makecell{CYs with inherited \\ $h^{1,1}_- = h^{2,1}_+ = 0$ involutions} & 11 & 66 & 267 & 1,033 & 3,623 & 12,253 & 17,253 \\ [0.8em]
\hline 
\hline 
 &  &  &  &  &  &  & \\ [-0.8em]
\makecell{models in ensemble} & 2 & 263 & 3,348 & 29,898 & 231,676 & 1,565,380 & 1,830,567 \\ [0.3em]
\hline 
\end{tabular} 
}
\caption{Number of CY orientifolds obtained in our scan.  
}\label{tab:ScanData}
\end{table}

\subsection{Construction of an ensemble}\label{sec:const_ensemble}
 
In this section, we define a model as a compactification of type IIB string theory in which we have specified:
\begin{enumerate}
    \item An orientifold of a CY threefold.
    \item A four-cycle chosen as a candidate to host D7-branes realizing QCD.\footnote{As explained in Section \ref{sec:stsetup}, although we do find explicit orientifolds, in this work we do not actually engineer the SM on D7-branes: we merely find four-cycles of suitable size.}
    \item A point in moduli space at which we compute the axion effective theory: namely, a point in K\"ahler moduli space, as well as values for $g_s$ and $W_0$.
\end{enumerate}
We enforce that the CY threefolds $X$ that we study satisfy the following properties:
\begin{enumerate}
    \item \label{enum:cy_conditions_1} $X$ arises from a favorable polytope,
    \item \label{enum:cy_conditions_2} $2 \leq h^{1,1}(X) \leq 7$, and
    \item \label{enum:cy_conditions_3} $X$ admits an $h^{1,1}_- = 0$ orientifold involution of O3/O7 type, inherited from the ambient toric variety, as detailed in Section \ref{sec:CYandOrientifold}. 
\end{enumerate}

We define the set
\begin{equation}
    \label{eq:ensemble}
    \mathcal{S} = \Big\{ \text{CY FRST class } X \text{ from KS dataset} \; \; \Big| \; \; X \text{ satisfies items \ref{enum:cy_conditions_1}-\ref{enum:cy_conditions_3}}\Big\}\,,
\end{equation} 
which will comprise the set of CY threefold topologies  in our ensemble. 

\begin{figure}
    \centering
    \includegraphics[width=0.7\linewidth]{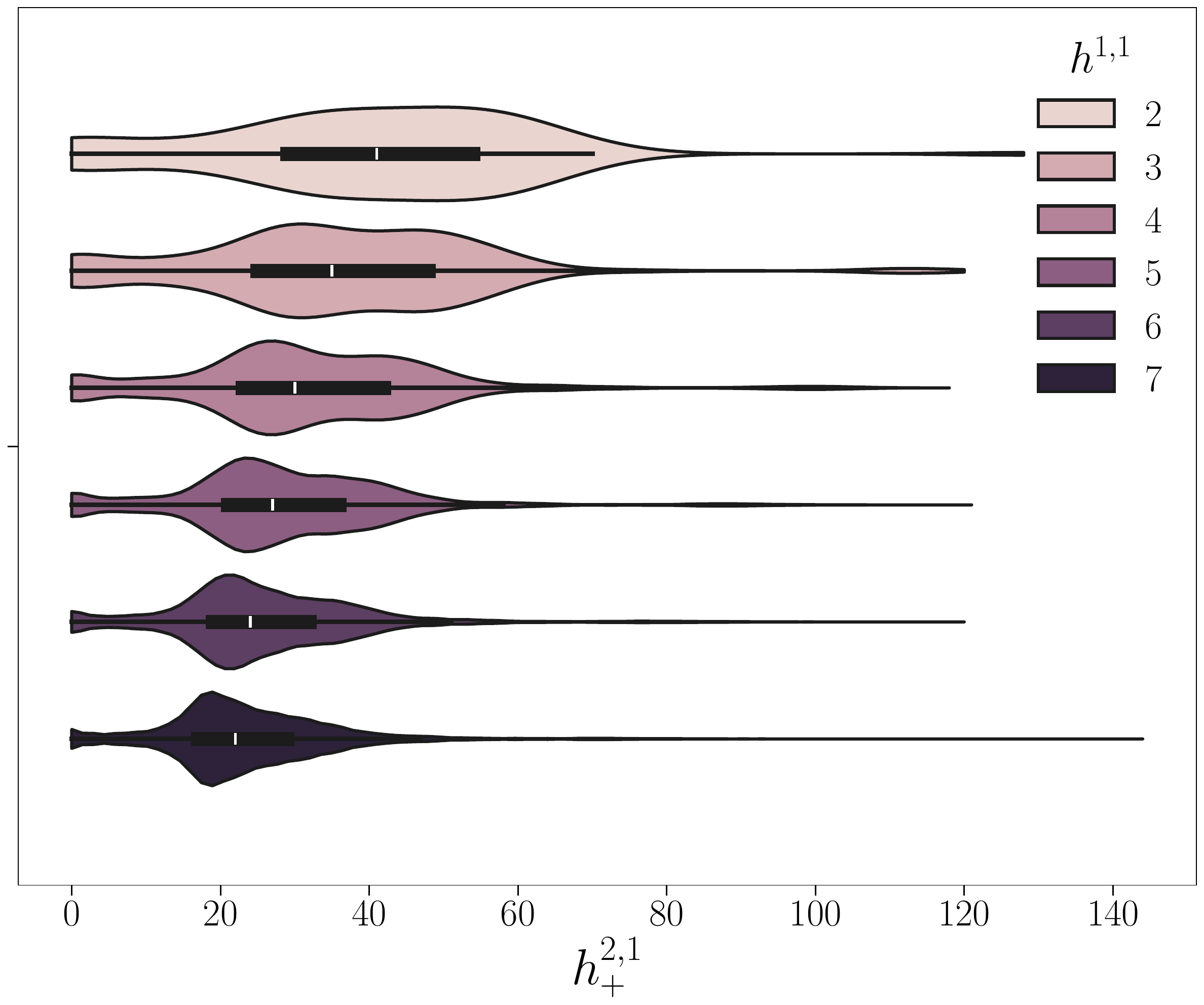}
    \caption{Distribution of $h^{2,1}_+$ for all inherited orientifolds of all two-face inequivalent geometries with $2 \leq h^{1,1} \leq 7$.}
    \label{fig:h21trend}
\end{figure}

To identify orientifolds with $h^{1,1}_- = 0$, we first generated all inherited orientifolds. This is the first time the methods of \cite{Moritz:2023jdb} have been applied at scale, so we take some time now to briefly discuss the statistics of the orientifolds we identified. Our numerical results are tabulated in \cref{tab:ScanData}, and in the range of Hodge numbers we study, we observe the following orientifold properties:
\begin{enumerate}
    \item Only $\sim$ 64\% of triangulations in this range admit an orientifold.
    \item $h^{1,1}_{-} = 0$ orientifolds are ubiquitous. $\sim$ 97\% of the orientifolds have this property and were previously classified in \cite{Crino:2022zjk}.
    \item $h^{2,1}_+ = 0$ orientifolds are rare: $\sim$ 5\% of the generated orientifolds have this property. In particular, in this range of Hodge numbers,
    the inherited orientifolds with $h^{2,1}_+ = 0$ 
    are precisely the orientifolds of \emph{trilayer polytopes} \cite{Moritz:2023jdb} resulting from the associated trilayer involution.   
    This involution is compatible with all triangulations of trilayer polytopes.
\end{enumerate}

We recall that our
effective theories contain $h^{2,1}_+$ U$(1)$ vector multiplets. 
In Fig.~\ref{fig:h21trend}, we show the distribution of $h^{2,1}_+$ for different values of $h^{1,1}$. We observe a clear trend of  $h^{2,1}_+$ decreasing as $h^{1,1}$ increases. This behavior can be attributed to the distribution of reflexive polytopes in the KS database: namely, that $h^{1,1}$ and $h^{2,1}$ are inversely related.\footnote{We comment that Fig.~\ref{fig:h21trend} and \Cref{tab:ScanData} perform slightly different counts: the former includes all orientifolds for all CYs, whereas the latter enumerates CYs with at least one orientifold satisfying a certain property. For example, if a CY admits $N$ orientifolds, it would contribute $N$ times to the figure but only one time to the third row of the table.}
Thus, at small $h^{1,1}$ values where fuzzy axions are most common, the number of possible dark photons is large.
We discuss the physics of dark photons in Section \ref{sec:darkphotons}.

Finally, regarding K\"ahler moduli space, we recall that we desire a point fulfilling the conditions of Section \ref{sec:KC}.
For most examples below, we achieve this by distinguishing a specific point $\vec{t}_0$ in the K\"ahler cone --- the tip of the stretched K\"ahler cone --- and for each choice of prime toric divisor hosting QCD, selecting all points on the ray generated by this form (i.e. points of the form $\lambda \vec{t}_0$) that satisfy the above three criteria.   

We summarize the algorithm described in this section in Algorithm \ref{alg} and detail the number of polytopes, the size of $\mathcal{S}$, and the number of eventual models in Table \ref{tab:ScanData}.

\begin{algorithm}[h]
\caption{Algorithm for generating the studied ensemble of axion models.}
\label{alg}
\begin{algorithmic}
\For{CY $X$ in $\mathcal{S}$ \eqref{eq:ensemble}} 
    \State let $\vec{t}_0$ denote the K\"ahler parameters at the tip of the stretched K\"ahler cone for $X$
    \For{$D \in$ $h^{1,1} + 4$ prime toric divisors of $X$} 
        \For{$\lambda \in \{\lambda \in \mathbb{R}_+ \; | \; \text{ at } \lambda \vec{t}_0, \text{ criteria of \S\ref{sec:KC} are satisfied for } D_\mathrm{QCD} = D \}$}
            \State append model with K\"ahler parameters $\lambda \vec{t}_0$ to ensemble
        \EndFor
    \EndFor
\EndFor
\end{algorithmic}
\end{algorithm}

\subsection{Examples}\label{sec:examples}

\subsubsection{$h^{1,1}=2$}\label{ex:h11_2}

\begin{figure}[t]
    \centering 
    \includegraphics[width=0.9\linewidth]{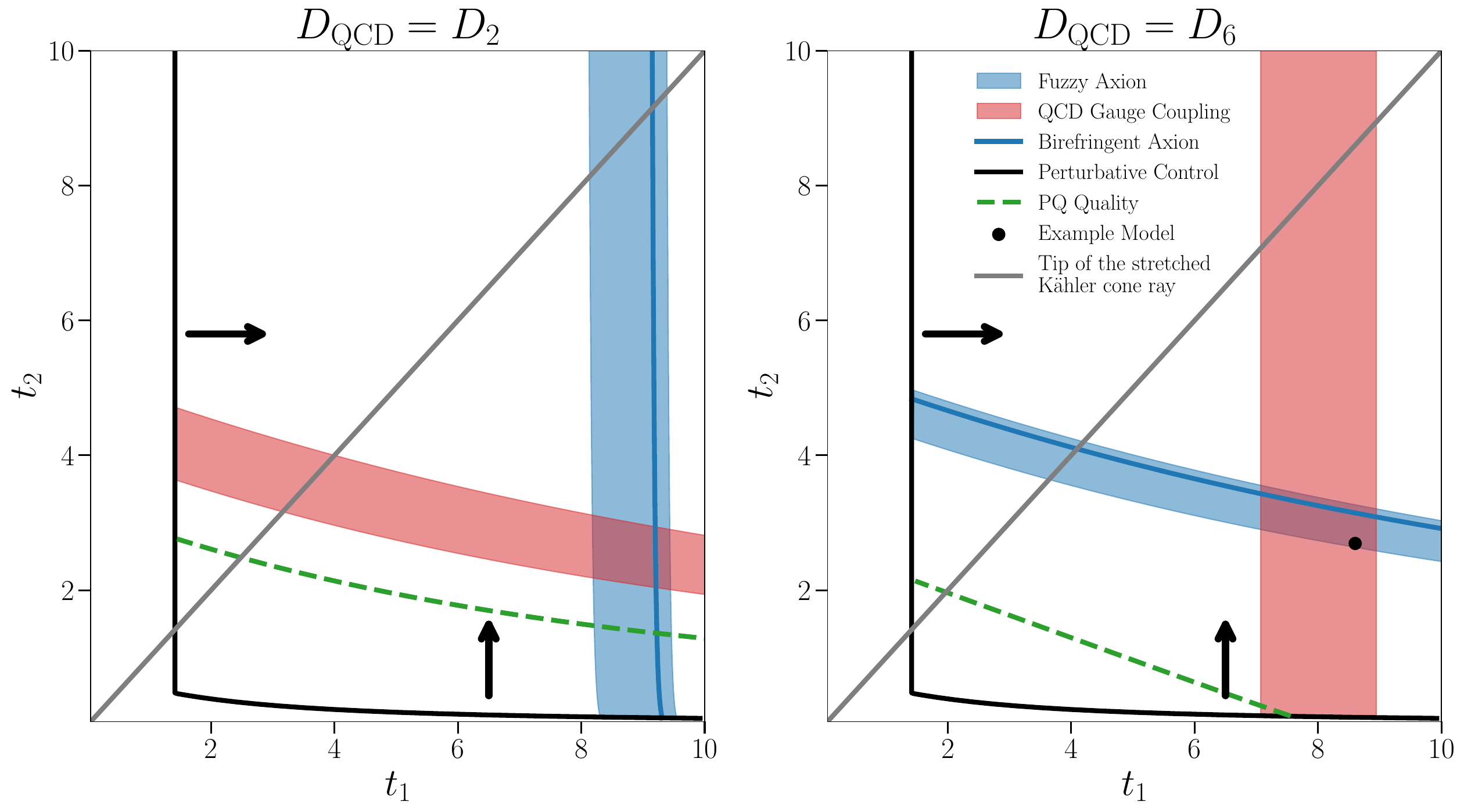} 
    \caption{
    K\"ahler cone for the geometry with $h^{1,1} = 2$ exhibited in \Cref{ex:h11_2}, for two choices of the prime toric divisor where QCD is hosted. Regions are shaded where fuzzy masses are achieved (blue) and where the correct QCD coupling is achieved (red). Contours are included denoting the boundary of perturbative control (black), the boundary of PQ quality (green), where an axion mass is $10^{-30}$ eV (dark blue; a prototypical value for realistic contributions to birefringence), and the ray generated by the tip of the stretched K\"ahler cone (gray).}
    \label{fig:h_2_cone_v2}
\end{figure}

We begin with an example that illustrates how the criteria laid out in Section \ref{sec:KC} correspond to particular loci in K\"ahler moduli space. By selecting a geometry with $h^{1,1} = 2$, the K\"ahler cone and these loci can be plotted in the plane without slicing or projecting, as we exploit in Fig.~\ref{fig:h_2_cone_v2}.
We consider a 4D reflexive polytope $\Delta^\circ \subset N$ with points given by the columns of
\begin{equation}\label{eq:firstpolytope}
\begin{pmatrix}
1 & -1 & 0 & 0 & -2 & 0 \\
0 & 3 & 0 & 0 & -1 & 1 \\
0 & -2 & 0 & 1 & 0 & 0 \\
0 & -1 & 1 & 0 & 0 & 0
\end{pmatrix}\,.
\end{equation}
A choice of height vector (specified in the supplementary material) induces a triangulation that defines
a toric variety 
in which the generic anticanonical hypersurface is a smooth CY threefold with $h^{1,1} = 2$ and $h^{2,1} = 132$. Such a hypersurface inherits an orientifold from the ambient variety, for which the Hodge numbers split as $(h^{1,1}_+, h^{1,1}_-, h^{2,1}_+, h^{2,1}_-) = (2, 0, 0, 132) $.
The toric rays associated to the divisors $D_1, \ldots, D_{6}$ are given by the columns of $\Delta^{\circ}$ in \eqref{eq:firstpolytope}.  
The QCD axion is associated to the prime toric divisor $D_6$, 
and the fuzzy axion is associated to the prime toric divisor $D_2$. 
We set $W_0 = 1$ and fix K\"ahler parameters $\mathbf{t}_\star$ (see supplementary materials for further details). 
Assuming moduli stabilization at $\mathbf{{t}}_\star$, the KK scale is $m_{KK} = 6.5 \cdot 10^{16}$ GeV, the QCD divisor has volume $\tau_\mathrm{QCD} = 37.0 = \alpha_{\rm QCD}(m_{\mathrm{UV}})^{-1}$, and the masses $m_a$ and decay constants $f_a$ of the $h^{1,1} = 2$ axions are 
\begin{equation}
\begin{aligned}
m_a &= (1.8 \cdot 10^{-9}, 5.0 \cdot 10^{-20})\; \mathrm{eV}, \\ 
f_a &= (3.2 \cdot 10^{15}, 1.2 \cdot 10^{16})\; \mathrm{GeV}\,.
\end{aligned}
\end{equation}
In particular, here the first axion is the QCD axion and the second axion is the fuzzy axion. The choice of initial misalignment angles $\theta_a$ that describes the observed dark matter abundance while maximizing the product of the angles is 
$\theta_a = (0.0063, 1)$. Hence, the total tuning is $\delta = 0.0063$. 
For such a choice, $\Omega_a/\Omega_\mathrm{DM} = (0.5, 0.5)$. 
We have deliberately chosen the point along this ray in K\"ahler moduli space such that the abundance is split evenly between the QCD and fuzzy axion.
Cosmologically, because we have no other axions (in particular, no heavy axions) it suffices to perform this initial angle tuning: we have no constraint on $T_R$.

Having studied one choice of K\"ahler parameters, we now turn our attention to the entire K\"ahler moduli space, as displayed in Fig.~\ref{fig:h_2_cone_v2}. 
An instanton charge matrix for the given geometry is
\begin{equation}\label{eq:glsmh112}
    Q = \begin{pmatrix}
        7 & 1 & 1 & 2 & 3 & 0 \\
        2 & 0 & 0 & 0 & 1 & 1 
    \end{pmatrix}\,. 
\end{equation}
In accordance with \cite{Gendler:2023kjt}, we restrict to the $h^{1,1} = 2$ most relevant instantons from among those given by the prime toric divisor classes: in this particular case, these will always be $D_2 = D_3$ and $D_6$, because all other prime toric divisors either have greater volume than both of these classes or is a multiple of one of them. Thus, it is natural to consider hosting QCD on either of these classes: in the above concrete example we consider the latter case, but both are depicted separately in Fig.~\ref{fig:h_2_cone_v2}. 

For each choice of the QCD divisor, we distinguish important regions in K\"ahler moduli space. In red we identify the region where the QCD divisor has a volume consistent with the experimentally observed value for the QCD gauge coupling $\alpha_\mathrm{QCD}$, or $25 \leq \tau_\mathrm{QCD} \leq 40$  (see \eqref{eq:gcoupling}). 
Divisor volumes are quadratic polynomials in the K\"ahler parameters (though they do not necessarily depend on all such parameters: we see in the left panel that the divisor volume $\tau_6$ is independent of $t_2$). 

In blue, we delineate the region where an axion is fuzzy, or has mass $m$ satisfying  $10^{-33} \; \mathrm{eV} \leq m \leq 10^{-18} \; \mathrm{eV}$. We recall that axion masses depend exponentially on the radial direction in the K\"ahler cone, quickly tending toward masslessness as one dilates. While there are two axions in this example, there is only one such region in each panel because the QCD axion cannot be ultralight. This is because its (zero temperature) 
mass is set by \eqref{eq:QCDaxion_mass}, and it would become ultralight only for 
QCD axion decay constants $f_\mathrm{QCD} \sim 10^{25}$ GeV, which is many orders of magnitude higher than what is encountered in the KS landscape (see Fig.~\ref{fig:h11_spread}). Thus, only the non-QCD axion in this example can be ultralight. We emphasize with a solid blue line the locus for which the fuzzy axion attains a mass of $10^{-30}$~eV, a value falling near the center of the preferred mass region for an axion-driven all-sky cosmic birefringence signal.

In black we exhibit the locus where the smallest prime toric divisor has unit volume: the region bounded by the walls of the K\"ahler cone and this locus is subject to strong $\alpha'$ corrections (see Section \ref{sec:KC}). 

With a dashed green line we display the locus where, following the discussion of \cite{Demirtas:2021gsq}, the quality problem in the EFT is just barely solved. Above this line, PQ quality is always solved. We postpone the discussion of the quality problem of the PQ symmetry breaking in our ensemble to \cref{sec:PQq}.
Meanwhile, in this example, we can appreciate the following. From \eqref{eq:glsmh112}, we see that if we set $D_\mathrm{QCD} = D_6$, the term in the axion potential that endangers the PQ solution is set by $D_5$, i.e. the divisor generating the scale $\Lambda_5^4$. It cannot be set by $D_2 = D_3$ or $D_4$ as their charges are orthogonal to $D_6$, or by $D_1$ because its action is always strictly larger than that of $D_5$. On the other hand, if $D_\mathrm{QCD} = D_2$, an analogous argument ensures that the relevant divisor that could spoil the PQ solution is $D_4$. Note that, because $D_4 = 2D_2 = 2D_3$, the boundaries of the red QCD coupling region and the PQ quality contour are parallel. In more complicated geometries, the divisor  that spoils PQ quality need not be a fixed prime toric divisor. \cref{fig:h_2_cone_v2} shows that in the region where we have both the fuzzy axion and the QCD axion, the quality problem is solved. 
 
We conclude by commenting on the gray curve 
in Fig.~\ref{fig:h_2_cone_v2}, which denotes the ray generated by the tip of the stretched K\"ahler cone (tip ray), our canonical choice of angular direction in the K\"ahler cone for the purposes of generating our ensemble of models (see Section~\ref{sec:const_ensemble}). In this example, we see that for both discussed choices for the QCD divisor, the unique region in moduli space where a fuzzy axion and a QCD axion can be simultaneously realized does not intersect the tip ray. This illustrates the sacrifice one makes by restricting to a dimension-one submanifold of moduli space. Indeed, in \Cref{tab:ScanData} we see only two models exist in our ensemble at $h^{1,1} = 2$, in spite of there being $36$ topologies and $6$ prime toric divisors per topology: the tip ray frequently misses the fuzzy $\cap$ QCD region of moduli space. 

\subsubsection{Cosmology illustration}\label{sec:toycosmo}

We now consider models whose spectrum includes heavy, light, and fuzzy axions, in order to illustrate the two reheating scenarios we introduced in Section \ref{se:choosecosmology}, 
as well as our prescription for setting the reheating temperature via $W_0$ \eqref{eq:TR_and_W0}. We consider a 4D reflexive polytope $\Delta^\circ \subset N$ with points given by the columns of
\begin{equation}
\begin{pmatrix}
0 & -1 & 0 & 0 & 2 & -1 & 1 & -1 & 0 & 1 \\
1 & 0 & 0 & 0 & -1 & -1 & 0 & 0 & -1 & -1 \\
0 & 1 & 0 & 1 & -1 & -1 & 0 & 0 & -1 & -1 \\
0 & -1 & 1 & 0 & 0 & 0 & 0 & 0 & 0 & 0
\end{pmatrix}\,.
\end{equation}
A choice of height vector (specified in the supplementary material) induces a triangulation
that defines
a toric variety 
in which the generic anticanonical hypersurface is a smooth CY threefold with $h^{1,1} = 6$ and $h^{2,1} = 46$. Such a hypersurface inherits an orientifold from the ambient fourfold, for which the Hodge numbers split as 
$(h^{1,1}_+, h^{1,1}_-, h^{2,1}_+, h^{2,1}_-) = (6, 0, 19, 27) $. 
Ordering the toric rays as above, the QCD axion is associated to the prime toric divisor $D_5$ and the fuzzy axion is associated to the prime toric divisor $D_{10}$.

We first select K\"ahler parameters $\mathbf{t}_\star$ 
and $W_0$ to realize the prompt reheating scenario, assuming that $W_0$ and $T_R$ are algebraically related as discussed at the end of Section~\ref{sec:prefactor}. In particular, we set $W_0 = 1.5 \cdot 10^{-5}$ and fix K\"ahler parameters $\mathbf{t}_\star$ (see supplementary materials for further details).
Assuming moduli stabilization at $\mathbf{{t}}_\star$, the KK scale is $m_{KK} = 5.3 \cdot 10^{16}$ GeV, the QCD divisor has volume $\tau_\mathrm{QCD} = 34.7$, and the masses $m_a$ and decay constants $f_a$ of the $h^{1,1} = 6$ axions are
\begin{equation}
\begin{aligned}
m_a &= (2.5 \cdot 10^{13}, 4.6 \cdot 10^{-2}, 1.5 \cdot 10^{-9}, 1.2 \cdot 10^{-16}, 1.0 \cdot 10^{-18}, 1.5 \cdot 10^{-21})\; \mathrm{eV}, \\ 
f_a &= (3.7 \cdot 10^{15}, 3.7 \cdot 10^{15}, 3.9 \cdot 10^{15}, 2.2 \cdot 10^{15}, 4.7 \cdot 10^{15}, 7.5 \cdot 10^{15})\; \mathrm{GeV}\,.
\end{aligned}
\end{equation}

We observe that we have populated all of the relevant regions of the axion mass spectrum. We have two heavy axions (those more massive than the QCD axion) with masses $25$ TeV and $46$ meV; the QCD axion, with mass $1.5 \cdot 10^{-9}$ eV; one light axion (those lighter than the QCD axion but heavier than the fuzzy window) with mass $1.2 \cdot 10^{-16}$ eV; and two fuzzy axions, one with mass $10^{-18}$ eV by construction which will dominate the fuzzy relic abundance, and another with mass $1.5 \cdot 10^{-21}$ eV 
which will provide a subleading contribution to the abundance. The induced reheating temperature from the choice of $W_0$ is $T_R = 3.3 \cdot 10^{3}$ GeV. In particular, $3H_R = 4.6 \cdot 10^{-2}$ eV, 
which is comparable to the mass of the lightest heavy axion. 
This realizes our prompt reheating scenario: we have used low-scale inflation to dilute our heavy axions, leaving only those at least as light as the QCD axion. In particular, because our lightest heavy axion was approximately meV scale, we were forced to select a very low reheating temperature: this exemplifies the correlation between heavy axions and the reheating temperature. The choice of initial misalignment angles $\theta_a$ for the four axions which roll after reheating are $\theta_a=(0.0044, 0.6615, 1, 1)$, yielding tuning measure $\delta = 0.0029$. For such a choice, these axions have abundances $\Omega_a/\Omega_\mathrm{DM} = (0.31, 0.31, 0.34, 0.04)$. Here, we see that the presence of a light axion aside from the QCD axion 
forced us to incur an additional fine-tuning penalty (in this case, an extra half-order of magnitude). This illustrates the role of the light axions in setting the total tuning in the prompt reheating scenario.  

We now consider a new choice of K\"ahler parameters and $W_0$ to realize the modulus domination, $w_R=0$, reheating scenario. 
In particular, we now set $W_0 = 1.2 \cdot 10^{-8}$ and fix novel K\"ahler parameters $\mathbf{t}_\star^\prime$ (see supplementary materials for further details).
If we assume moduli stabilization at $\mathbf{t}_\star^\prime$, the KK scale is $m_{KK} = 5.5 \cdot 10^{16}$ GeV, the QCD divisor has volume $\tau_\mathrm{QCD} = 33.5$, and the masses $m_a$ and decay constants $f_a$ of the $h^{1,1} = 6$ axions are
\begin{equation}
\begin{aligned}
m_a &= (1.8 \cdot 10^{12}, 1.2 \cdot 10^{-2}, 1.4 \cdot 10^{-9}, 1.1 \cdot 10^{-16}, 1.0 \cdot 10^{-18}, 1.9 \cdot 10^{-21})\; \mathrm{eV}, \\ 
f_a &= (3.8 \cdot 10^{15}, 3.9 \cdot 10^{15}, 4.0 \cdot 10^{15}, 2.2 \cdot 10^{15}, 4.9 \cdot 10^{15}, 7.8 \cdot 10^{15})\; \mathrm{GeV}\,.  
\end{aligned}
\end{equation}
The induced reheating temperature from this new choice of $W_0$ is $T_R = 1.3 \cdot 10^{-1}$ GeV. In particular, $3H_R = 3.1 \cdot 10^{-11}$ eV, 
so all non-fuzzy axions will start to roll before reheating but do not have their abundance totally diluted given that we now assume a modulus-dominated epoch. 

In this scenario, we choose the product of initial misalignment angles to be close to the maximum value $\delta\approx(\theta_{\rm QCD})^{N_{\rm heavy}+1}\approx 10^{-6}$. The choice of angles $\theta_a$ that minimizes the overall abundance for this value of $\delta$ is $\theta_a = (0.014, 0.014, 0.014, 0.52, 0.72, 1)$. The axions then have abundances $\Omega_a/\Omega_\mathrm{DM} = (0.19, 0.19, 0.19, 0.19, 0.19, 0.040)$. In this scenario the heaviest axion, with mass $m_a = 1.8 \cdot 10^{12}\text{ eV}$, 
is heavy enough that it could decay to two photons on a timescale shorter than the age of the Universe if $c_{a\gamma\gamma}\approx 1$: see 
\eqref{eq:decayrate}. The resulting decay would occur after BBN, and such decays are strongly constrained, requiring $\Omega_a/\Omega_\mathrm{DM} \ll 1$. 
Thus the heavy axion in this scenario would have to be tuned even more strongly if $c_{a\gamma\gamma}\approx 1$.

\subsubsection{Lightest fuzzy abundance}\label{sec:best_abundance}
 
\begin{figure}
    \centering
    \includegraphics[width=0.9\linewidth]{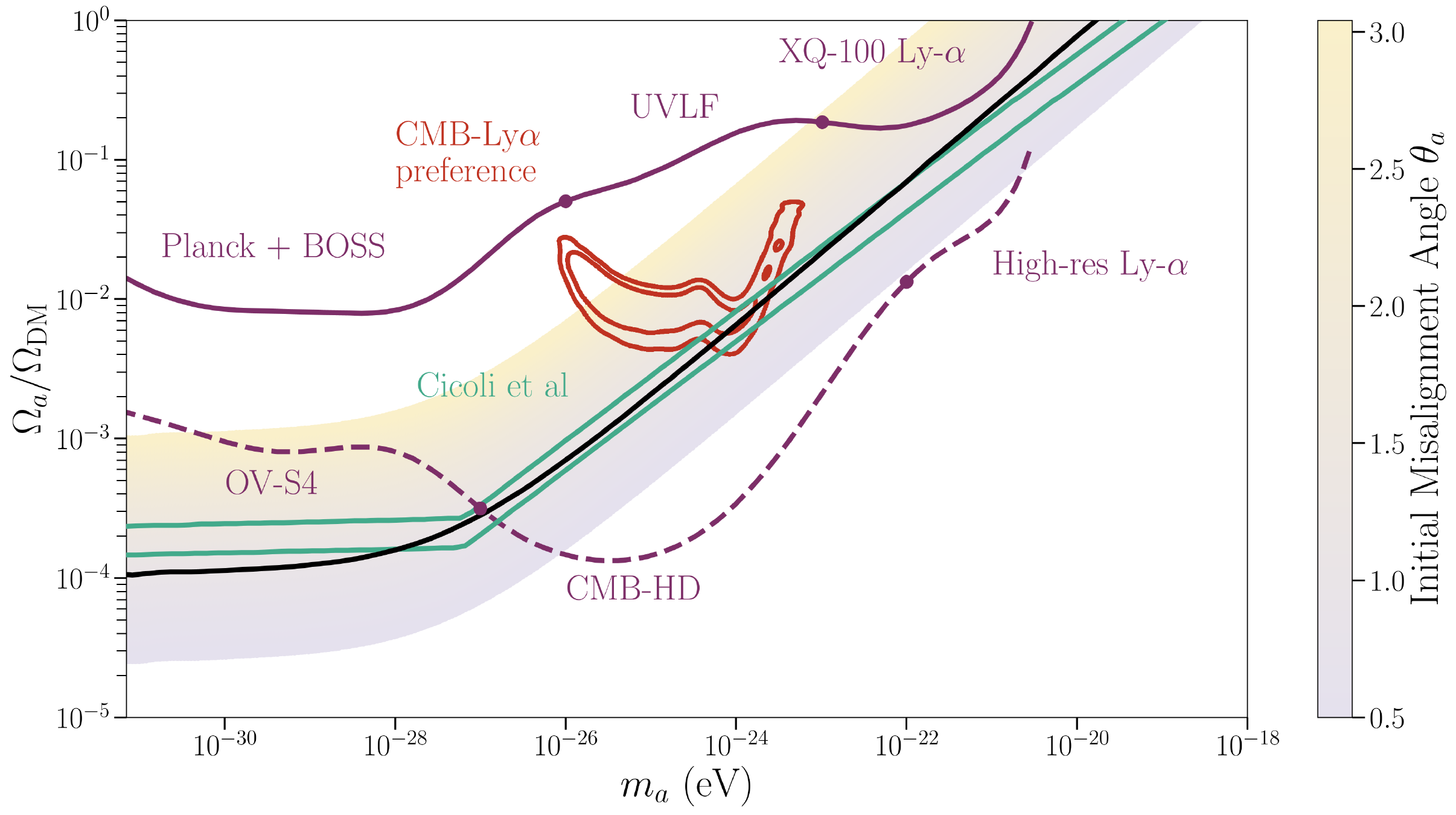}    
    \caption{Relic abundance and mass as a function of overall CY volume and the initial misalignment angle for the maximal abundance example exhibited in Section~\ref{sec:best_abundance}. Interpolations of dark matter constraints and forecasts are overlaid (see Section \ref{sec:observables} 
    for details), along with the abundance prediction from \cite{Cicoli:2021gss} and the preferred region from \cite{Rogers:2023upm}.}
    \label{fig:max_abundance}
\end{figure}

In this section, we present the geometry featuring the largest fuzzy axion decay constant we have identified in our setting. 
Correspondingly, this example contains the lightest fuzzy axion that can constitute 100\% of the observed dark matter. We consider a 4D reflexive polytope $\Delta^\circ \subset N$ with points given by the columns of
\begin{equation}
\begin{pmatrix}
1 & -1 & -1 & 0 & 0 & 0 & 1 & 1 & 0 & 0 & 0 \\
1 & -1 & 1 & 0 & 0 & 1 & 0 & -1 & -1 & 1 & -1 \\
-1 & 2 & -1 & 0 & 1 & 0 & 0 & 0 & 0 & -1 & 1 \\
-1 & 0 & 1 & 1 & 0 & 0 & 0 & 0 & -1 & 0 & 0
\end{pmatrix}\,.
\end{equation}
A choice of height vector (specified in the supplementary material) induces a triangulation
that defines
a toric variety 
in which the generic anticanonical hypersurface is a smooth CY threefold with $h^{1,1} = 7$ and $h^{2,1} = 47$. Such a hypersurface inherits an orientifold from the ambient fourfold, for which the Hodge numbers split as $(h^{1,1+}, h^{1,1-}, h^{2,1+}, h^{2,1-}) = (7, 0, 9, 16)$. Ordering the toric rays as above, the QCD axion is associated to the prime toric divisor $D_1$ and the fuzzy axion is associated to the prime toric divisor $D_2$.

In this example, the K\"ahler parameters $\mathbf{t}_\star$ are obtained by exploring the K\"ahler cone for points with optimal fuzzy misalignment abundance. Hereby, we are leveraging the computational framework provided by the auto-differentiation and optimization libraries \textsc{jax}~\cite{jax2018github} and \textsc{optax}~\cite{deepmind2020jax}. By employing the \textsc{adam}~\cite{kingma2014adam} optimizer, we minimize a carefully designed loss function that encodes the desired geometric properties described in 
Section~\ref{sec:KC}. This approach allows us to systematically navigate the K\"ahler cone and identify optimal K\"ahler parameters efficiently where the untuned misalignment abundance matches the observed value. 
 
We set $W_0 = 1$ and fix K\"ahler parameters $\mathbf{t}_\star$ as determined by our optimization procedure (see supplementary materials for further details).
Assuming moduli stabilization at $\mathbf{{t}}_\star$, the KK scale is $m_{KK} = 6.9 \cdot 10^{16}$ GeV, the QCD divisor has volume $\tau_\mathrm{QCD} = 35.6$, and the masses $m_a$ and decay constants $f_a$ of the $h^{1,1} = 7$ axions are
\begin{equation}
\begin{aligned}
m_a &= (5.9 \cdot 10^{23}, 3.9 \cdot 10^{19}, 5.0 \cdot 10^{4}, 3.8 \cdot 10^{0}, 2.2 \cdot 10^{-7}, 7.7 \cdot 10^{-10}, 1.7 \cdot 10^{-20})\; \mathrm{eV}\,, \\ 
f_a &= (8.0 \cdot 10^{15}, 4.6 \cdot 10^{15}, 3.5 \cdot 10^{15}, 4.8 \cdot 10^{15}, 5.1 \cdot 10^{15}, 7.4 \cdot 10^{15}, 2.2 \cdot 10^{16})\; \mathrm{GeV}\,. 
\end{aligned}
\end{equation} 
In particular, here the sixth axion is the QCD axion and the seventh axion is the fuzzy axion. Choosing a unit initial misalignment angle for the fuzzy axion, its abundance satisfies $\Omega_a/\Omega_\mathrm{DM} = 1$. This model yields the lightest axion we have found whose untuned misalignment abundance achieves the observed value. However,  
the heavier axions in this model will in general overproduce DM.

Alternatively, we can explore more phenomenologically realistic parameters, again letting $W_0$ and $T_R$ be algebraically related as in Section~\ref{sec:prefactor}, setting $W_0 = 4.8 \cdot 10^{-8}$, and considering new K\"ahler parameters $\mathbf{t}_\star^\prime$. At this point in K\"aher moduli space the masses $m_a$ and decay constants $f$ of the $h^{1,1} = 7$ axions are
\begin{equation}
\begin{aligned}
m_a &= (2.0 \cdot 10^{20}, 2.6 \cdot 10^{16}, 3.0 \cdot 10^{2}, 4.2 \cdot 10^{-2}, 6.9 \cdot 10^{-9}, 7.2 \cdot 10^{-10}, 3.2 \cdot 10^{-21})\; \mathrm{eV}, \\ 
f_a &= (8.5 \cdot 10^{15}, 4.9 \cdot 10^{15}, 3.7 \cdot 10^{15}, 5.1 \cdot 10^{15}, 5.5 \cdot 10^{15}, 7.9 \cdot 10^{15}, 2.4 \cdot 10^{16})\; \mathrm{GeV}\,. 
\end{aligned}
\end{equation}
The induced reheating temperature from the choice of $W_0$ is $T_R = 1.4$ GeV (see Section \ref{sec:stsetup}). In particular, $3H_R = 6.9 \cdot 10^{-9}$ eV, 
so only the QCD axion and the fuzzy axion do not have their misalignment abundance diluted away according to our prompt reheating $w_R = -1^{+}$ scenario. 
The choice of initial misalignment angles $\theta_a$ for these two axions that yields the observed dark matter abundance while maximizing the product of the angles is $\theta_a = (0.0036, 1)$: that is, the total misalignment tuning is $\delta = 0.0036$. For such a choice, each of these axions 
has an abundance of $\Omega/\Omega_\mathrm{DM} = 0.5$. 

Finally, we can consider one more choice of K\"ahler parameters and $W_0$ to again realize the modulus domination, $w_R=0$, reheating scenario. 
In particular, we set $W_0 = 1.0 \cdot 10^{-9}$ and fix novel K\"ahler parameters $\mathbf{t}_\star^{\prime\prime}$ (see supplementary materials for further details).
If we assume moduli stabilization at $\mathbf{t}_\star^{\prime\prime}$, the KK scale is $m_{KK} = 7.5 \cdot 10^{16}$ GeV, the QCD divisor has volume $\tau_\mathrm{QCD} = 32.7$, and the masses $m_a$ and decay constants $f_a$ of the $h^{1,1} = 7$ axions are
\begin{equation}
\begin{aligned}
m_a &= (3.3 \cdot 10^{19}, 5.0 \cdot 10^{15}, 1.1 \cdot 10^{2}, 1.8 \cdot 10^{-2}, 3.9 \cdot 10^{-9}, 7.1 \cdot 10^{-10}, 3.0 \cdot 10^{-21})\; \mathrm{eV}, \\ 
f_a &= (8.7 \cdot 10^{15}, 5.0 \cdot 10^{15}, 3.8 \cdot 10^{15}, 5.2 \cdot 10^{15}, 5.6 \cdot 10^{15}, 8.0 \cdot 10^{15}, 2.4 \cdot 10^{16})\; \mathrm{GeV}\,. 
\end{aligned}
\end{equation}
The induced reheating temperature from this new choice of $W_0$ is $T_R = 7.1$ MeV, just above $T_\mathrm{BBN}$. In particular, $3H_R = 6.7 \cdot 10^{-14}$ eV, 
so all non-fuzzy axions will start to roll before reheating and contribute similarly towards the total abundance (modulo heavier axions potentially decaying into photons). 
In this scenario, we are forced to choose an even higher tuning penalty of $\delta = 10^{-9} \approx(\theta_{\rm QCD})^{N_{\rm heavy}+1}$ because there is one more heavy axion than in the previous example. The choice of angles $\theta_a$ that minimizes the overall abundance for this value of $\delta$ is $\theta_a = (0.024, 0.041, 0.054, 0.039, 0.037, 0.025, 0.54)$. The relic abundance is then evenly distributed among all of the axions, each having abundances $\Omega_a/\Omega_\mathrm{DM} \approx 0.14$.

This maximal fuzzy axion misalignment abundance is compared against theoretical predictions along with phenomenological constraints, forecasts, and targets in Fig.~\ref{fig:max_abundance}. The shaded region is produced by dilating the K\"ahler parameters along the ray generated by $\mathbf{t}_\star$ (thereby varying the mass, and subdominantly varying the decay constant) and the initial misalignment angle, which together determine the misalignment abundance. This model is compared to the cosmological constraints (solid) and forecasts (dashed) discussed in Section~\ref{sec:observables}. We observe that there are regions along this ray in moduli space that are already excluded by observations for $\theta_a\approx 1$. In particular, at $m_a=10^{-22}\text{ eV}$ we find models that are excluded by the Lyman-$\alpha$ forest flux power spectrum (Ly$\alpha$) measurements of \cite{Kobayashi:2017jcf} for large values of $\theta_a$. 
These are precisely the constraints illustrated schematically in Fig.~\ref{fig:pk_plot}. A large region of parameter space of this model is accessible to future observations. Furthermore, we illustrate the results of \cite{Rogers:2023upm}, which leads to a preferred region of $(m_a,\Omega_a)$. In the preferred region, fuzzy axions can explain a possible tension between the matter power spectrum as determined by the eBOSS Lyman-$\alpha$ forest compared to that determined by the CMB, by introducing the characteristic `step-like' feature in $P(k)$~\cite{Amendola:2005ad,Arvanitaki:2009fg,Marsh:2010wq,Rogers:2023ezo}.

On the more formal front, we see reasonable agreement between our explicit model and the predictions for four-form axions given in \cite{Cicoli:2021gss}, with discrepancies between the slopes of the respective regions being given by the different ways that the loci in moduli space are constructed. 
We note in particular that predictions of \cite{Cicoli:2021gss} based on moduli stabilization nicely fit as a subset of the model, actually overlapping with the untuned misalignment abundance line. On the cosmological front, we see that for heavy masses in the fuzzy regime and large misalignment angles, this model is excluded by the constraints from the Lyman-$\alpha$ forest, while for masses $\gtrsim 10^{-27}$ eV the model will be probed by future cosmological observations from Lyman-$\alpha$ and CMB-HD. Finally, we see that for intermediate masses and $\gtrsim 1$ initial misalignment angle, this model realizes the region in parameter space identified in \cite{Rogers:2023ezo} for its capacity to alleviate tensions between \textit{Planck} CMB observations and eBOSS Lyman-$\alpha$ forest data.

\subsubsection{Hierarchical decay constants\label{ex:fiber}}

We now exhibit a geometry that realizes the mechanism of Section \ref{sec:fiber} for hierarchical decay constants.
This example features a fuzzy axion and a QCD axion, without the need for contrived reheating or fine-tuned initial conditions to avoid overproduction of axion DM. We consider a 4D reflexive polytope $\Delta^\circ \subset N$ with points given by the columns of
\begin{equation}
\begin{pmatrix}
0 & 0 & 0 & 0 & -2 & 1 & -1 \\
1 & -1 & 0 & 0 & -1 & 0 & 0 \\
0 & -1 & 0 & 1 & 0 & 0 & 0 \\
0 & -1 & 1 & 0 & 0 & 0 & 0
\end{pmatrix}\,.
\end{equation}
In particular, an instanton charge matrix for the given geometry is
\begin{equation}
    \begin{pmatrix}
        1 & 1 & 1 & 1 & 0 & 0 & 0 \\
        0 & -1 & -1 & -1 & 1 & 2 & 0 \\
        0 & 0 & 0 & 0 & 0 & 1 & 1
    \end{pmatrix}\,.
\end{equation}
A choice of height vector (specified in the supplementary material) induces a triangulation
that defines
a toric variety  
in which the  
generic anticanonical hypersurface is a smooth CY threefold with $h^{1,1} = 3$ and $h^{2,1} = 69$. Such a hypersurface inherits an orientifold from the ambient fourfold, for which the Hodge numbers split as $(h^{1,1}_+, h^{1,1}_-, h^{2,1}_+, h^{2,1}_-) = (3, 0, 23, 46) $.

We distinguish the following three divisors, expressed in the basis determined by the above charge matrix.
\begin{equation}
    \begin{aligned}
        D_a &= (0, 2, 1)\,, \\
        D_b &= (6,-2,-1)\,, \\
        D_c &= (0, 1, 0)\,.
    \end{aligned}
\end{equation}
In particular, $D_a$ and $D_c$ are prime toric. The volumes of these divisors are given by the following homogeneous quadratic polynomials in the K\"ahler parameters, adopting the basis dual to the chosen basis of divisors:
\begin{equation}
    \begin{aligned}
        \tau_a &= 2t_1^2\, , \\
        \tau_b &= 4t_1(t_1 + 6t_3)\, , \\
        \tau_c &= (t_2 - 2t_3)^2 \, .
    \end{aligned}
\end{equation}
The CY volume $\mathcal{V}$ can be expressed in terms of these divisor volumes as
\begin{equation}
    \mathcal{V} = \frac{1}{12\sqrt{2}}\sqrt{\tau_a}\tau_b - \frac{1}{3}\tau_c^{3/2}\,.
\end{equation}
In this way, we have achieved the desired volume factorization discussed in Section~\ref{sec:fiber}.\footnote{While we have not studied the necessary and sufficient conditions for such a factorization, we comment that we identified candidate geometries by following \cite{Cicoli:2011it} and searching for CYs with  K3  fibrations and diagonal del Pezzo divisors. In this section we study such a geometry, with $D_a$ the class of the K3 fiber and $D_c$ a diagonal del Pezzo. In particular, $D_c$ has the Hodge diamond of dP$_7$.} Thus, through a careful choice of K\"ahler parameters, we can render $\tau_b \gg 1$ while keeping $\tau_a,\tau_c\sim\mathcal{O}(1)$. In particular, we consider the limit $t_3 \gg 1$ with $t_1 \sim t_2/2t_3 \sim 1$. In this case, all prime toric divisor volumes are proportional to $t_3^2$ aside from the prime toric divisors $D_a$, $D_c$, and $D_7 = (0,0,1)$. 
We choose to host QCD on $D_7$: fixing K\"ahler parameters $\mathbf{t}_\star$ (see supplementary materials for further details) 
we find that the axion basis consists of $D_7$, $D_c$, and the much larger $D_2$. We set $W_0 = 1$ and, assuming moduli stabilization at $\mathbf{{t}}_\star$, the KK scale is $m_{KK} = 8.8 \cdot 10^{11}$ GeV and the QCD divisor has volume $\tau_\mathrm{QCD} = 29.2$. We also comment that the Weil-Petersson measure evaluates to $6.9 \cdot 10^{-18}$ at $\mathbf{{t}}_\star$, while for comparison at the tip of the stretched K\"ahler cone it achieves the value $4.7 \cdot 10^{-4}$.  In this sense, as discussed in the Introduction, the models in our general ensemble are `typical', while regions with decay constant hierarchy are `atypical'. 

At $\mathbf{{t}}_\star$ the masses $m_a$ and decay constants $f_a$ of the $h^{1,1} = 3$ axions are\footnote{The lightest axion has mass $\ll 10^{-100}$ eV, so we can safely treat it as massless.}
\begin{equation}
\begin{aligned}
m_a &= (9.2 \cdot 10^{-6}, 7.1 \cdot 10^{-19}, 0)\; \mathrm{eV}, \\ 
f_a &= (6.2 \cdot 10^{11}, 6.3 \cdot 10^{15}, 2.8 \cdot 10^{8})\; \mathrm{GeV}\,.
\end{aligned}
\end{equation}
Indeed, we identify a decay constant hierarchy of $\mathcal{O}(10^4)$ between the fuzzy and QCD axion, which results in their relic abundances --- in the absence of any initial misalignment angle tuning --- being $\Omega_\mathrm{fuzzy}/\Omega_{\rm DM} = 0.52$ and $\Omega_\mathrm{QCD}/\Omega_{\rm DM} = 0.48$, respectively. 
The massless axion contributes negligible relic abundance. Cosmologically, because we have no other axions (in particular, no heavy axions) we have no constraint on $T_R$ in this model. 

We stress that the K\"ahler parameters $\mathbf{t}_\star$ were not identified using the optimization methods employed in the previous section.
While one could indeed apply those methods to an analogous loss function designed for maximizing decay constant hierarchies, several properties of $\mathbf{t}_\star$ --- e.g., its close proximity to the walls of the K\"ahler cone and the magnitude of $\|\mathbf{t}_\star\|$ (the distance measure seen by the optimization algorithm) --- render this limit difficult for an out-of-the-box optimization algorithm to discover when initialized at more generic points in the moduli space.
We also used Markov Chain Monte Carlo
with the Weil-Petersson measure, and similarly failed to find large hierarchies in this way. 
Instead, $\mathbf{t}_\star$ was found more manually via the geometric construction detailed in Section~\ref{sec:fiber}. 
The failure of brute force
methods underscores the special and fine tuned nature of the hierarchy regions, but also the
utility of geometric reasoning over brute force. We will address tuning measures on moduli
space versus tuning measures on axion initial conditions in a future work.

\subsubsection{Birefringence} \label{sec:birefringence_model}

Cosmic birefringence, i.e., the rotation of the polarization angle of the CMB, can be induced by an axion with $H_0\lesssim m_a\lesssim H_{\rm CMB}$ coupled to $F_{\mu\nu}\tilde{F}^{\mu\nu}$ of electromagnetism~\cite{Carroll:1989vb,Arvanitaki:2009fg}.
In Ref.~\cite{Gendler:2023kjt}
we outlined in detail the physics necessary for a large birefringence angle caused by a $C_4$ axion with the SM realized on D7-branes on 4-cycles. In particular, birefringence is impossible in a GUT and furthermore requires electromagnetism to be realized on a cycle with volume around 40 or larger, or equivalently instanton actions $S\gtrsim 200$~\cite{Alvey:2021hjp,Gendler:2023kjt}. Here, we demonstrate that it is possible to have birefringence, a fuzzy axion with large abundance, and the QCD axion all at once in a model with low $h^{1,1}$.  
The novel feature of realizing birefringence at the same time as a fuzzy axion means that typically the birefringent axion will have large $f_a$, and thus a large abundance also.  

We consider a 4D reflexive polytope $\Delta^\circ \subset N$ with points given by the columns of
\begin{equation}
\begin{pmatrix}
-1 & 0 & 0 & 0 & 1 & -1 & 1 \\
-1 & 0 & 0 & 1 & -1 & 0 & 0 \\
-1 & 0 & 1 & 0 & 0 & 0 & 0 \\
0 & 1 & 0 & 0 & -1 & 0 & 0
\end{pmatrix}\,.
\end{equation}
A choice of height vector (specified in the supplementary material) induces a triangulation
that defines
a toric variety 
in which the generic anticanonical hypersurface is a smooth CY threefold with $h^{1,1} = 3$ and $h^{2,1} = 69$. Such a hypersurface inherits an orientifold from the ambient fourfold, for which the Hodge numbers split as $(h^{1,1}_+, h^{1,1}_-, h^{2,1}_+, h^{2,1}_-) = (3, 0, 29, 40) $. 
Ordering the toric rays as above, the QCD axion is associated to the prime toric divisor $D_7$ and the fuzzy axion is associated to the prime toric divisor $D_6$. 
We set $W_0 = 1$ and fix K\"ahler parameters $\mathbf{t}_\star$ (see supplementary materials for further details).
Assuming moduli stabilization at $\mathbf{{t}}_\star$, the KK scale is $m_{KK} = 8.8 \cdot 10^{16}$ GeV, the QCD divisor has volume $\tau_\mathrm{QCD} = 33.5$, and the masses $m_a$ and decay constants $f_a$ of the $h^{1,1} = 3$ axions are
\begin{equation}
\begin{aligned}
m_a &= (1.1 \cdot 10^{-9}, 8.9 \cdot 10^{-19}, 1.2 \cdot 10^{-32})\; \mathrm{eV}, \\ 
f_a &= (5.3 \cdot 10^{15}, 5.9 \cdot 10^{15}, 9.6 \cdot 10^{15})\; \mathrm{GeV}\,.
\end{aligned}
\end{equation}
In particular, here the first 
axion is the QCD axion and the remaining two are fuzzy. The lightest is the candidate for birefringence. 
For these $(m_a,f_a)$ pairs the relative DM abundances with $\theta_a=1$ are: 
\begin{equation}
    \Omega_a/\Omega_\mathrm{DM} =(2.3\times 10^4, 0.50, 3.3\times 10^{-5})\, .
\end{equation}
The choice of initial misalignment angles $\theta_a$ for the two heavier axions that gives the observed dark matter abundance while maximizing the product of the angles is $\theta_a = (0.0046, 1)$: that is, $\delta = 0.0046$. For such a choice, $\Omega_a/\Omega_\mathrm{DM} = (0.5, 0.5)$. Cosmologically, because we have no other axions (in particular, no heavy axions) it suffices to perform this initial angle tuning: we have no constraint on $T_R$.

We can realize a cosmic birefringence signature from this model via the lightest axion as follows. Let us host 
QED\footnote{We reiterate that we are not constructing the SM explicitly in this work: we are only identifying suitable four-cycles on which one could aim to place seven-branes hosting the SM gauge group factors.}
on the prime toric divisor $D_1$, which is the divisor associated to the axion with mass $\approx 10^{-32}$ eV. 
Thus, following \cite{Gendler:2023kjt}, we can compute that the axion-photon coupling is $g_{a\gamma\gamma} = 1.2 \cdot 10^{-19}$ GeV$^{-1}$ 
and the relevant quantity $\beta = g_{a\gamma\gamma}f_a \theta_a = 1.2 \cdot 10^{-3} \cdot \theta_a$.\footnote{Following \cite{Gendler:2023kjt}, note that we neglect possible increases in $c_{a\gamma\gamma}$ caused by fermion-induced anomalies or other IR physics~\cite{DiLuzio:2016sbl}. The only IR contribution we include is the pion mixing contribution for the QCD axion~\cite{Srednicki:1985xd}.}

The birefringence angle of the CMB is measured to be $\beta = (5.2\pm 1.7)\cdot~10^{-3}$ (68\% C.L.), and is believed to be consistent with a cosmological origin~\cite{Diego-Palazuelos:2022dsq}. Our model can give $\beta$ within the 68\% C.L. region for $\theta_a\approx 3$. The relative abundance of the birefringent axion will then be $\Omega_a/\Omega_\mathrm{DM} = 3\cdot 10^{-4}$, a factor of two below the forecast sensitivity of CMB-S4 to the OV effect in the mass range of birefringence. The effect of the birefringent axion on structure formation thus appears just out of reach. However, as we noted already, the low mass OV forecast shown in Fig.~\ref{fig:max_abundance} is only available for CMB-S4~\cite{Dvorkin:2022bsc}. The OV effect manifests in high multipoles  in the temperature anisotropies, which will be measured with better signal to noise by CMB-HD~\cite{Farren:2021jcd}. It is thus conceivable that CMB-HD will have the sensitivity to detect the birefringent axion in OV in this model, and warrants further investigation following \cite{Farren:2021jcd}. This particular example also has a high abundance of the fuzzy axion with $m_a\approx 10^{-20}\text{ eV}$, which may be testable from star cluster dynamics~\cite{Marsh:2019bjr,Dalal:2022rmp}. The QCD axion in this model with $m_a=1.1\cdot 10^{-9}\text{ eV}$ is accessible to the proposed direct detection experiment DM-Radio~\cite{DMRadio:2022pkf}. Thus, in this example, it may be possible to simultaneously measure effects of all three axions and determine their masses, abundances, and decay constants, which would provide striking evidence for the axiverse.

\subsubsection{Degenerate axions}\label{ex:degen}

Finally, we consider a model featuring axions with nearly degenerate masses, which introduces subtleties in the calculation of axion observables and motivates our study of fuzzy axion halos in Section~\ref{sec:halo_simulation}.
We consider a 4D reflexive polytope $\Delta^\circ \subset N$ with points given by the columns of
\begin{equation}
\begin{pmatrix}
-1 & 0 & 0 & 0 & 0 & 0 & 0 & 1 \\
0 & -1 & 0 & 0 & 0 & 0 & 1 & 0 \\
0 & 0 & -1 & 0 & 0 & 1 & 0 & 0 \\
0 & 0 & 0 & -1 & 1 & 0 & 0 & 0
\end{pmatrix}\,.
\end{equation}
A choice of height vector (specified in the supplementary material) induces a triangulation
that defines
a toric variety 
in which the 
generic anticanonical hypersurface is a smooth CY threefold with 
$h^{1,1} = 4$ and $h^{2,1} = 68$. In particular, this is a CICY --- the so-called `tetraquadric' in $(\mathbb{P}^1)^4$ --- which has been employed as a CY compactification case study in diverse string phenomenology contexts \cite{buchbinder2014moduli, long2023non, hendi2024learning}. Such a hypersurface inherits an orientifold from the ambient fourfold, for which the Hodge numbers split as $(h^{1,1}_+, h^{1,1}_-, h^{2,1}_+, h^{2,1}_-) = (4, 0, 25, 43) $. 
Ordering the toric rays as above, the QCD axion is associated to the prime toric divisor $D_2$ and the fuzzy axion is associated to the prime toric divisor $D_4$. 
We set $W_0 = 1$ and fix K\"ahler parameters $\mathbf{t}_\star$ (see supplementary materials for further details). 
Assuming moduli stabilization at $\mathbf{{t}}_\star$, the KK scale is $m_{KK} = 1.0 \cdot 10^{17}$ GeV, the QCD divisor has volume $\tau_\mathrm{QCD} = 36.0$, and the masses $m_a$ and decay constants $f_a$ of the $h^{1,1} = 4$ axions are
\begin{equation}
\begin{aligned}
m_a &= (7.6 \cdot 10^{-10}, 3.3 \cdot 10^{-20}, 3.3 \cdot 10^{-20}, 2.6 \cdot 10^{-20})\; \mathrm{eV}, \\ 
f_a &= (7.5 \cdot 10^{15}, 7.9 \cdot 10^{15}, 7.9 \cdot 10^{15}, 7.9 \cdot 10^{15})\; \mathrm{GeV}\,.
\end{aligned}
\end{equation}
In particular, here the first axion is the QCD axion and the remaining are fuzzy. The choice of initial misalignment angles $\theta_a$ which describes the observed dark matter abundance while maximizing the product of the angles is $\theta_a = (0.0038, 1, 1, 1)$. For such a choice, $\Omega_a/\Omega_\mathrm{DM} = (0.5, 0.17, 0.17, 0.16)$. We have deliberately chosen the point along this ray in K\"ahler moduli space such that the dark matter is split 50-50 between the QCD axion and the fuzzy triplet. Cosmologically, because we have no other axions (in particular, no heavy axions) it suffices to perform this initial angle tuning: we have no constraint on $T_R$. We consider novel phenomenology of models with near-degenerate fuzzy axions in Section~\ref{sec:halo_simulation}.   

As discussed in Section~\ref{sec:effpotential}, these mass degeneracies are a sign that our perturbative scheme for the calculation of masses and decay constants is in jeopardy. Indeed, for the lightest three axions, the ratios $\Lambda_3^4/\Lambda_2^4$ and $\Lambda_4^4/\Lambda_3^4$ assumed to be much less than one take on the values $0.98$ and $0.65$, respectively. For this example, we use the high numerical precision Python library \textsc{mpmath} to compute the masses and decay constants for this model by brute force, explicitly computing the eigenvalues of the Hessian and deducing the decay constants from the 
quartic self-coupling as discussed in \cite{Mehta:2021pwf}. We find that the masses computed using the hierarchical approximation differ from the brute force calculation by no more than $5\%$, while the decay constants differ by no more than $35\%$. 

It is worth commenting that in the hierarchical limit \eqref{eq:lambdahier},
each mass eigenstate is canonically paired with a sinusoidal term in the potential,
allowing decay constants to be determined by the relation $m^2 = \Lambda^4/f^2$.
In the presence of mass degeneracies,
one has to select a new definition for decay constants: this can be done, for example, via the axion quartic self-couplings (as in \cite{Mehta:2021pwf}) or the geometric field range (as in \cite{Demirtas:2018akl}). We chose to employ the former definition when computing the decay constants with high precision numerical diagonalization above.

\subsection{Statistics of the fuzzy axiverse}\label{sec:stats}

\begin{figure}
    \centering
    \includegraphics[width=0.9\linewidth]{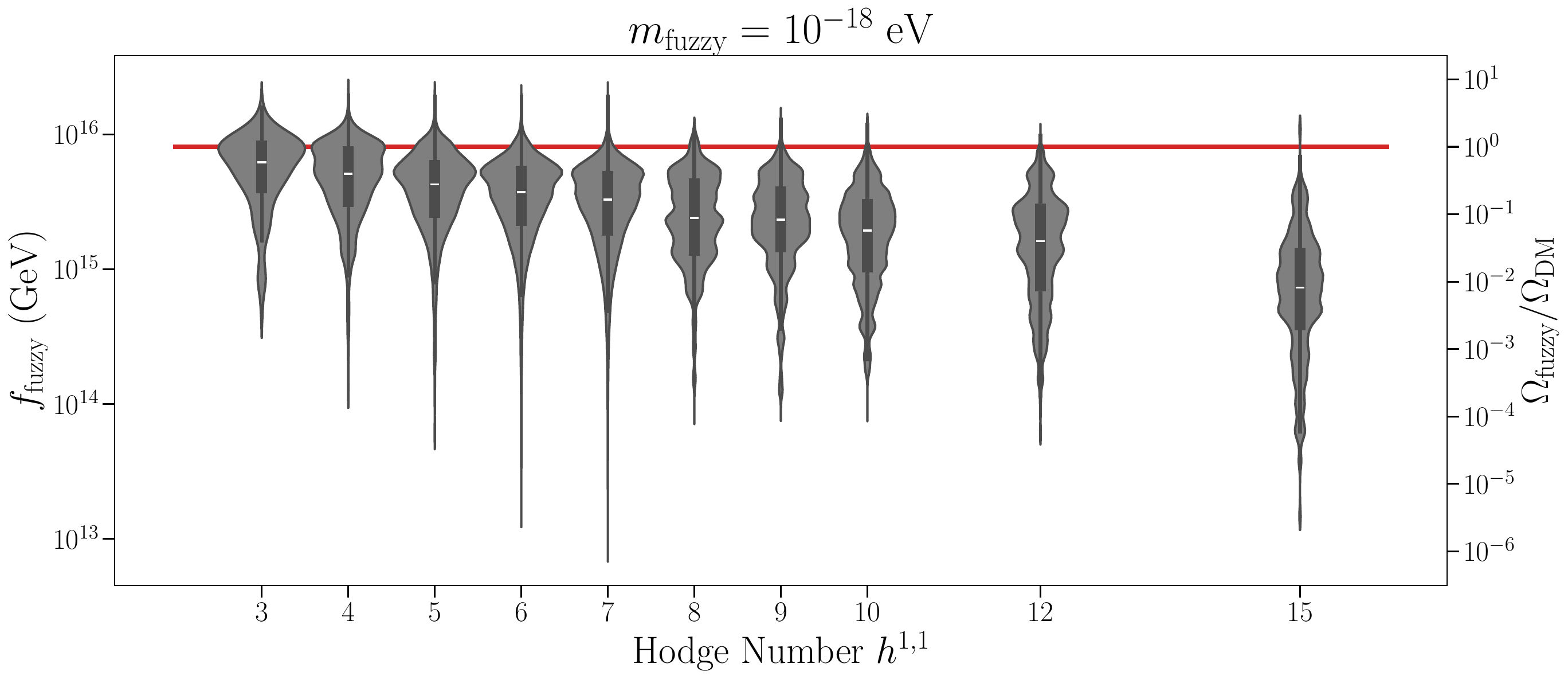} 
    \caption{Distribution of fuzzy decay constants and misalignment relic abundance for different values of $h^{1,1}$. For $h^{1,1} \leq 7$, we exhibit the distribution for the ensemble constructed in Section \ref{sec:const_ensemble}, while for $h^{1,1} > 7$ we coarsely sample polytopes and FRST classes from the KS database.}
    \label{fig:h11_spread}
\end{figure}

In the previous section we showed a small number of examples, including one optimized to maximize the relic abundance of the fuzzy axion. We now turn our attention to the statistics of fuzzy axions in our ensemble, whose construction was detailed in Section \ref{sec:const_ensemble}. 

All models in our ensemble feature an axion with mass $10^{-18}$ eV by construction, but such axions are not always fuzzy: their decay constants may not be large enough to result in non-negligible misalignment relic abundance. Indeed, as mentioned in the Introduction, models giving rise to axions with large enough decay constants typically have small $h^{1,1}$: see Fig.~\ref{fig:h11_spread}. In this plot, we exhibit the distribution of decay constants in our ensemble along with models constructed according to Algorithm \ref{alg} for a coarse subsampling of polytopes and FRST classes at fixed values of $h^{1,1} > 7$.\footnote{For models with $h^{1,1} > 7$, the prefactor induced by the choice $g_s \sim 0.5$ was inserted by hand according to the discussion in Section~\ref{sec:prefactor}.} For the remainder of this section, we will report primarily on subsets of our $2 \le h^{1,1} \le 7$ 
ensemble featuring truly fuzzy axions: that is, a top percentile of our models, when sorted by the relic abundance of the candidate fuzzy axions. 

\begin{figure}[b]
    \centering
    \includegraphics[width=0.9\linewidth]{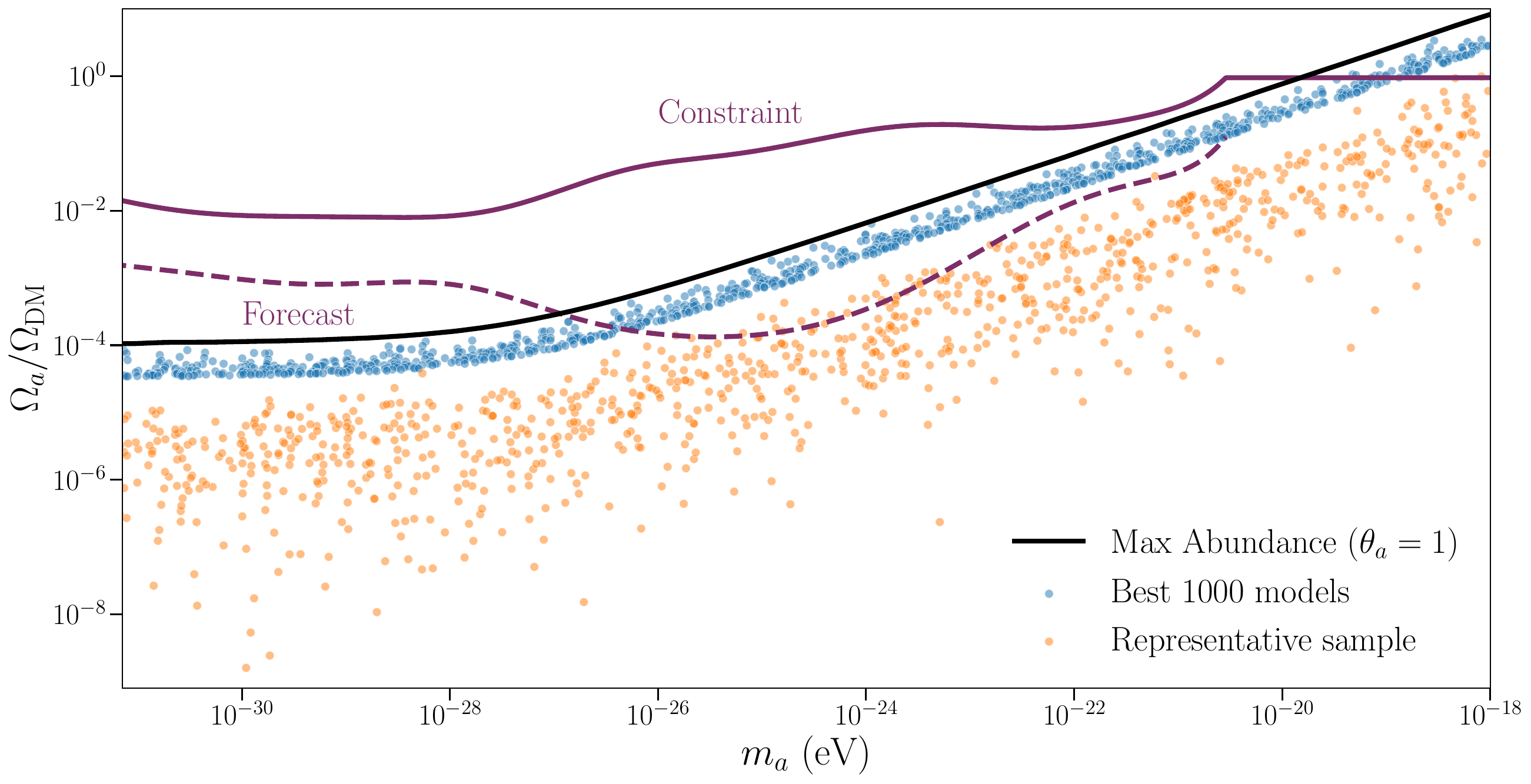}
    \caption{Distribution of masses and misalignment relic abundances for the 1000 models from our ensemble with the largest relic abundance from fuzzy axions ($99.95$ percentile, blue) and a random, representative sample from our ensemble (orange). 
    The spread in masses is by construction (we randomly dilate the Kahler parameters to achieve a log-uniform distribution of masses in the fuzzy range) while the distribution of relic abundances for fixed mass is indicative of the ensemble.
    Overlaid are the same interpolations of dark matter constraints and forecasts from Fig. \ref{fig:max_abundance}, along with the $\theta_a = 1$ contour from the best abundance example from that same figure.}
    \label{fig:population_scatter}
\end{figure}

First, to exhibit the largest relic abundances achieved by fuzzy axions in our ensemble, we select models whose fuzzy abundance falls above the 99.95th percentile (for a total of 1000 models) and randomly dilate along the ray defined by the tip of the stretched K\"ahler cone to place the fuzzy axion at a random mass in the range $[10^{-32} \text{ eV}, 10^{-18} \text{ eV}]$. We additionally apply the same procedure to a random selection of 1000 models, to portray the characteristic relic abundances of the entire ensemble. The results are displayed in Fig.~\ref{fig:population_scatter}. 

Dilation along a ray in the K\"ahler cone moves the models along fairly well-defined contours in $m_a$-$\Omega_a$ 
space: the axion mass decreases exponentially as a function of dilation while the decay constant only decreases polynomially, so to leading order, dilation varies the mass with fixed $f$. The position of any given axion along such a contour in Fig.~\ref{fig:population_scatter} has been generated randomly by construction. The scatter of models perpendicular to these contours is generated by the statistical scatter of CY orientifolds in the KS database. We observe that the scatter is consistent with the fully optimized model: points in the distribution lie close to, but not above, the optimized case. The full population gives rise to much smaller values of $\Omega_a$. We comment that (almost) arbitrarily small values of $\Omega_a$ are achievable in the KS database by passing to geometries with larger values of $h^{1,1}$ due to the aforementioned mean pattern of decreasing decay constants as $h^{1,1}$ increases. 

Additionally, we study the two reheating scenarios that we describe in Section \ref{se:choosecosmology}: namely, prompt reheating and modulus domination. In particular, 
we select $\mathcal{O}(20,000)$ models falling above the $90$th percentile of fuzzy relic abundance among our ensemble, ensuring that this abundance is $\gtrsim 50\%$ of the observed value. With these models, we investigate the distribution of the initial misalignment tuning product $\delta$ (defined in Section \ref{se:choosecosmology} as the product of the initial misalignment angles) and reheating temperatures $T_R$ required to implement each of the two scenarios while avoiding overproduction of dark matter from heavy axions. Let us briefly recall how we select these parameters in each scenario.

In the prompt reheating scenario, we set $T_R$ such that $3H_R$ is the mass of the lightest heavy axion and enforce the observed relic abundance to set the product of initial misalignment angles $\delta$. Qualitatively, the characteristic reheating temperature is set by the number of heavy axions, as they are distributed effectively log-uniformly between the QCD axion mass and the KK scale, meaning more draws results in the lightest heavy axion mass (which sets $T_R$) being smaller. We omit models with no heavy axions from the right panel, as $T_R$ is unconstrained for such models. Analogously, the characteristic initial misalignment tuning is set by the number of light axions: the light axion relic abundance is primarily set by the heaviest light axions (e.g., the QCD axion), and the more light axions 
there are, 
the more populated 
the heavy end of the light axion mass range becomes.

\begin{figure}[b]
    \centering
    \includegraphics[width=0.9\linewidth]{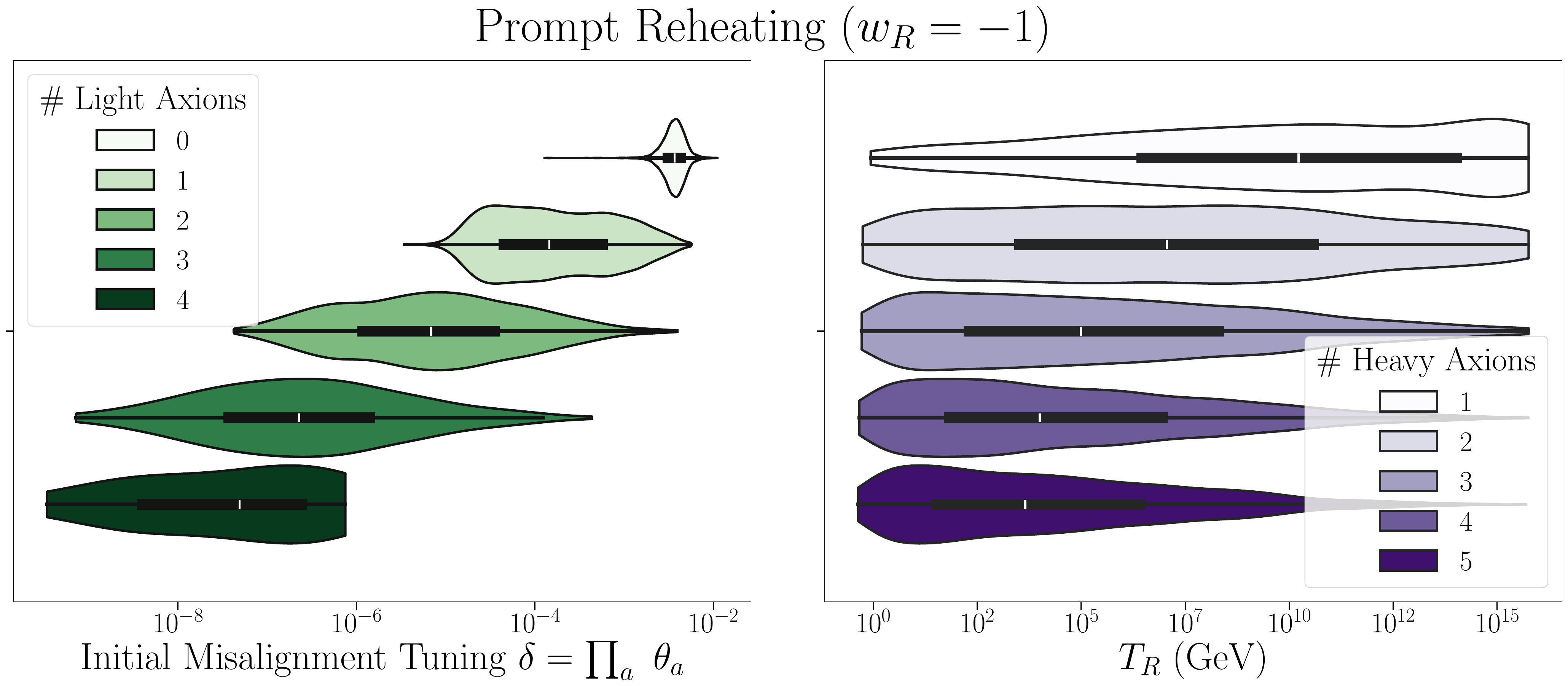}
    \caption{Variation of the free cosmological parameters --- the reheating temperature $T_R$ and the total initial misalignment tuning $\delta$ --- for the prompt reheating ($w = -1$) scenario (see Section~\ref{se:choosecosmology}) across a random  
    subset of $\mathcal{O}(20,000)$ models from above the $90$th percentile of fuzzy relic abundance in our ensemble of models.}
    \label{fig:prompt_reheating}
\end{figure}

In Fig. \ref{fig:prompt_reheating} we plot the distributions of the reheating temperatures $T_R$ and the initial misalignment tuning product $\delta$ for our chosen subset of the ensemble, split by number of heavy and light axions, respectively.

\begin{figure}[h!]
    \centering
    \includegraphics[width=0.9\linewidth]{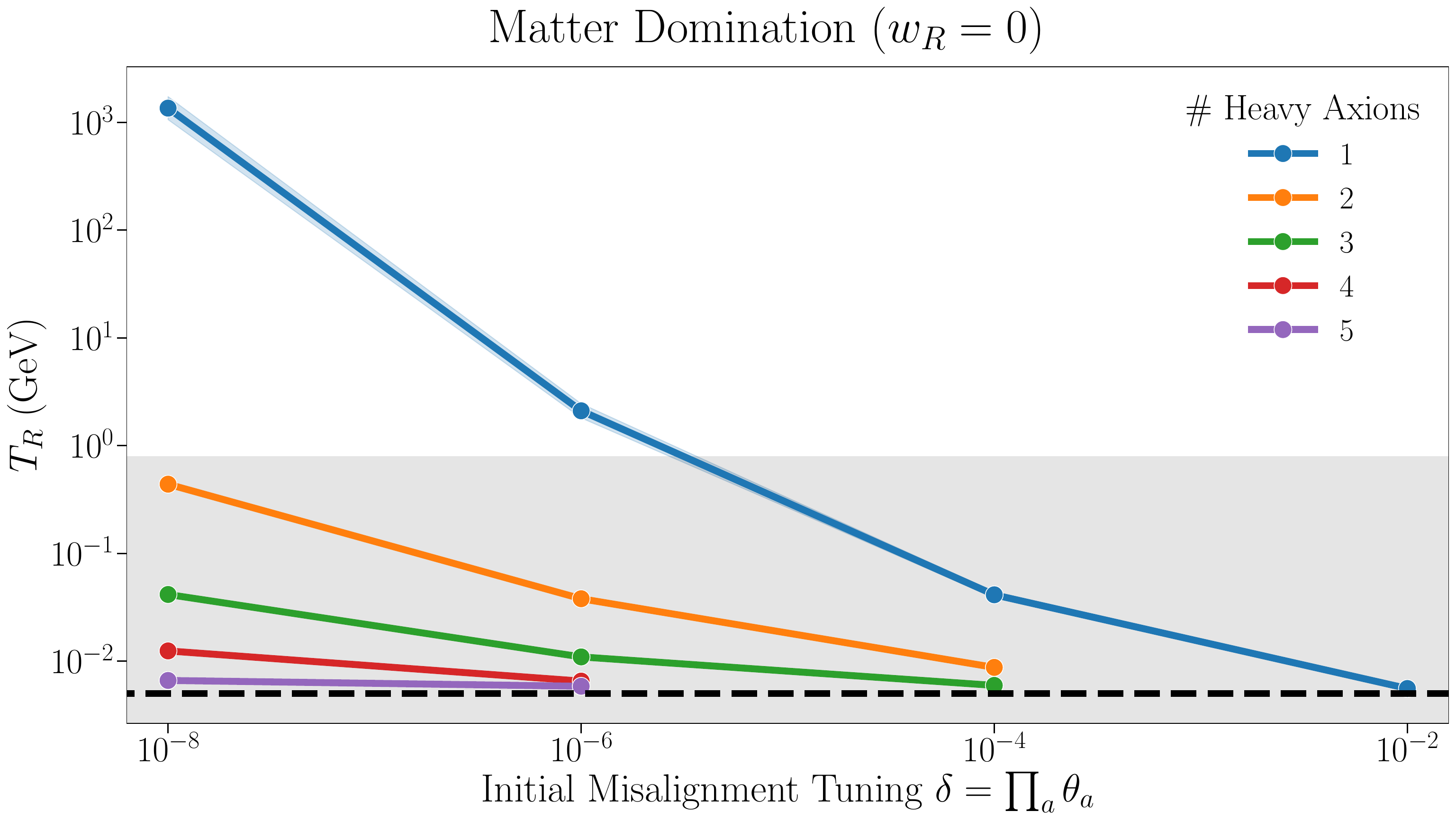}
    \caption{Average value of the free cosmological parameter $T_R$ as a function of the total initial misalignment tuning $\delta$ in the modulus domination ($w_R = 0$) scenario (see Section~\ref{se:choosecosmology}) across a random $\mathcal{O}(20,000)$ models 
    from above the $90$th percentile of fuzzy relic abundance in our ensemble of models. 
    $T_\mathrm{BBN}$ (dashed black) is overlaid, and the region with $T_R < T_{\mathrm{osc,QCD}}$ for the average QCD axion in our ensemble is shaded. Markers denote the sampled values of $\delta$ and error bands depict the standard error. For fixed values of $\delta$, most models fail to achieve the observed relic abundance for any $T_R > T_\mathrm{BBN}$: we omit such models and only report the statistics for those with phenomenologically viable $T_R$.}
    \label{fig:moduli_domination}
\end{figure}

In the modulus domination scenario, we consider varied fixed values of the initial misalignment tuning $\delta$ and for each fixed value $T_R$ is determined by enforcing the observed relic abundance. In this scenario, the characteristic $T_R$ for fixed $\delta$ is set by the number of heavy axions, as their abundance is now parametrically larger than that of the light axions. In Fig.~\ref{fig:moduli_domination} we plot contours expressing the average $T_R$ as a function of $\delta$ in the modulus domination scenario, differentiated by the number of heavy axions. For fixed $\delta$, we omit models for which there does not exist $T_R > T_{\rm BBN}$ yielding the observed relic density: we note that given sufficiently many heavy axions, all models are omitted for sufficiently large $\delta$. That is, in this subset, to ensure $T_R > T_{\rm BBN}$ one is obligated to take on a significant initial misalignment tuning penalty.

\newpage
\section{New fuzzy phenomenology}\label{sec:new_pheno}

In this section we provide new computations of two aspects of late Universe phenomenology associated with fuzzy axions. In Section \ref{sec:halo_simulation} we briefly discuss and provide 3D simulation results for three component wave DM (i.e. three fuzzy axions) with nearly degenerate masses, and highlight some of the key features of this case. In Section \ref{sec:fuzzy+darkphotons} we present new results of an analysis of CMB and galaxy survey data providing constraints on the dark radiation content in the presence of an ultralight axion.

\subsection{A mixed fuzzy axion halo}\label{sec:halo_simulation}

In this section, we briefly reiterate some of the main aspects related to DM halo phenomenology, in the context of multi-component wave dark matter~\cite{Amin:2022pzv,Jain:2022agt,Jain:2023ojg,Chen:2023bqy,Gosenca:2023yjc,vanDissel:2023vhu,Luu:2023dmi,Glennon:2023jsp,Huang:2022ffc,Street:2022nib}. We introduce Schr\"odinger (complex) fields $\Psi_a$ that are related to the corresponding (real) scalar classical fields $\phi_a$ by
\begin{align}
    \phi_a({\bm x},t) = \frac{1}{\sqrt{2m_a}}\,e^{-im_at}\,\Psi_a({\bm x},t) + \rm{c.c.}\,,
\end{align}
and where $\phi_a$ are minimally coupled to gravity. 
Since the DM is cold/non-relativistic, starting from the Klein-Gordon set plus Einstein's equations and upon taking the non-relativistic limit (which is to say that $\Psi_a$ has considerable support only over wave modes $k \ll m_a$~\cite{Marsh:2015wka,Namjoo:2017nia,Salehian:2020bon,Salehian:2021khb,Adshead:2021kvl,Jain:2021pnk}), we obtain the following set of Schr\"odinger equations coupled via the Newtonian gravitational potential $\Phi$:
\begin{equation}
\begin{split}
\label{eq:Schrodinger_Posisson}
    i\frac{\partial}{\partial t}\Psi_a &= -\frac{\nabla^2}{2m_a}\Psi_a + m_a\Phi\Psi_a\,,\\
    \nabla^2\Phi &= 4\pi G(\rho-\bar{\rho})\qquad{\rm where}\qquad \rho = \sum_{a} m_a|\Psi_a|^2\,, 
\end{split}
\end{equation}
and $\bar{\rho}$ is the spatially averaged energy density. Here, we worked only within a  Minkowski background and weak gravity setting, i.e. $g_{\mu\nu} = \eta_{\mu\nu} + h_{\mu\nu}$ with $h \ll 1$, and where $\Phi = h_{00}/2 = {\rm Tr}[h_{ij}]/2$. In order to include Hubble expansion, one can replace $\partial_t \rightarrow \partial_t + 3H/2$ along with $\nabla^2 \rightarrow a^{-2}(t)\nabla^2$ where $a(t)$ is the scale factor. However, in relevance to large scale structure formation in the late Universe, the Hubble horizon is much larger than the typical de Broglie scale associated with fuzzy dark matter particles. 

 \begin{figure*}[t]
    \centering
    \includegraphics[width=1\linewidth]{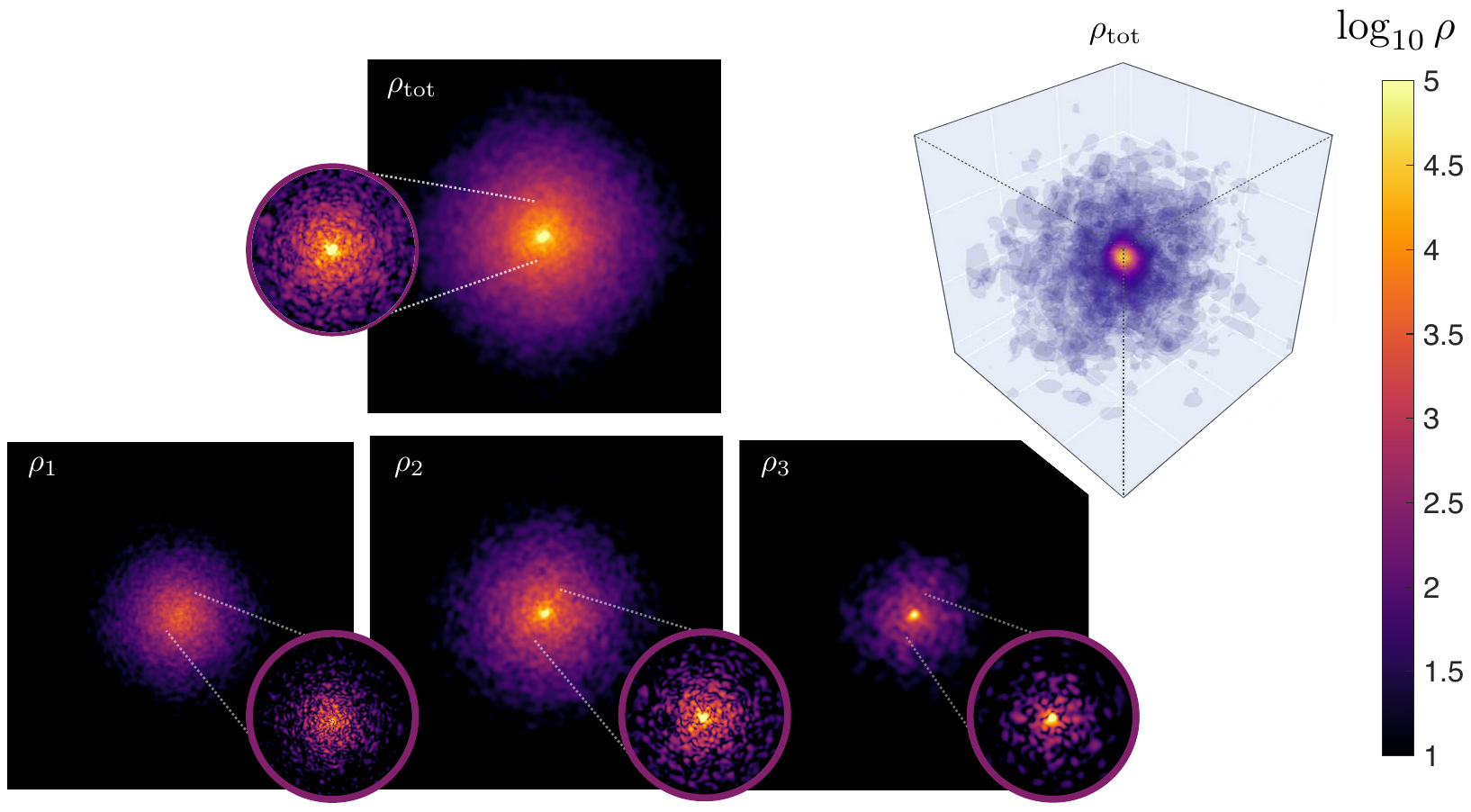}    
    \caption{Simulation snapshots of a DM halo composed of three fuzzy axion fields with mass ratios $\{1,0.75,0.5\}$.  
    The simulation uses a $192^3$ lattice with parameters $L_{\rm box} = 75\,{\rm kpc}\times (\tilde{M}\tilde{m}^2)^{-1}$, and $\bar{\rho}_{\rm tot} = 96\,M_{\odot}/{\rm kpc}^3 \times (\tilde{M}^4\tilde{m}^6)$. Square panels show the projected densities along the $z$-axis, while circular insets display densities in the $z=0$ slice. The reduced granularity in the interference pattern of the total density, compared to the individual densities, reflects smearing. Solitons are visible in the two lighter axion fields (second and third). The $3$D rendering of the total density is shown in the cube.}
\label{fig:halo_3comps}
\end{figure*}

As such, it is sufficient to work with Eqns.~\eqref{eq:Schrodinger_Posisson} in order to capture the non-linear gravitational dynamics and late time properties of DM halos. This Schr\"odinger-Poisson system has certain scaling symmetries, which significantly increase the generality of our results. Besides particle masses of all the different axions, there are equally many total mass scales in the problem, equal to the total masses contained within each axion field in a target volume of space (e.g. an isolated DM halo). Let us define a common particle mass scale $m$ with respect to which all the axion masses are defined, $m_a = \zeta_a m$. Let us also define a common total mass scale $M$, with respect to which the total mass within each axion field is expressed, $M_a = \eta_a M$. Then, the combined total mass from all axion fields is simply $M_{\rm tot} = M\sum_a\eta_a$. With these two mass scales $m$ and $M$, we rescale the space-time coordinates and the fields as
\begin{align}
\label{eq:rescaled_variables}
    t = \frac{\mpl^4}{M^2m^3}\tilde{t} \qquad {\bm x} = \frac{\mpl^2}{Mm^2}\tilde{\bm x} \qquad \Phi = \frac{M^2m^2}{\mpl^4}\tilde{\Phi} \qquad \Psi_a = \frac{M^2m^2\sqrt{m}}{\mpl^3}\,\tilde{\Psi}_a\,,
\end{align}
fetching the following dimensionless Schr\"odinger-Poisson system
\begin{align}
\label{eq:Schrodinger_Posisson_rescaled}
    i\frac{\partial}{\partial \tilde{t}}\tilde{\Psi}_a &= -\frac{1}{2\zeta_a}\tilde{\nabla}^2\tilde{\Psi}_a + \zeta_a\tilde{\Phi}\tilde{\Psi}_a\nonumber\\
    \tilde{\nabla}^2\tilde{\Phi} &= \frac{1}{2}(\tilde{\rho} - \bar{\tilde{\rho}})\qquad{\rm where}\qquad \tilde{\rho} = \sum_a\tilde{\rho}_a = \sum_{a} \zeta_a|\tilde{\Psi}_a|^2\,,
\end{align}
and $\int{\rm d}^3\tilde{x}\,\tilde{\rho}_a(\tilde{\bm x},\tilde{t}) = \eta_a$.
The scaling symmetry is now manifest via \eqref{eq:rescaled_variables}: if we have a solution set $\{\psi_a({\bm x}, t)\}$ for a system with axion particle masses $m_a = \zeta_a m$ and respective total masses $M_a = \eta_a M$, then $\{\psi^{(\beta\gamma)}_a({\bm x},t) = \beta^{5/2}\gamma^2\,\psi_a(\gamma\beta^2{\bm x}, \gamma^2\beta^3t)\}$ is the corresponding solution set for a system with axion masses $m_a = \beta\zeta_a m$ and respective total masses $M_a = \gamma\eta_a M$. For astrophysical purposes, we define the following parameters 
\begin{align}
    \tilde{m} := \frac{m}{2.3 \cdot 10^{-21}\,{\rm eV}}\,, \qquad \tilde{M} := \frac{M}{6.4 \cdot 10^3\,M_{\odot}}\,,
\end{align}
and evolve the coupled system of equations~\eqref{eq:Schrodinger_Posisson_rescaled} using a split-Fourier technique~\cite{Springel:2005mi,Schive:2009hw,Mocz:2015sda,Schwabe:2016rze,Mocz:2017wlg,Zhang:2016uiy,Edwards:2018ccc,Nori:2018hu,Jain:2022agt,Jain:2023qty}. We begin with random Gaussian initial conditions for different species such that the total simulation box size is larger than the Jeans scale for all species. That is, in terms of the physical scales, $L_{\rm box} > \ell_{J, a} := \bar{v}_a(\pi/G\bar{\rho})^{1/2}$ where $\bar{v}_a := \sqrt{\langle{\bm k}^2\rangle}/m_a$ 
is the typical `velocity' associated with fluctuations in the field $\Psi_a$. Soon after, we observe formation of halos in different regions of the simulation box, which then collapse under gravity (hierarchical structure formation). For a concrete example, we perform simulations for two different cases: one with only a single axion comprising all of the DM, and the second with three different axions comprising all of the DM, with not so disparate individual masses and DM fractions. We take $\{\zeta_a\} = \{1,0.75,0.5\}$, and $\{\eta_a\} = \{0.3,0.5,0.2\} \times 6328$ for the three axion case, while $\zeta = 1$ and $\eta = 6328$ for the single axion case. The multi-axion case takes a 30\%, 50\%, 20\% composition of the total relic abundance among fuzzy axions, with no significant cold component. Removing the cold component entirely is done for computational simplicity, and could be achieved in practice by further tuning of the QCD axion misalignment angle to $\mathcal{O}(10^{-3})$.

\begin{figure*}[t]
    \centering
    \includegraphics[width=1\linewidth]{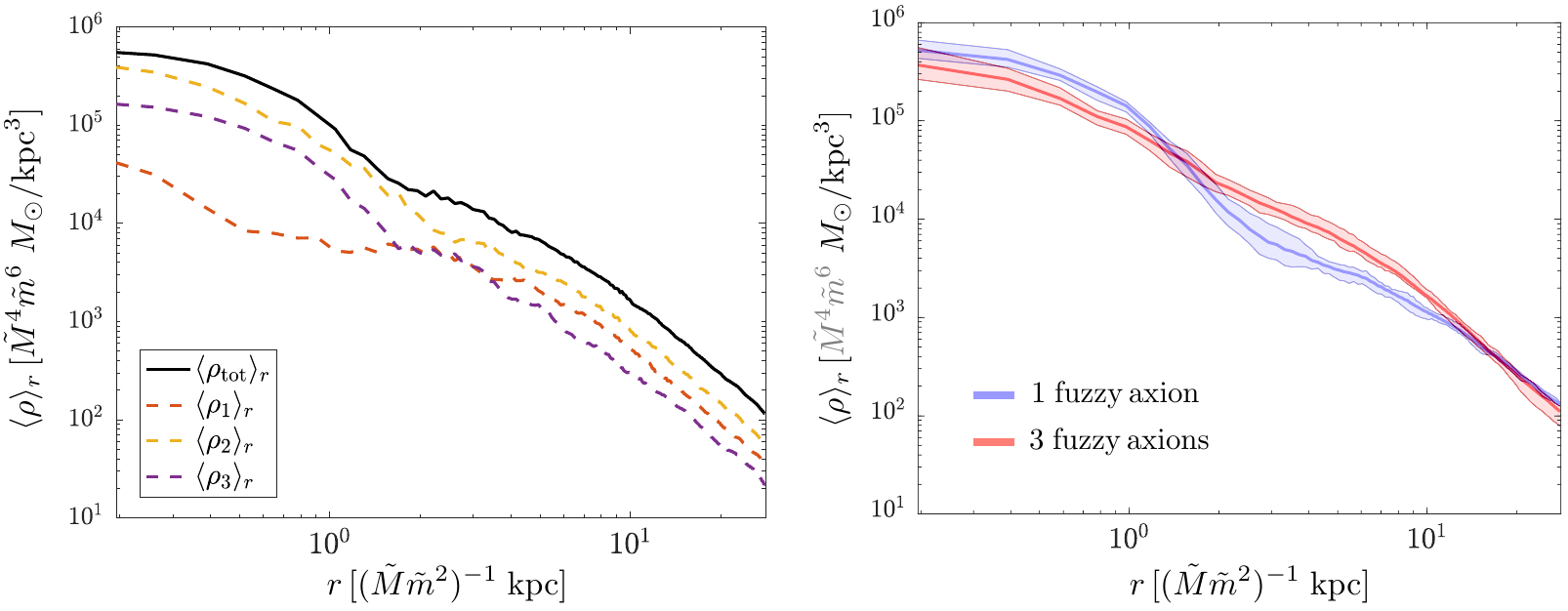}    
    \caption{Radially averaged mass densities as a function of radial distance from the center of the DM halo. The left panel shows the densities for the three individual axion fields and the total density from the simulation in Fig.~\ref{fig:halo_3comps}. The right panel compares the total mass density for 
    the 
    single fuzzy axion and three fuzzy axion cases, based on 10 simulations for each scenario. Shaded bands represent the range across simulations, while solid lines denote the averages.}
\label{fig:radial_profiles}
\end{figure*}

We highlight three key features of multi-component wave dark matter halos: (a) reduced interference effects relative to single-component halos~\cite{Amin:2022pzv,Gosenca:2023yjc,vanDissel:2023vhu}; (b) the emergence of nested solitons across distinct components~\cite{Jain:2023ojg,vanDissel:2023vhu,Luu:2023dmi}; and (c) a smoother core-to-halo transition in the radially averaged mass density profile~\cite{Amin:2022pzv,Jain:2023ojg,Gosenca:2023yjc}. In Fig.~\ref{fig:halo_3comps}, we provide density fields / halo structure obtained from a simulation of three fuzzy axions. Additionally, Fig.~\ref{fig:radial_profiles} provides radially averaged density profiles for this simulation, along with a comparison between single-component and three-component halos for an ensemble of $10$ simulations for each. Further details are available in the figure captions.
 
In the case where multiple axion masses are close together, reduced interference compared to a single axion case will change the heating and cooling effects caused by the turbulent halo environment~\cite{Hui:2016ltb,Bar-Or:2018pxz,El-Zant:2019ios}. In particular, the strongest constraints on fuzzy dark matter, which require $m_a\gtrsim 10^{-19}\text{ eV}$ if $\Omega_a/\Omega_d\approx 1$, arise from such heating effects on old star clusters in the Eridanus II~\cite{Marsh:2019bjr} and Segue I~\cite{Dalal:2022rmp} ultrafaint dwarf galaxies. Having multiple fields with masses near $10^{-19}\text{ eV}$ (such as Example \ref{ex:degen}) reduces interference as shown in Fig.~\ref{fig:halo_3comps} and would relax these constraints. Re-evaluating such constraints is beyond the scope of this work.

Fig.~\ref{fig:radial_profiles} shows that multiple fuzzy axions give rise to nested solitons. The core radius of each soliton remains on the same scale as the single axion case, and indeed the overall profile becomes slightly flatter. This result suggests that the constraint on the fuzzy axion mass $m_a\gtrsim 10^{-21}\text{ eV}$ if $\Omega_a/\Omega_d\approx 1$ from the cuspy inner profile of the Leo II dwarf spheroidal galaxy~\cite{Zimmermann:2024xvd} is not relaxed by including multiple near-degenerate axions.

\subsection{New CMB constraints on fuzzy axions and dark radiation}
\label{sec:fuzzy+darkphotons}

As discussed earlier, the orientifold projection of CYs in the KS dataset generically predicts the existence of dark photons. 
We do not assess at present whether these massless spin-1 degrees of freedom could acquire masses (through mechanisms such as the St\"uckelberg or Higgs mechanism) that could make them potential dark matter candidates. In any event, they may be coupled to the visible and/or dark sectors in the 4D effective theory, raising the possibility of their production in the early Universe as \emph{dark radiation}. 

An additional abundance of dark radiation in cosmology is captured by the parameter $N_{\rm ur}$, the number of ultrarelativistic species (also called $N_{\rm eff}$, the effective number of massless neutrinos). Previous studies have considered cosmological observables in the presence of varying $N_{\rm ur}$ and fuzzy axions~\cite{Marsh:2011bf,Hlozek:2016lzm} and found that it is possible for the two parameters, $N_{\rm ur}$ and $\Omega_a h^2$, to have non-trivial degeneracies.  
Refs.~\cite{Marsh:2011bf,Hlozek:2016lzm}, however, did not work with real cosmological data, but only with Fisher matrix forecasts for future data. No subsequent studies of ultralight axions using real data (e.g., Refs.~\cite{Hlozek:2014lca,Hlozek:2017zzf,Lague:2021frh}) have considered the case of simultaneous variation of $N_{\rm ur}$ and $\Omega_a h^2$. We carry out this analysis below, in order to ascertain whether there emerges a new maximum posterior with $\Omega_ah^2>0$ and $N_{\rm ur}>2$ (indicating a preference for the combined model), or if limits on $N_{\rm ur}$ are significantly tightened/weakened in the presence of fuzzy axions. 

\begin{figure*}[t]
    \centering
    \includegraphics[width=1\linewidth]{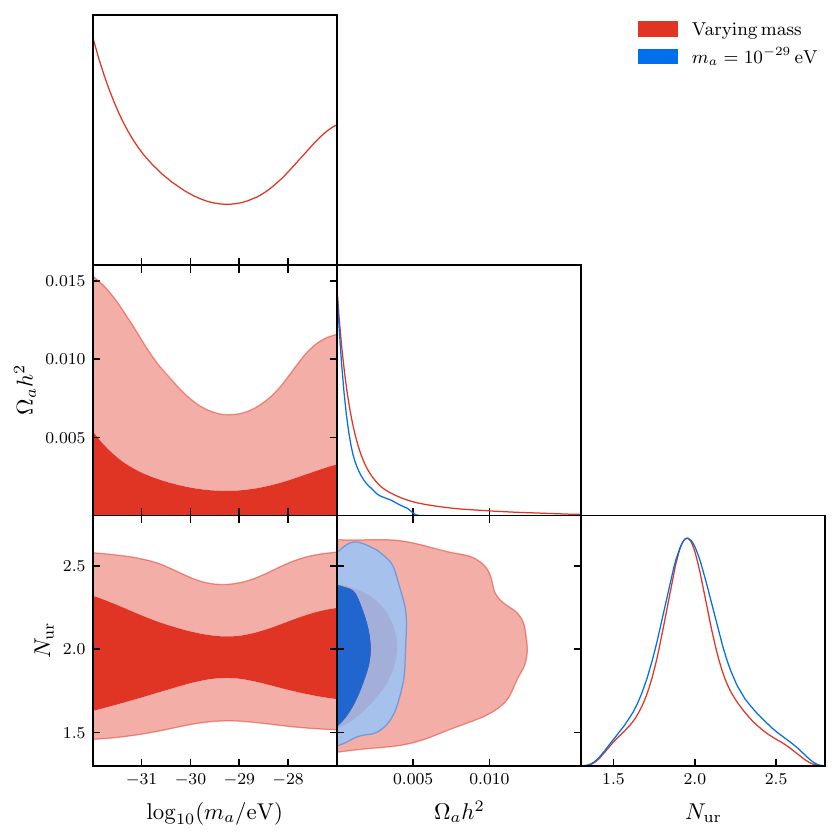}    
    \caption{Joint posteriors on axion and dark photon parameter space, derived from cosmological observables. $N_{\rm ur}$ is the number of ultrarelativistic species, assuming a single massive neutrino, which takes the value $N_{\rm ur}=2.04$ in the SM. The red contours show the results of an analysis where the axion mass is varied with a log-flat prior in the dark energy-like regime most strongly constrained by CMB data. The blue contours show an analysis with fixed axion mass. Additional cosmological and nuisance parameters appropriate to the used likelihoods are varied and marginalized, but we do not show the distributions as they are unchanged by the addition of axions. See the text for more details.}
\label{fig:N_ur_triangle}
\end{figure*}

We sample the posterior distribution of the axion mass, the axion energy density, and the number of ultrarelativistic species using a combination of \textit{Planck} 2018 cosmic microwave background temperature, polarization, and lensing data~\cite{Planck:2019nip}, BAO, 
and supernovae magnitudes from the Pantheon+ survey~\cite{Brout:2022vxf}, employing the inference software \textsc{MontePython}~\cite{Audren:2012wb} (see the documentation for details of the BAO likelihoods). We used the nested sampler \textsc{MultiNest}~\cite{Feroz:2008xx} and the Boltzmann solver \textsc{AxiCLASS}~\cite{Smith:2019ihp,Poulin:2018dzj} (based on \textsc{class}~\cite{2011JCAP...07..034B}), varying the full cosmological model and nuisance calibration parameters for \textit{Planck} temperature and polarization data, and the absolute supernova magnitude. 
We allow for one massive neutrino with a fixed baseline mass as determined by oscillation experiments. The massless neutrinos of the SM contribute $N_{\rm ur}=2.04$. We vary $N_{\rm ur}$ and $\Omega_a h^2$ with uniform priors. 

The \emph{Planck} CMB anisotropies provide the strongest constraint on fuzzy axions with $m_a<10^{-25}\text{ eV}$. We adopt a log-flat prior on $10^{-32}\leq m_a\leq 10^{-27}\text{ eV}$, placing axions in the `dark energy like' regime and facilitating efficient sampling of the joint posterior on $(m_a,\Omega_a h^2)$ without encountering difficult to sample changes in the degeneracy directions~\cite{Hlozek:2014lca}.  
The results are shown in red in Fig.~\ref{fig:N_ur_triangle}. The bottom right panel displays the joint posterior on $(\Omega_a h^2,N_{\rm ur})$ marginalized over the axion mass. We find no strong degeneracy between the axion density and the number of ultrarelativistic species.

To investigate this further, we conduct a second analysis with $m_a=10^{-29}\text{ eV}$. The results are shown in blue in Fig.~\ref{fig:N_ur_triangle}. As in the previous case, we find no strong degeneracy between the axion density and the number of ultrarelativistic species. The constraint on $N_{\rm ur}$ at $\Omega_a h^2=0$ corresponding to the CDM limit remains unchanged in the presence of a non-zero axion density. This result contrasts with the approximate Fisher matrix forecasts in Refs.~\cite{Marsh:2011bf,Hlozek:2016lzm}, which observed a degeneracy between DE-like axions and ultrarelativistic species. 

This discrepancy may stem from several factors. 
The Fisher matrix forecasts assumed a fiducial cosmology with a non-zero axion fraction being the preferred maximum posterior. It is a fact that Fisher matrix degeneracies can change when the fiducial value changes, for example to zero axion fraction as preferred by the maximum posterior in our analysis of real data. Fisher forecasts also assume Gaussianity of posteriors, which is never exactly respected by real data, and is furthermore violated for a one-sided posterior peaked at zero.

The analysis in this section demonstrates that the possible presence of fuzzy axions in the Universe does not affect the inferred value of $N_{\rm ur}$, leaving it consistent with the SM within 95\% confidence limits of order 0.3-0.5 depending on the likelihoods used~\cite{Planck:2018vyg}. Likewise, allowing for additional ultrarelativistic species does not change the strength of cosmological limits on the fuzzy axion density or lead to a preference for non-zero density in DE-like axions. We conclude that it will therefore be relatively safe to separately investigate the possible constraints imposed on string theory models with non-zero $h^{2,1}_+$, which may come accompanied by fuzzy axions, without having to consider the effect of one on the other in leading accuracy. If the dark photons are able to come into thermal equilibrium, or are non-thermally produced in any significant number leading to large $N_{\rm ur}\gtrsim 2.5$, this can be used to further constrain the landscape.

\section{Discussion}
\label{sec:discussion}

\subsection{Comments on cosmology}

In this work, we have taken a simplified approach towards the cosmological history of the Universe prior to reheating, considering just the matter-dominated case, and the prompt reheating case, with an instantaneous transition to the standard radiation-dominated cosmology at $T_{R}$. The prompt reheating case is very simple, and isocurvature bounds on the QCD axion alone favor $H_{I}\lesssim 10^9\text{ GeV}$~\cite{WMAP:2008lyn,Hertzberg:2008wr,Planck:2018jri}.\footnote{We recall that in our models the QCD axion is a closed string axion, which is necessarily in the `pre-inflation' scenario~\cite{Reece:2024wrn} (assuming inflation is a 4D EFT), and thus it necessarily has isocurvature perturbations.} Taking $H_{I}$ significantly below this would seem to involve unnecessary fine-tuning of the inflationary potential (see Ref.~\cite{Marsh:2019bjr} and references therein for discussions of ultra-low-scale inflation).

The $w_R=0$  
reheating scenario comes with its own novel predictions, such as early structure formation~\cite{Erickcek:2011us,Eggemeier:2021smj}. 
If the matter dominated epoch ends with perturbative reheating, then there is necessarily a small out-of-equilibrium radiation bath present as well. In this case, it is unclear what effect this will have on the QCD axion effective potential, since $m(T)$ is normally computed (via instantons or on the lattice) assuming a QCD bath in thermal equilibrium at temperature $T$. This is particularly relevant when $T_{R} \lesssim T_{\rm osc}\approx 0.2\,\text{GeV}$.

Baryogenesis without a thermal bath of sufficiently high temperature above the electroweak scale would also appear to be difficult (direct decays being one possibility, e.g.~\cite{Dimopoulos:1987rk,Elor:2018twp,Asaka:2019ocw}). Thus, in a complete model, there will be tensions on the one hand from the axion abundance driving $H_{I}$ and $T_{R}$ low, while other observational and model-building pressures drive these parameters up. 

Our cosmological model for simplicity considered only the vacuum misalignment population of axions, since this is the relevant channel for fuzzy DM. Freeze-out/freeze-in production of axions was considered in Ref.~\cite{Gendler:2023kjt}. At the low values of $h^{1,1}$ and consequent high values of $f_a$ in the present work, thermal production of axions is likely to be suppressed.  

There are still other DM candidates beyond axions which we have made no attempt to model. If the gravitino is also the lightest supersymmetric particle and is therefore a stable DM candidate, then thermally produced gravitinos~\cite{Bolz:2000fu} could have an overabundance problem if the reheating temperature required to dilute the abundance of the lightest heavy axion is above the freeze-out temperature of the gravitino. Another possibility is that misaligned axions could decay to dark non-Abelian sectors, producing confined DM (e.g. glueballs) at the expense of axions. We have not modelled dark gauge sectors in this work, although such sectors are required by tadpole cancellation in a CY orientifold, and are abundantly present in F-theory. For discussion of dark matter from confining dark sectors, see e.g. Refs.~\cite{Boddy:2014yra,Soni:2016gzf,Halverson:2016nfq,Acharya:2017szw,Hertzberg:2019bvt,Hertzberg:2019prp}.

Our broad conclusion of the need for a modified thermal history and relatively low temperature reheating is consistent with the original assumptions in Ref.~\cite{Arvanitaki:2009fg}, which prefer prompt reheating with $H_{I}\sim 0.1\text{ GeV}\Rightarrow T_{R}\sim 10^8 \text{ GeV}$. It would be interesting to thoroughly investigate how overabundance problems may lead to a preference for a particular reheating temperature or reheating equation of state in string theory. In the context of a detailed model for reheating driven by a modulus, as we have discussed, it may be possible to connect an inference on $T_{R}$ to an inference on the SUSY scale~\cite{Iliesiu:2013rqa,Coughlan:1983ci,Acharya:2008bk}.\footnote{We thank Bobby Acharya for stressing to us the possible importance of this point.}

\subsection{PQ quality}\label{sec:PQq}
The Peccei-Quinn~\cite{Peccei:1977hh} (PQ) quality problem is induced by explicit CP-violating terms that can shift the QCD axion minimum away from  the value $\theta_a\approx 0$ required by the neutron electric dipole moment measurement~\cite{Abel:2020pzs}. In the string theory context, such terms arise from stringy instantons, as well as high-energy QCD instantons (see \cite{Demirtas:2021gsq} for details). Compliance with bounds on the neutron electric dipole moment entails: 
\begin{align}
    \Delta\theta_{\rm QCD}+\Delta\theta_{\rm stringy}\lesssim 10^{-10}.
    \end{align} 
Thus, for the QCD axion to have high enough PQ quality to solve the strong CP problem, the scale of stringy CP-violating terms must be much lower than the contribution of non-perturbative QCD effects to the potential.

The largest contribution generically comes from high-energy QCD instanton effects, which can break CP and impact the quality of the QCD axion. However,~\cite{Demirtas:2021gsq} showed that under reasonable UV assumptions, such a contribution shifts the minimum of the QCD axion potential by
\begin{equation}
    \Delta \theta_{\rm QCD}\sim 10^{-12}\left(\frac{1\,\rm TeV}{M_{\rm SUSY}}\right)^3\,,
\end{equation}
where $M_{\rm SUSY}$ is the SUSY-breaking scale. Therefore, for $M_{\rm SUSY}>1$ TeV, $\Delta \theta_{\rm QCD}$ is within the experimental bound.

We turn now to the stringy instanton contributions,  $\Delta\theta_{\rm stringy}$. These are given by the potentials generated by Euclidean D3-branes wrapping 
divisors: here, following \cite{Demirtas:2021gsq}, we consider superpotential terms from prime toric divisors. 
To assess the magnitude of this effect, we compute
\begin{align}
    \Delta \theta_{\mathrm{stringy}} = \frac{\Lambda^4_{\mathrm{PQ}}}{\chi(0)_{\mathrm{QCD}}}\,,
\end{align}
where $\Lambda^4_{\mathrm{PQ}}$ is the largest CP-violating instanton scale.

One finding of the present paper is that models exhibiting a fuzzy axion also favor QCD axions with high PQ quality. 
The reason is that we find a typical ordering of scales in our models:
\begin{equation}
\Lambda_{\rm PQ}^4 <\Lambda_{\rm lightest \, axion}^4\leq \Lambda_{\rm fuzzy}^4\,,
\end{equation}
which means that
\begin{equation}\label{eq:PQscenario1}
    \frac{\Lambda^4_{\rm PQ}}{\chi(0)_{\mathrm{QCD}}} \lesssim 10^{-10} \,\,\Leftrightarrow\,\,  \frac{m_{\rm lightest\, axion}}{m_\mathrm{QCD}} \lesssim 10^{-5}\,,
\end{equation}
assuming roughly comparable decay constants. We see that 
when \eqref{eq:PQscenario1} holds, 
the quality problem is always solved by requiring a fuzzy axion. On the other hand, if $\Lambda_{\mathrm{PQ}}^4>\Lambda_{\rm lightest \, axion}^4$, it is not guaranteed that the quality problem is solved. 

\begin{figure}
    \centering
    \includegraphics[width=0.9\linewidth]{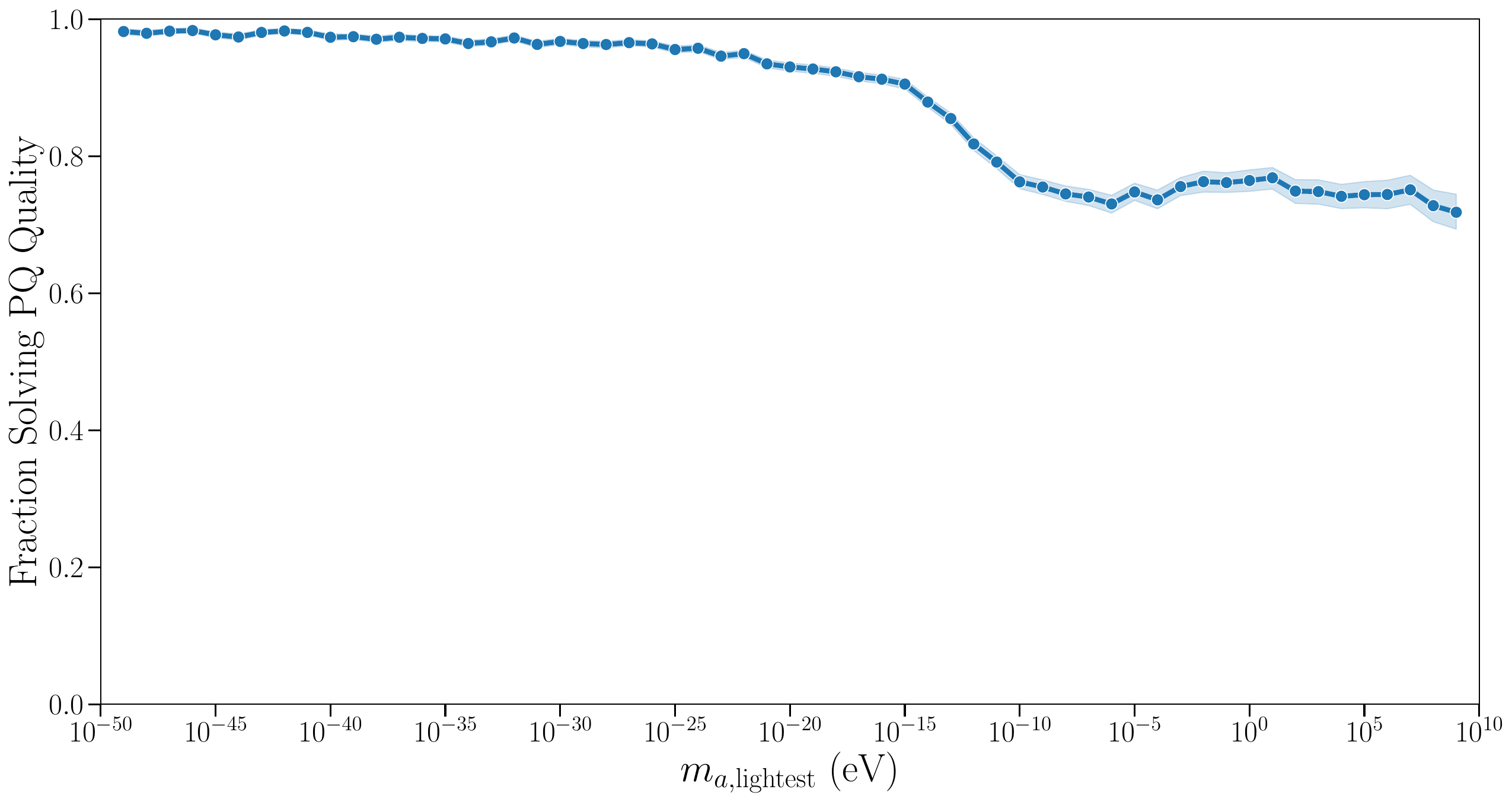}    
    \caption{Fraction of models with solved PQ quality as a function of the lightest axion mass. These data correspond to 2000 random CYs from our ensemble, sampled at 10 uniformly spaced values of the QCD divisor volume.}
\label{fig:pq_quality}
\end{figure}

In Fig.~\ref{fig:pq_quality}, we exhibit this correlation between the lightest axion mass and PQ quality. To achieve this, we have taken $2000$ random CYs from our ensemble and, for each, computed axion observables for $10$ uniformly sampled values for the volume of the QCD divisor (thereby no longer requiring an axion with mass $10^{-18}$ eV). We then bin the models by rounding $\log_{10}(m_a/\mathrm{eV})$ to the nearest integer and plotting the fraction of models in each mass bin 
that solve PQ quality, shading by the standard error. As expected from \eqref{eq:PQscenario1},  
we see a transition in PQ quality once the lightest axion crosses above the critical mass of 
$10^{-13}$ eV, which aligns with the expected QCD axion mass $m_\mathrm{QCD}\sim 10^{-8}$ eV for $f_a\sim 10^{15}$ GeV in the canonical band.

\subsection{Fuzzy axions and the Weak Gravity Conjecture}
In this section, we aim to clarify the relationship of fuzzy axion dark matter to the axionic Weak Gravity Conjecture 
and compare our results to the prior 
literature.

The WGC for axions \cite{Arkani-Hamed:2006emk} 
is the claim\footnote{See e.g.~\cite{Palti:2019pca} for a review of related conjectures.} that in the theory of an axion with decay constant $f_a$,
there should exist an instanton with action $S_{\text{inst}}$ obeying
\begin{align}
    S_{\text{inst}} f_{a} \leq c M_{\mathrm{pl}}\,,
    \label{eq:axionwgc}
\end{align}
where $c$ is an unknown $\mathcal{O}(1)$ number.  

It was suggested in \cite{Alonso:2017avz,Hebecker:2018ofv, Cicoli:2021gss} that an $\mathcal{O}(1)$ abundance of fuzzy axion dark matter may violate the bound \eqref{eq:axionwgc} in the case that $c=1$. The reason was as follows: suppose that we have an axion Lagrangian of the form 
\begin{align}
    \mathcal{L}_{\text{ax}} = -\frac{1}{2}  (\partial a)^2 -\mpl^4 B e^{-S_{\text{inst}}} \cos (a/f_a)\,, 
    \label{eq:axionLT}
\end{align} where $B$ is a constant.
Using \eqref{eq:axionLT}, the conjecture \eqref{eq:axionwgc} can be rewritten as
\begin{align}
    \log \left(\frac{B M_{\text{pl}}^4}{m^2 f_a^2} \right) f_a \leq c M_{\text{pl}}\,.
    \label{eq:axion_inequality}
\end{align}
The abundance of misalignment-produced QCD axion dark matter is given by \eqref{eqn:fuzzy_relic_abundance}. For the reference values  $\theta_a=1$, $m_a = 10^{-22}$ eV, and $\Omega_a h^2 = 0.12$,  one finds a decay constant of $f_a=8\cdot 10^{16} \, \text{GeV}$. Using these  values in \eqref{eq:axion_inequality} and temporarily taking 
$B=1$, we have
\begin{align}
    S_{\text{inst}} f_a = 7.83 M_{\text{pl}}\,.
    \label{eq:Sfvalue}
\end{align} 
Thus one finds that 
an ultralight axion with $m_a=10^{-22}\text{ eV}$ and $\theta_a=1$ that constitutes the entirety of the observed dark matter abundance would violate the $c=1$ axionic WGC (for $B=1$).

However, there are ambiguities in this conclusion that should be mentioned.
First, the value of $c$ in \eqref{eq:axionwgc} has not been completely 
established (however, see \cite{Harlow:2022ich}). 
Moreover, there is an inherent ambiguity in applying the axionic WGC, at least as currently formulated, to the theory given in \eqref{eq:axionLT}. The axion WGC has been presented as a statement about the relationship between axion field ranges and instanton actions. However, at the level of the low energy theory \eqref{eq:axionLT}, low-energy experiments cannot distinguish between the `bare' instanton action $S_{\text{inst}}$ and an `effective' instanton action  $S_{\mathrm{eff}}$ that includes the prefactor:
\begin{align}
    S_{\mathrm{eff}} \coloneqq S_{\text{inst}}-\log(B)\,.
\end{align}
Because the axionic WGC lacks a sharp bottom-up motivation independent of being a generalization of the (much better understood) WGC for black holes,
it is not clear to us what the correct interpretation is in the case that $B$ is not $\mathcal{O}(1)$ (as it often is not, for example in cases where $B$ contains a factor of the flux superpotential, $W_0$).

Even if one takes $c=B=1$, physical parameters can be tuned so that \eqref{eq:axion_inequality} is satisfied.  For example, with $m_a = 10^{-18}\text{ eV}$, $\theta_a=1$ and $B=1$ one finds $S_{\mathrm{inst}} f_a = 0.74 \ M_{\text{pl}}$. Likewise, tuning $\theta_a \approx \pi$ can produce anharmonic corrections to \eqref{eqn:fuzzy_relic_abundance} that increase the abundance. It is also worth noting that current bounds on fuzzy dark matter (Figure~\ref{fig:population_scatter} and e.g. \cite{Zimmermann:2024xvd}) do not actually allow fuzzy axions to compose $100\%$ of the dark matter abundance at $m_a=10^{-22}\text{ eV}$. 

In any event, our computation of the instanton actions and axion decay constants is quite definite, at the level of the approximations explained in Section \ref{sec:stsetup}.  Our results could therefore be used to test any version of the axionic WGC that is sufficiently definite to be falsifiable.
We leave this as a task for the future.

\subsection{Dark photons}\label{sec:darkphotons}

Dark photons are an intriguing possibility arising in our compactifications.  In this section we make a few preliminary comments, deferring a complete treatment to a future work.

We explained in Section \ref{sec:CYandOrientifold}
that
the Hodge number $h^{2,1}_+$ counts the number of U(1) gauge fields arising from closed string vector multiplets.  
These gauge fields can be thought of as dark photons from the moduli sector, and are distinct from possible dark photons from hidden sector gauge groups on D7-branes or D3-branes.
We will not consider dark photons from D-branes, and so in the remainder we use the term
`dark photons' exclusively for U(1) gauge fields counted by $h^{2,1}_+$.
 
In the KS database, polytopes with low $h^{1,1}$ generally have high $h^{2,1}$, and 
moreover we have found that orientifold projections inherited from the ambient variety maintain this inverse correlation at the level of $h^{1,1}_+$ and $h^{2,1}_+$: see Fig.~\ref{fig:h21trend}. In particular, our 
ensemble is dominated by those at $h^{1,1} = 7$ (see \Cref{tab:ScanData}) and the median $h^{2,1}_+$ value for orientifolds of such geometries is $22$. 
We thus find that models with low $h^{1,1}_+$ --- the models in which $C_4$ axions can most readily have large enough $f_a$ to produce fuzzy dark matter, and hence the focus of the present paper ---  generically include a large number of dark photons.
  
The couplings of dark photons in string theory are not 
completely understood (see, however, \cite{Plauschinn:2008yd,Goodsell:2009xc,Cicoli:2011yh, Camara:2011jg,Reece:2018zvv}).  
An intriguing direction for future work is the study of mechanisms that could generate masses for the dark photons, as well as potential St\"uckelberg couplings between dark photons and axions. 
In the remainder of this section we will \emph{assume} that some of the dark photons remain massless, and we comment on some phenomena that result. 

\subsubsection*{Gravitational Production of Dark Spin-1 Particles}

We first consider purely gravitational production of dark massless spin-1 particles, i.e.~production through interactions mediated by the graviton. At leading order, the relevant interaction vertex is $h_{\mu\nu} T^{\mu\nu}/\mpl$, where $h_{\mu\nu}$ represents the graviton and $T^{\mu\nu}$ is the energy-momentum tensor. 
Here, $h_{\mu\nu}$ is normalized to have mass dimension one, making it consistent with standard conventions in perturbative gravity. This interaction facilitates processes like ${\rm SM} + {\rm SM} \rightarrow \gamma_D + \gamma_D$, where   $\gamma_D$ is the dark photon. The rate of such processes can be estimated as $\Gamma \sim g\times(k^2/\mpl^2)^2\times k/12\pi^3$, where $g$ represents the total number of SM degrees of freedom, $k$ is the typical wave-number or energy scale of particles in the SM thermal bath, and $k \sim T$, the temperature of the bath. The ratio of this production rate to the Hubble rate $H$ is given by  
\begin{align}
    \frac{\Gamma}{H} \sim g\,10^{-13} \left(\frac{T}{10^{15}\,{\rm GeV}}\right)^3.
\end{align}
The energy density of dark photons produced via this mechanism can be roughly characterized as $\rho \sim \rho_{\rm eq}\times(\Gamma/H)$ where $\rho_{\rm eq}$ is the equilibrium energy density. Even for a reheating temperature as high as $10^{15}$ GeV (close to the maximum value allowed by constraints on the inflationary tensor to scalar ratio), the dark photon energy density is negligibly small, giving $N_{\rm ur} \sim 10^{-13}$.

\subsubsection*{Constraints from Reheating Mechanisms}

The mechanism of reheating has significant implications for the production of dark spin-1 particles. Consider scenarios where the inflaton is coupled to the SM sector via renormalizable operators.  
For example, even if the inflaton $\phi$ is a singlet under all SM gauge groups, there can be renormalizable inflaton-SM couplings like $\phi H^2$ or $\phi^2 H^2$, where $H$ is the SM Higgs. Couplings to the dark photons, however, could arise only through non-renormalizable operators, such as dimension-5 interactions of type $\sim \phi FF/M$ or $\sim \phi F\tilde{F}/M$, where $F$ is the field strength of the dark photon and $M$ is a high mass scale.  
Such a setup can even be technically natural within the dark sector. If the inflaton has direct renormalizable couplings to the SM but only higher-dimensional couplings to the dark sector, as above, the reheating process preferentially populates the SM bath~\cite{Hertzberg:2019prp}. This minimizes potential issues with the effective number of ultrarelativistic species, $N_{\rm ur}$, which could arise if the inflaton populated the dark sector too efficiently. On the other hand, with comparable scales for the dimension-5 operators coupling $\phi$ to the SM photons and to the dark photons (without any renormalizable couplings as above), a significant contribution to $N_{\rm ur}$ might emerge. As we showed in Section~\ref{sec:fuzzy+darkphotons}, this is strongly constrained,  even accounting for possible degeneracies with fuzzy axions.

In scenarios where the dark photons thermalize with the SM, or are non-thermally produced in sufficient quantities such that $N_{\rm ur} \gtrsim 2$, they could impose additional constraints on the high-energy landscape. However, careful model building can mitigate these concerns, allowing for consistent predictions of both $N_{\rm ur}$ and dark sector abundances~\cite{Adshead:2016xxj}.

\section{Conclusions}\label{sec:conclusions}

We have investigated the prevalence and the phenomenology of fuzzy axion dark matter in compactifications of type IIB string theory on orientifolds of CY hypersurfaces.  
Our main finding is that a detectable relic abundance of fuzzy axion dark matter can indeed occur in this setting, provided that the total number of axions is sufficiently small ($h^{1,1} \lesssim 10$), but it is generally accompanied by a potential overabundance of heavier axion dark matter, and possibly also by dark photons.  

Without any remedy, the resulting models are excluded by the heavy axion overabundance.
However, we showed that fine-tuning initial displacements, and/or modifying the reheating epoch, can reduce the abundance of heavy axions to acceptable levels and produce fuzzy dark matter models that are allowed by present constraints and can be tested by future measurements. 

We also showed  
that the issue of heavy axion overproduction
can be circumvented  by a geometric mechanism,
in compactifications with volume forms that factorize as in 
\eqref{eq:fibervolume}, which can occur in cases with K3 fibrations.
We gave an explicit example with decay constant hierarchies that are sufficient to avoid overproduction.
This geometric mechanism is possible only along special loci in moduli space, and in a region that is atypical according to the Weil-Petersson measure.  
 
Fig.~\ref{fig:population_scatter} shows a key result of this work: string theory targets for ultralight axion cosmology.  
Out of our $1.8$ million
type IIB Calabi-Yau hypersurface models with $N \le 7$ axions,
roughly the top 0.1\% 
give rise to fuzzy axion 
dark matter abundances
that may be accessible to future observations.\footnote{The fraction of accessible models increases drastically if $\theta_a\approx 3$.} Thus, \emph{detection of fuzzy axions from string theory is possible}, and could inform us about our place in the landscape. Furthermore, we showed
that such 
models could also include a birefringent axion with a possibly detectable large scale structure signal, as well as a detectable QCD axion. The present work has thus made significant steps forward in the program of using axion physics to determine `cohomologies from cosmology'~\cite{Arvanitaki:2009fg}.

The heavy axions in our models typically impose constraints on cosmological parameters such
as 
$w_R$, $T_R$, $H_{I}$, and $\theta_a$. The cosmology we found for each model is specific to the particular location in K\"ahler moduli space: in our ensemble we considered only models along a fixed ray giving a fuzzy axion and acceptable UV gauge couplings. 
Upon sampling away from this ray in the K\"ahler cone, the axion masses redistribute themselves, but maintain similar $f_a$. Thus, our findings about reheating and fine-tuning (Figs.~\ref{fig:prompt_reheating},~\ref{fig:moduli_domination}) should be seen as indicative rather than definitive. Nonetheless, the conclusion that  an abundance of fuzzy axion dark matter generically demands contrived
reheating, 
and/or fine tuning of initial displacements --- which we have for the first time established systematically  in a large ensemble --- seems robust and of significance.

In this work we 
deployed for the first time in a large ensemble
the orientifold algorithm that was
devised by Moritz in \cite{Moritz:2023jdb}.
We systematically constructed orientifolds inherited from the ambient variety for \emph{all} CYs with $2\leq h^{1,1}\leq 7$ that arise from triangulations of favorable polytopes in the KS database~\cite{Kreuzer:2000xy}. 
In this sense, our dataset for finding fuzzy axions was  topologically exhaustive.  
We found that for orientifolds in this range of Hodge numbers,
$h^{2,1}_+$ is generally non-zero, and is often large (see Fig.~\ref{fig:h21trend}). 
Thus, axions in this corner of the landscape are often accompanied by dark photons.

There are a number of important directions for future work.
We intend to investigate cosmological axion constraints across K\"ahler moduli space and for $h^{1,1}>7$, and also include the dynamics of moduli stabilization.  Our analysis was restricted to axions from the Ramond-Ramond four-form $C_4$, but one should also consider orientifolds with $h^{1,1}_->0$, in which there are also axions from the two-forms $B_2$ and $C_2$.   We remarked that dark photons are prevalent in our examples, but we deferred the task of characterizing the masses, couplings, and cosmological abundance of the dark photons.

Self-interactions of axions, which we have omitted in this work, present exciting opportunities for new physics, such as their role in the reshuffling of axions’ zero mode energy densities \cite{Cyncynates:2021xzw,Murai:2023xjn}, enabling production even of photophilic axions from other axions present in the cosmological bath \cite{Jain:2024dtw}, nucleation of solitons in dark matter halos \cite{Jain:2023tsr}, and triggering of Axinovae events \cite{Levkov:2016rkk,Helfer:2016ljl}.  Characterizing such phenomena for string theory axions would be worthwhile.

Finally, our treatment of the visible sector was very limited.
An essential task for the future is to engineer explicit and fully realistic visible sectors
in this class of compactifications, and understand in detail the couplings of axions to the Standard Model,
which could provide a crucial observational handle on these scenarios.

\acknowledgments

We are indebted to Michele Cicoli for suggesting a mechanism to find examples with hierarchical decay constants.
We thank Jim Halverson, Jacob M.~Leedom, Jakob Moritz, Alexander Westphal, and Harrison Winch for useful discussions. ES is especially grateful for conversations with Sebastian Vander Ploeg Fallon regarding numerical calculations of axion observables. 
We are grateful to several institutions for their  hospitality during the course of this work: Cornell U.~(FC, MJ, and NR), King's College London (ES), and
Harvard U.~and the Swampland Initiative (NR).
DJEM is supported by an Ernest Rutherford Fellowship from the Science and Technologies Facilities Council (Grant No. ST/T004037/1). DJEM, FC, MJ, and LM are supported by a Leverhulme Trust Research Project (Grant No. RPG-2022-145). The Dunlap Institute is funded through an endowment established by the David Dunlap family and the University of Toronto. The work of NG
was supported in part by a grant from the Simons
Foundation (602883,CV), the DellaPietra Foundation, and by the NSF grant PHY-2013858. The research of LM and ES is supported in part by NSF grant PHY-2309456. This material is based upon work supported by the National Science Foundation Graduate Research Fellowship under Grant No. 2139899. NR is supported by a Leverhulme Trust Research Project Grant RPG-2021-423. This article is based upon work from COST Action COSMIC WISPers CA21106, supported by COST (European Cooperation in Science and Technology). This work made use of the open source software 
\textsc{CYTools}~\cite{Demirtas:2022hqf},
\textsc{GetDist}~\cite{Lewis:2019xzd},
\textsc{jax}~\cite{jax2018github},  
\textsc{matplotlib}~\cite{Hunter:2007},  
\textsc{numpy}~\cite{ harris2020array1},
\textsc{optax}~\cite{deepmind2020jax}, 
\textsc{scikit-learn}~\cite{scikit-learn},
\textsc{scipy}~\cite{2020SciPy-NMeth1},
and \textsc{seaborn}~\cite{Waskom2021}.  

\bibliographystyle{JHEP}
\bibliography{refs}

\end{document}